\documentclass[aps, prd, reprint,10pt, notitlepage, a4paper,floats, floaqutfix,amsmath, amssymb, amsfonts,superscriptaddress,showpacs, showkeys,nofootinbib,longbibliography]{revtex4-2}

\pdfoutput=1

\usepackage{epsfig}
\usepackage{graphics}
\usepackage{graphicx}
\usepackage{amsmath,amsfonts,amssymb,amsthm}
\usepackage[usenames,dvipsnames]{xcolor}
\usepackage{wasysym}
\usepackage{times}
\usepackage{mathptmx}
\usepackage{gensymb}
\usepackage{appendix}
\usepackage{listings}
\usepackage{url}
\usepackage[normalem]{ulem}
\usepackage{alltt}
\usepackage[colorlinks]{hyperref}
\usepackage{cleveref}
\usepackage{longtable}
\usepackage{enumitem}
\setlist{nosep}
\usepackage{color}
\usepackage{calc}
\usepackage{tensor}
\usepackage{bm}
\usepackage{times}
\usepackage{multirow}
\usepackage[varg]{txfonts}
\usepackage{float}
\usepackage{dcolumn}
\usepackage[nolist,nohyperlinks]{acronym}
\usepackage{xspace}
\usepackage[english]{babel}
\usepackage[abs]{overpic}
\usepackage{pict2e}
\usepackage[caption=false]{subfig} 
\allowdisplaybreaks[1]
\usepackage[utf8]{inputenc}
\usepackage{gensymb}
\usepackage{bm}
\usepackage{stackengine}
\usepackage{boldline,multirow}
\usepackage{braket}
\usepackage{longtable}
\usepackage{makecell}

\usepackage{tabularx}
\usepackage{rotating}
\usepackage{booktabs}

\graphicspath{{pics/}}

\newcommand{\AEI}{Max Planck Institute for Gravitational Physics (Albert Einstein Institute), Am M\"uhlenberg 1, Potsdam 14476, Germany}
\newcommand{\NBIA}{Niels Bohr International Academy, Niels Bohr Institute, Blegdamsvej 17, 2100 Copenhagen, Denmark}
\newcommand{\Maryland}{Department of Physics, University of Maryland, College Park, MD 20742, USA}
\newcommand{\PI}{Perimeter Institute for Theoretical Physics, 31 Caroline Street North, Waterloo, ON N2L 2Y5, Canada}
\newcommand{\URI}{Department of Physics, East Hall, University of Rhode Island, Kingston, RI 02881, USA}
\newcommand{\URICCR}{Center for Computational Research, Tyler Hall, University of Rhode Island, Kingston, RI 02881, USA}
\newcommand{\SB}{Department of Physics, University of California, Santa Barbara, CA 93106, USA}
\newcommand{\JHU}{William H. Miller III Department of Physics and Astronomy, Johns Hopkins University, 3400 North Charles Street, Baltimore, Maryland, 21218, USA}
\newcommand{\Toulouse}{Laboratoire des 2 Infinis - Toulouse (L2IT-IN2P3), Universit\'{e} de Toulouse, CNRS, UPS, F-31062 Toulouse Cedex 9, France}
\newcommand{\Caltech}{Theoretical Astrophysics 350-17, California Institute of Technology, Pasadena, California 91125, USA}
\newcommand{\Fullerton}{Nicholas and Lee Begovich Center for Gravitational Wave Physics and Astronomy, California State University Fullerton, Fullerton CA 92831 USA}
\newcommand{\Syracuse}{Department of Physics, Syracuse University, Syracuse NY 13244, USA}
\newcommand{\Cornell}{Cornell Center for Astrophysics and Planetary Science, Cornell University, Ithaca, New York 14853, USA}
\newcommand{\Coimbra}{CFisUC, Department of Physics, University of Coimbra, 3004-516 Coimbra, Portugal}
\newcommand{\CIERA}{Center for Interdisciplinary Exploration and Research in Astrophysics (CIERA), Northwestern University, 1800 Sherman Ave, Evanston, IL 60201, USA}

\definecolor{dodgerblue}{HTML}{1E90FF}
\definecolor{viennared}{HTML}{DA0A14}

\hypersetup{citecolor=dodgerblue}

\def\mr{\mathrm}

\def\cross{\times}
\DeclareMathOperator{\Order}{\mathcal{O}}
\def\mF{\mathcal{F}}

\newcommand{\smr}[0]{\nu} 

\newcommand{\pySEOBNR}{\texttt{pySEOBNR}}

\newcommand{\highlight}[1]{\textcolor{black}{#1}}

\begin{document}

\title{
	Laying the foundation of the effective-one-body waveform models \texttt{SEOBNRv5}: \\improved accuracy and efficiency for spinning non-precessing binary black holes
}

\author{Lorenzo Pompili}\email{lorenzo.pompili@aei.mpg.de}
\affiliation{\AEI}
\author{Alessandra Buonanno}
\affiliation{\AEI}
\affiliation{\Maryland}
\author{Héctor Estellés}
\affiliation{\AEI}
\author{Mohammed Khalil}
\affiliation{\PI}
\affiliation{\AEI}
\affiliation{\Maryland}
\author{\\Maarten van de Meent}
\affiliation{\AEI}
\affiliation{\NBIA}
\author{Deyan P. Mihaylov}     
\affiliation{\AEI}
\author{Serguei Ossokine}
\affiliation{\AEI}
\author{Michael Pürrer}        
\affiliation{\URI}
\affiliation{\URICCR}
\affiliation{\AEI}
\author{Antoni Ramos-Buades}
\affiliation{\AEI}
\author{\\Ajit Kumar Mehta}
\affiliation{\SB}
\affiliation{\AEI}
\author{Roberto Cotesta}       
\affiliation{\JHU}
\author{Sylvain Marsat}        
\affiliation{\Toulouse}
\author{Michael Boyle}          
\affiliation{\Cornell}
\author{Lawrence E. Kidder}          
\affiliation{\Cornell}
\author{\\Harald P. Pfeiffer}              
\affiliation{\AEI}
\author{Mark A. Scheel}            
\affiliation{\Caltech}
\author{Hannes R. R{\"u}ter}              
\affiliation{\Coimbra}
\author{Nils Vu}              
\affiliation{\Caltech}
\author{Reetika Dudi}              
\affiliation{\AEI}
\author{\\Sizheng Ma}             
\affiliation{\Caltech}
\author{Keefe Mitman}             
\affiliation{\Caltech}
\author{Denyz Melchor}            
\affiliation{\Fullerton}
\author{Sierra Thomas}            
\affiliation{\Fullerton}
\affiliation{\Syracuse}
\author{Jennifer Sanchez}            
\affiliation{\CIERA}

\date{\today}

\begin{abstract}
	We present \texttt{SEOBNRv5HM}, a more accurate and faster inspiral-merger-ringdown gravitational waveform model for quasi-circular,
	spinning, nonprecessing binary black holes within the effective-one-body (EOB) formalism.
	Compared to its predecessor, \texttt{SEOBNRv4HM}, the waveform model
	i) incorporates recent high-order post-Newtonian results in the inspiral, with improved resummations,  
	ii) includes the gravitational modes $(\ell, |m|)=(3,2),(4,3)$, in addition to the $(2,2),(3,3),(2,1),(4,4),(5,5)$ modes already implemented in \texttt{SEOBNRv4HM},
	iii) is calibrated to larger mass-ratios and spins using a catalog of 442 numerical-relativity (NR) simulations and 13 additional waveforms from black-hole perturbation theory,
	iv) incorporates information from second-order gravitational self-force (2GSF) in the nonspinning modes and radiation-reaction force.	
	Computing the unfaithfulness against NR simulations, we find that for the dominant $(2,2)$ mode the 
	maximum unfaithfulness in the total mass range $10 \mbox{--} 300 M_{\odot}$ is below $10^{-3}$ for $90 \%$ of the cases 
	($38 \%$ for \texttt{SEOBNRv4HM}). When including all modes up to $\ell = 5$ we find 
	$98 \%$ ($49 \%$) of the cases with unfaithfulness below $10^{-2}$ ($10^{-3}$), while  
	these numbers reduce to $88 \%$ ($5 \%$) when using \texttt{SEOBNRv4HM}.
	Furthermore, the model shows improved agreement with NR in other dynamical quantities (e.g., the angular momentum flux and binding energy), 
	providing a powerful check of its physical robustness. 
	We implemented the waveform model in a high-performance Python package (\texttt{pySEOBNR}), which 
	leads to evaluation times faster than \texttt{SEOBNRv4HM} by a factor $10$ to $50$, depending on the configuration, and provides the 
	flexibility to easily include spin-precession and eccentric effects,
	thus making it the starting point for a new generation of EOBNR waveform models (\texttt{SEOBNRv5})
	to be employed for upcoming observing runs of the LIGO-Virgo-KAGRA detectors.
\end{abstract}

\maketitle


\section{Introduction}

Gravitational-wave (GW) astronomy has rapidly advanced since the first
detection of GWs from a binary black-hole (BBH) merger in 2015 \cite{LIGOScientific:2016aoc},
recording about ten events in the initial and second observing runs~\cite{LIGOScientific:2018mvr,Venumadhav:2019lyq} and about 
one hundred events in the third observing run~\cite{LIGOScientific:2020ibl,LIGOScientific:2021usb,LIGOScientific:2021djp,Nitz:2021uxj,Olsen:2022pin} 
of the LIGO-Virgo detectors~\cite{LIGOScientific:2014pky,VIRGO:2014yos,aLIGO:2020wna,Tse:2019wcy,Virgo:2019juy}. With upcoming upgrades of existing detectors and new facilities on the ground, such as  
Einstein Telescope~\cite{Punturo:2010zz} and Cosmic Explorer~\cite{Reitze:2019iox,Evans:2021gyd}, and the space-based mission 
LISA \cite{amaro2017laser}, it is expected that the 
merger rates of compact binaries will significantly increase.  Accurately
modeling the GWs emitted by binary systems is essential to take full
advantage of the discovery potential of ever more sensitive GW detectors, 
enriching our knowledge of astrophysics, cosmology, gravity and fundamental physics. 

Numerical relativity (NR) simulations~\cite{Pretorius:2005gq,Campanelli:2005dd,Baker:2005vv}
can provide the most accurate waveforms, but they are computationally
expensive, which makes it important to develop waveform models that
combine analytical approximation methods with NR results. 
The most commonly used approaches to build complete
inspiral-merger-ringdown (IMR) waveform models of compact 
binaries are the NR surrogate, phenomenological and
effective-one-body (EOB) families. NR surrogate models
~\cite{Blackman:2015pia,Blackman:2017dfb,Blackman:2017pcm,Varma:2018mmi,Varma:2019csw,Williams:2019vub,Rifat:2019ltp,Islam:2021mha,Islam:2022laz,Yoo:2022erv} 
interpolate NR waveforms in a reduced order representation, thus they provide us with the most accurate models for higher multipoles
~\cite{Varma:2018mmi} and spin precession \cite{Blackman:2017pcm,
  Varma:2019csw}, but they are limited to the region of parameter
space where NR simulations exist. 
Furthermore, their length currently restricts their use to binaries with total masses $\gtrsim 60 M_\odot$, 
unless the NR surrogates are hybridized to an inspiral approximant, such as EOB or PN waveforms~\cite{Varma:2018mmi,Yoo:2022erv}. Inspiral-merger-ringdown
phenomenological models ({\tt IMRPhenom})~\cite{Pan:2007nw,Ajith:2007qp,Ajith:2009bn,Santamaria:2010yb,Hannam:2013oca,Husa:2015iqa,Khan:2015jqa,Dietrich:2017aum,London:2017bcn,Khan:2018fmp,Khan:2019kot,Dietrich:2019kaq,Thompson:2020nei,Pratten:2020fqn,Pratten:2020ceb,Garcia-Quiros:2020qpx,Estelles:2020osj,Estelles:2020twz,Estelles:2021gvs,Hamilton:2021pkf} 
combine post-Newtonian (PN) and EOB waveforms for the inspiral with
fits to NR results for the late inspiral and merger-ringdown parts of
the waveform, and aim to be as fast as possible for data-analysis purposes. The EOB
formalism~\cite{Buonanno:1998gg,Buonanno:2000ef,Damour:2000we,Damour:2001tu,Buonanno:2005xu}
combines information from several analytical approximation methods
with NR results.  It maps the dynamics of a compact binary to that of
a test mass (or test spin) in a deformed Schwarzschild (or Kerr) 
background, with the deformation parameter being the symmetric mass ratio.  
EOB waveform models of BBHs have been constructed for nonspinning
\cite{Buonanno:1998gg,Buonanno:2000ef,Damour:2000we,Buonanno:2006ui,Buonanno:2007pf,Damour:2007yf,Damour:2008gu,Buonanno:2009qa,
  Pan:2011gk,Damour:2012ky,Damour:2015isa,Nagar:2019wds},
spinning~\cite{Damour:2001tu,Buonanno:2005xu,Damour:2007vq,
  Damour:2008qf,Pan:2009wj,Damour:2008te,Barausse:2009xi,Barausse:2011ys,Nagar:2011fx,Damour:2014sva,Balmelli:2015zsa,Khalil:2020mmr,Taracchini:2012ig,Taracchini:2013rva,Bohe:2016gbl,Cotesta:2018fcv,Pan:2013rra,Babak:2016tgq,Ossokine:2020kjp,Nagar:2018plt,Nagar:2018zoe,Akcay:2020qrj,Gamba:2021ydi},
and eccentric
binaries~\cite{Bini:2012ji,Hinderer:2017jcs,Chiaramello:2020ehz,Nagar:2021gss,Khalil:2021txt,Ramos-Buades:2021adz,Albanesi:2022xge}.
Matter effects have also been incorporated in EOB models in Refs.~\cite{Bernuzzi:2014owa,Hinderer:2016eia,Steinhoff:2016rfi,Akcay:2018yyh,Steinhoff:2021dsn,Matas:2020wab,Gonzalez:2022prs}.
To reduce the computational cost of EOB waveforms, surrogate or
reduced-order models have been developed in Refs.~\cite{Field:2013cfa,Purrer:2014fza,Purrer:2015tud,Lackey:2016krb,Lackey:2018zvw,Cotesta:2020qhw,Gadre:2022sed,Tissino:2022thn,Khan:2020fso,Thomas:2022rmc}. 
Parameter-estimation codes based on machine-learning methods, notably neural posterior estimation, are also available to speed up inference studies~\cite{Dax:2021tsq,Dax:2022pxd}. 
More specifically, there are currently two state-of-the-art families
of EOB waveform models: \texttt{SEOBNR} (e.g., see
Refs.~\cite{Bohe:2016gbl,Cotesta:2018fcv,Ossokine:2020kjp,Cotesta:2020qhw,Ramos-Buades:2021adz,Mihaylov:2021bpf})
and \texttt{TEOBResumS} (e.g., see
Refs.~\cite{Nagar:2018zoe,Nagar:2019wds,Nagar:2020pcj,Gamba:2021ydi,Riemenschneider:2021ppj,Chiaramello:2020ehz}). Here, we focus on the former.

The expected increase in sensitivity during the fourth observing run (O4)~\cite{LVKO4} of the LIGO-Virgo-KAGRA (LVK) Collaboration~\cite{LIGOScientific:2014pky,VIRGO:2014yos,KAGRA:2020tym}, which is planned to start in May 2023, will likely allow us to observe events in unexplored regions of parameter space 
with high spins and large mass ratios. In these regions of parameter space state-of-the-art waveform models tend to disagree \cite{Bohe:2016gbl, Ossokine:2020kjp, Pratten:2020igi, Colleoni:2020tgc, Estelles:2021gvs}, as they are mostly calibrated to 
NR simulations having both moderate spins, say $\lesssim0.5$, and comparable mass ratios, say $1\mbox{--}4$,  
and waveform modeling systematics could be comparable to 
statistical errors.  In order to improve the accuracy of EOB models, one takes advantage of the strong-field information from NR simulations, and also includes the latest results from the main analytical approximation methods, that is PN, post-Minkowskian and gravitational self-force theory~\cite{Damour:2016gwp,Damour:2017zjx,Antonelli:2019ytb,Damgaard:2021rnk,Khalil:2022ylj,Damour:2022ybd,Damour:2016gwp,Damour:2017zjx,Antonelli:2019ytb,Damgaard:2021rnk,Khalil:2022ylj,Damour:2022ybd,VandeMeentv5}.

Within the \texttt{SEOBNR} family of EOB models, we present \texttt{SEOBNRv5HM}~\footnote{\texttt{SEOBNRv5HM} is publicly available through the python package \texttt{pySEOBNR} \href{https://git.ligo.org/waveforms/software/pyseobnr}{\texttt{git.ligo.org/waveforms/software/pyseobnr}}. Stable versions of \texttt{pySEOBNR} are published through the Python Package Index (PyPI), and can be installed via ~\texttt{pip install pyseobnr}.}
, a new IMR  multipolar waveform 
model for quasi-circular, spinning, nonprecessing BBHs.
In \texttt{SEOBNRv5HM} we employ the most recent PN results for the three main components of the dynamics and gravitational radiation: 
the Hamiltonian \cite{Balmelli:2015zsa,Khalil:2020mmr,Khalilv5}, the radiation-reaction (RR) force and waveform modes~\cite{Henry:2022dzx}. 
Furthermore, we directly incorporate information from second-order self-force (2GSF) \cite{VandeMeentv5,Warburton:2021kwk,Wardell:2021fyy} in the modes and RR force.
\texttt{SEOBNRv5HM} includes the gravitational modes $(\ell, |m|)=(3,2),(4,3)$, in addition to the $(\ell, |m|)=(2,2),(3,3),(2,1),(4,4),(5,5)$ 
modes already implemented in \texttt{SEOBNRv4HM} \cite{Cotesta:2018fcv}, and models the mode-mixing in the merger-ringdown of the $(3,2),(4,3)$ modes.
We calibrate \texttt{SEOBNRv5HM} to 442 numerical-relativity (NR) waveforms, all produced with the pseudo-Spectral Einstein code (\texttt{SpEC}) 
of the Simulating eXtreme Spacetimes (\texttt{SXS}) collaboration 
\cite{SXS:catalog,Boyle:2019kee,Chu:2015kft,Blackman:2017dfb,Hemberger:2013hsa,Scheel:2014ina,Lovelace:2014twa,Abbott:2016apu,Bohe:2016gbl,Blackman:2015pia,Lovelace:2016uwp,Varma:2018mmi,Abbott:2016nmj,Varma:2019csw,Kumar:2015tha,Mroue:2013xna,Yoo:2022erv}, 
except for a simulation with mass ratio and (dimensionless spins) $q = 8, ~\chi_1 = 0.85, ~\chi_2 = 0.85$ produced with the \texttt{Einstein Toolkit} code \cite{EinsteinToolkit:2022_11, Cotesta:2018fcv}.
We also incorporate information from 13 waveforms computed by solving the Teukolsky equation in the framework of BH 
perturbation theory~\cite{Barausse:2011kb,Taracchini:2014zpa}, with mass ratio $q = 10^{3}$ and dimensionless spins values in the range $-0.99 \leq \chi \leq 0.99$.
This greatly extends the NR calibration coverage with respect to \texttt{SEOBNRv4} \cite{Bohe:2016gbl}, which used 140 NR waveforms, 
especially towards larger mass-ratios and spins. Indeed, we include several NR simulations with mass ratios between 10 and 20,
in a region of parameter space where no simulations were available when \texttt{SEOBNRv4} was developed. 

We validate the model by computing the unfaithfulness against NR simulations, 
and by comparing several dynamical quantities, such as the angular-momentum flux and binding energy, 
providing an important check of its physical robustness and giving confidence of its reliability when extrapolating it 
outside the NR calibration region. Computational efficiency is also a key aspect of waveform models, as Bayesian parameter estimation of GW events with stochastic sampling techniques 
typically requires millions of waveform evaluations. For this purpose, we implemented \texttt{SEOBNRv5HM} in a flexible, high-performance Python package (\texttt{pySEOBNR}~\cite{Mihaylovv5}), which leads to evaluation times faster than \texttt{SEOBNRv4HM}. 
We then show that the \texttt{SEOBNRv5HM} model can be employed for GW parameter estimation with standard stochastic samplers thanks to its high computational efficiency. 
We perform Bayesian inference studies using \texttt{SEOBNRv5HM} by injecting synthetic NR signals in zero noise, and by reanalysing GW events from previous observing runs. 
Further speedup in waveform evaluation time of about an order of magnitude can be obtained by surrogate models. 
We build a frequency domain reduced order model of \texttt{SEOBNRv5}, following Ref.~\cite{Cotesta:2020qhw}.


This work is part of a series of articles \cite{Khalilv5,Mihaylovv5,VandeMeentv5,RamosBuadesv5} describing the \texttt{SEOBNRv5} family for O4~\cite{LVKO4}, and it is organized as follows. After an introduction to the notation used in this paper, in Sec.~\ref{sec:ham_eom} 
we describe the \texttt{SEOBNRv5} aligned-spin Hamiltonian and equations of motion. In Sec.~\ref{sec:waveforms} we outline the construction of the multipolar waveform modes, highlighting improvements and differences with respect to \texttt{SEOBNRv4HM}, 
and in Sec.~\ref{sec:calibration} we illustrate how \texttt{SEOBNRv5HM} is calibrated against 442 NR simulations. 
In Sec.~\ref{sec:performance} we compare the accuracy of \texttt{SEOBNRv5HM}, and of other state-of-the-art 
waveform models, against NR simulations, and investigate the regions of parameter space where 
they exhibit the largest differences from NR waveforms and from each other. 
We also present comparisons against NR results for the angular-momentum flux and binding energy of 
\texttt{SEOBNRv4} and \texttt{SEOBNRv5}. In Sec.~\ref{sec:PE} we study the model's accuracy in Bayesian 
inference analyses, by performing a synthetic NR injection in zero noise and by analyzing GW events observed by the LVK detectors. 
In Sec.~\ref{sec:ROM} we outline the performance of a frequency-domain reduced-order model of \texttt{SEOBNRv5}.
Section \ref{sec:conclusions} summarizes our main conclusions and discusses future work.
Finally, Appendices~\ref{app:HamCoeffs} and \ref{app:modes} provide the complete expression for the Hamiltonian and the multipolar waveform modes used for this work. 
In Appendices~\ref{appx:NQC_fits} and~\ref{appx:coeff_fits}, we provide all expressions for the fits to NR simulations entering the construction of the waveform modes. The Appendix~\ref{appx:robustness} presents some tests of the robustness of the calibration pipeline to NR waveforms, and in Appendix~\ref{appx:NQC_flux} we check the potential impact of including additional corrections in the RR force for a specific binary configuration.
Finally, in Appendix~\ref{appx:pheonthm}, we extend the comparison of Sec.~\ref{sec:performance} to the state-of-the-art time-domain phenomenological model \texttt{IMRPhenomTHM} \cite{Estelles:2020osj,Estelles:2020twz}.


\section*{Notation}
We use natural units in which $c = G =1$. We consider a binary with masses $m_1$ and $m_2$, with $m_1 \geq m_2$, and define the following combinations of the masses:
\begin{equation}
\begin{gathered}
	M\equiv m_1 + m_2, \qquad \mu \equiv \frac{m_1m_2}{M}, \qquad \nu \equiv \frac{\mu}{M},\\
	\delta \equiv\frac{m_1 - m_2}{M},  \qquad q \equiv \frac{m_1}{m_2}.	
\end{gathered}
\end{equation}
For binaries with nonprecessing spins of magnitude $S_1$ and $S_2$, we define the dimensionless spins
\begin{gather}
	\chi_{\mr i} \equiv \frac{a_{\mr i}}{m_{\mr i}} = \frac{S_{\mr i}}{m_{\mr i}^2},
\end{gather}
where $\mr i = 1,2$, and define the following spin variables:
\begin{equation}
\label{spinComb}
\begin{gathered}
	\chi_S \equiv \frac{\chi_1 + \chi_2}{2}, \qquad \chi_A \equiv \frac{\chi_1 - \chi_2}{2},\\
	\chi_{\mathrm{eff}} \equiv \frac{\left(m_1 \chi_1+m_2 \chi_2\right)}{m_1+m_2}, \\
	a_\pm \equiv a_1 \pm a_2 = m_1 \chi_1 \pm m_2 \chi_2.	
\end{gathered}
\end{equation}

The relative position and momentum vectors, in the binary's center-of-mass frame, are denoted $\bm{r}$ and $\bm{p}$, with
\begin{equation}
\bm p^2 = p_r^2 + \frac{L^2}{r^2}, \quad
p_r= \bm{n}\cdot\bm{p}, \quad
\bm{L}=\bm{r}\cross\bm{p},
\end{equation}
where $\bm{n}\equiv\bm{r}/r$, and $\bm{L}$ is the orbital angular momentum with magnitude $L$.
Since in this work we discuss nonprecessing (or aligned-spin) BHs, we consider equatorial orbits, and use 
polar-coordinates phase-space variables $(r,\phi,p_r,p_\phi)$, where the angular momentum reduces to $L = p_\phi$.

The orbital frequency is denoted $\Omega$, and we define the dimensionless frequency parameter $v_\Omega \equiv (M\Omega)^{1/3}$. We also often use $u \equiv M/r$ instead of $r$.

\section{The SEOBNR\lowercase{v}5 aligned-spin Hamiltonian and equations of motion}
\label{sec:ham_eom}

In the EOB formalism \cite{Buonanno:1998gg,Buonanno:2000ef,Damour:2000we,Damour:2001tu,Buonanno:2005xu}, the two-body dynamics
is mapped onto the effective dynamics of a test body in a deformed Schwarzschild
or Kerr background, with the deformation parametrized by the symmetric mass-ratio $\nu$.
The energy map relating the effective Hamiltonian $H_{\rm{eff}}$ and the two-body EOB
Hamiltonian $H_{\rm{EOB}}$ is given by
\begin{equation}
\label{eq:HEOB}
	H_{\mathrm{EOB}}=M \sqrt{1+2 \nu \left(\frac{H_{\mathrm{eff}}}{\mu}-1\right)}\,.
\end{equation}
The generic-spin Hamiltonian we use in \texttt{SEOBNRv5} is based on that of a \emph{test mass} in a deformed Kerr background \cite{Damour:2001tu,Damour:2008qf,Pan:2009wj,Nagar:2011fx,Damour:2014sva,Balmelli:2015zsa,Khalil:2020mmr,Khalilv5}.
In contrast, the \texttt{SEOBNRv4}~\cite{Bohe:2016gbl,Cotesta:2018fcv,Ossokine:2020kjp} Hamiltonian was based on the one of a \emph{
test spin} in a deformed Kerr background \cite{Barausse:2009aa,Barausse:2011ys,Vines:2016unv}.

The \texttt{SEOBNRv5} Hamiltonian includes most of the 
5PN nonspinning contributions, together with spin-orbit
(SO) information up to the next-to-next-to-leading order (NNLO), spin-spin (SS) information to NNLO,
as well as cubic- and quartic-in-spin terms at leading order (LO), corresponding to all PN information up to 4PN order for precessing spins.
More details about the derivation of the generic-spin Hamiltonian, together with the full expressions, are given in Ref.~\cite{Khalilv5}.
Here, we summarize the structure of the aligned-spin Hamiltonian, and its zero-spin limit, highlighting where NR calibration parameters enter the expressions.

\subsection{Nonspinning effective Hamiltonian}
\label{sec:zerospin_H}
The effective Hamiltonian for nonspinning (noS) binaries can be written as
\begin{equation}
\label{Heffzero}
H_\text{eff}^\text{noS} = \sqrt{p_{r_*}^2 + A_\text{noS}(r) \left[\mu^2 + \frac{p_\phi^2}{r^2} + Q_\text{noS}(r,p_{r_*})\right]},
\end{equation}
where we use the tortoise-coordinate $p_{r_*}$ instead of $p_r$, since it improves the stability of the equations of motion during the 
plunge and close to merger~\cite{Damour:2007xr,Pan:2009wj}.
For nonspinning binaries, $r_*$ is defined by
\begin{equation}
\frac{dr_*}{dr} = \frac{1}{\xi(r)}, \qquad
\xi(r) \equiv A_\text{noS}(r)\sqrt{\bar{D}_\text{noS}(r)},
\end{equation}
with the conjugate momentum $p_{r_*}$ given by
\begin{equation}
\label{prstar}
p_{r_*} =  p_r \,\xi(r).
\end{equation}

For the potentials $A_\text{noS}(r)$ and $\bar{D}_\text{noS}(r)$, we use the 5PN results of Refs.~\cite{Bini:2019nra,Bini:2020wpo}, which are complete except for two quadratic-in-$\nu$ coefficients.
The 5PN Taylor-expanded $A_\text{noS}$ is given by
\begin{align}
\label{eq:Ataypm}
A_\text{noS}^\text{Tay}(u) &= 1 - 2 u + 2\nu u^3 + \nu \left(\frac{94}{3}-\frac{41 \pi ^2}{32}\right) u^4 \nonumber \\
&\quad + \Bigg[\nu  \left(\frac{2275 \pi ^2}{512}-\frac{4237}{60}+\frac{128 \gamma_E }{5}+\frac{256 \ln 2}{5}\right) \nonumber\\
&\qquad
+ \left(\frac{41 \pi ^2}{32}-\frac{221}{6}\right) \nu ^2 +\frac{64}{5} \nu \ln u\Bigg] u^5 \nonumber\\
&\quad + \left[\nu a_6 + \left(-\frac{144 \nu ^2}{5}-\frac{7004 \nu }{105}\right) \ln u\right] u^6,
\end{align}
where $u \equiv M/r$, and we replace the coefficient of $u^6$, except for the log part, by the parameter $a_6$, which is calibrated to NR simulations.

The 5PN Taylor-expanded $\bar{D}_\text{noS}(r)$ potential is given by Eq.~\eqref{DpmTay} in Appendix~\ref{app:HamCoeffs}.
The 5.5PN contributions to $A_\text{noS}(r)$ and $\bar{D}_\text{noS}(r)$ are known from Refs.~\cite{Damour:2015isa,Bini:2020wpo}; 
however, since we Pad\'e resum these potentials (as explained in Sec.~\ref{sec:calibration}), we find it more convenient to stop at 5PN.
For the $Q_\text{noS}(r)$ potential, we use the full 5.5PN expansion, which is also expanded in eccentricity to $\Order(p_r^8)$, as given by Eq.~\eqref{QpmTay}.

The calibration parameter $a_6$ is a function of $\nu$; to determine its value in the limit $\nu \to 0$, we use the GSF results of Refs.~\cite{Barack:2010tm,Akcay:2012ea,Isoyama:2014mja} for the frequency shift of the innermost stable circular orbit (ISCO), which is given by
\begin{align}
\begin{aligned}
M \Omega_\text{ISCO}^\text{1SF} &= 6^{-3/2} (1 + C_\Omega / q), \\
C_\Omega &= 1.25101539 \pm 4\times 10^{-8}.
\end{aligned}
\end{align}
The ISCO can be computed from the EOB Hamiltonian by solving $(\partial H/\partial r)_{|_{p_r=0}} = 0 = (\partial^2 H/\partial r^2)_{|_{p_r=0}}$ for $r$ and $p_\phi$. We find the value of $a_6$ that gives the best agreement with $\Omega_\text{ISCO}^\text{1SF}$ is 
\begin{equation}
\label{eq:a6_TPL}
\left. a_6\right|_{\nu\to 0} \simeq 39.0967.
\end{equation}
The fit we use for $a_6(\nu)$ is given by Eq.~\eqref{eq:a6_fit} below.


\subsection{Aligned-spin effective Hamiltonian}
For aligned-spins, the effective Hamiltonian reduces to the equatorial Kerr Hamiltonian in the test-particle limit (TPL), with the Kerr spin $a$ mapped to the binary's spins via $a = a_1 + a_2 \equiv a_+$. 
To include 4PN information for spinning binaries and arbitrary mass-ratios, we use the following ansatz~\cite{Khalilv5}:
\begin{align}
\label{HeffAnzAlign}
H_\text{eff}^\text{align} &\equiv  H_{\rm odd} + H_{\rm even}\,, \nonumber \\
& =\frac{Mp_\phi \left(g_{a_+} a_+ + g_{a_-} \delta a_-\right) + \text{SO}_\text{calib}
+ G_{a^3}^\text{align}}{r^3+a_+^2 (r+2M)} \nonumber\\
&
+ \Bigg [A^\text{align} \Bigg(
\mu^2 + p^2 + B_{np}^\text{align} p_r^2 + B_{npa}^\text{Kerr\,eq} \frac{p_\phi^2 a_+^2}{r^2} + Q^\text{align}
\Bigg)\Bigg ]^{1/2}, 
\end{align}
where the first term on the right-hand side only includes the odd-in-spin contributions (in the numerator), while the second term (square root) includes the even-in-spin contributions.

The gyro-gravitomagnetic factors $g_{a_+}(r)$ and $g_{a_-}(r)$ in the SO part of the Hamiltonian (\ref{HeffAnzAlign}) are sometimes chosen to be in a gauge such that they are functions of $1/r$ and $p_r^2$ only~\cite{Damour:2008qf,Nagar:2011fx}, though Refs.~\cite{Barausse:2009xi,Barausse:2011ys} made different choices.
In building the \texttt{SEOBNRv5} model, we find better agreement with NR waveforms when using a gauge in which $g_{a_+}(r)$ and $g_{a_-}(r)$ depend only on $1/r$ and $p_\phi^2/r^2$. The 4.5PN SO coupling was derived in Refs.~\cite{Antonelli:2020aeb,Antonelli:2020ybz,Mandal:2022nty,Kim:2022pou}, and can be included in the 
gyro-gravitomagnetic factors (see Eqs. (30a) and (30b) in Ref.~\cite{Khalilv5}).
However, when calibrating to NR simulations, we find that using a calibration term at 5.5PN has a small effect on the dynamics, and thus we only include the 3.5PN SO information (given in Eqs.~\eqref{gyros}) with a 4.5PN SO calibration term of the form
\begin{equation}
	\label{eq:SO_calib}
	\text{SO}_\text{calib} =\nu d_\text{SO} \frac{M^4}{r^3} p_\phi a_+.
\end{equation}
Furthermore, the function $G_{a^3}^\text{align}(r)$ in Eq.~(\ref{HeffAnzAlign}) contains S$^3$ corrections.
The nonspinning and SS contributions are included in $A^\text{align}(r)$, $B_{np}^\text{align}(r)$ and $Q^\text{align}(r)$, with no S$^4$ corrections needed since the Kerr Hamiltonian reproduces all even-in-spin leading PN orders for binary BHs~\cite{Vines:2016qwa}.
Explicit expressions for the functions in the Hamiltonian are provided in Appendix~\ref{app:HamCoeffs} (and also in Ref.~\cite{Khalilv5}).

When using tortoise-coordinates for spinning binaries, a convenient choice for $\xi(r)$ is
\begin{equation}
\xi(r) = \frac{\sqrt{\bar{D}_\text{noS}}\left(A_\text{noS} + a_+^2/r^2\right)}{1 + a_+^2/r^2},
\end{equation}
which is similar to what was used in \texttt{SEOBNRv4}~\cite{Pan:2009wj,Taracchini:2012ig} except for the different resummation and PN orders in $A_\text{noS}$ and $\bar{D}_\text{noS}$.
In the $\nu\to 0$ limit, $\xi$ reduces to the Kerr value $(dr/dr_*) = (r^2-2Mr+a_+^2)/(r^2+a_+^2)$.

\subsection{Equations of motion and radiation-reaction force}
The equations of motion for aligned-spins, in terms of $p_{r_*}$, are given by Eqs.~(10) of Ref.~\cite{Pan:2011gk}, and read
\begin{align}
\label{eq:eom_prst}
\begin{aligned}
\dot{r} &= \xi \frac{\partial H}{\partial p_{r_*}}, \quad
&\dot{p}_{r_*} &= -\xi \frac{\partial H}{\partial r} + \frac{p_{r_*}}{p_\phi} \mF_\phi, \\
\dot{\phi} &= \frac{\partial H}{\partial p_\phi}, \quad
&\dot{p}_\phi &= \mF_\phi,
\end{aligned}
\end{align}
where the RR force $\mF_\phi$ is obtained by summing the GW modes in factorized form~\cite{Damour:2007xr,Damour:2007yf,Damour:2008gu,Pan:2010hz}, $h_{\ell m}^{\mathrm{F}}$, which we define in Sec.~\ref{subsec:hlm_inspiral_plunge}, that is

\begin{equation}
\label{RRforce}
\mF_\phi \equiv - \frac{\Omega}{8 \pi} \sum_{\ell=2}^8 \sum_{m=1}^{\ell} m^2\left|d_L h_{\ell m}^{\mathrm{F}}\right|^2,
\end{equation}
where $\Omega$ is the orbital frequency, and $d_L$ is the luminosity distance of the binary to the observer.

The equations of motion can be written more explicitly as follows:
\begin{subequations}
\begin{align}
\dot{r} & = \frac{M A^\text{align}}{2 H_{\rm{EOB}} H_{\rm even}}\left[\frac{2p_{r_*}}{\xi}(1+B_{np}^\text{align})+\xi\frac{\partial Q^\text{align}}{\partial p_{r_*}}\right], \\
\dot{\phi} &= \frac{M}{H_{\rm{EOB}}}\left[p_{\phi}\frac{\partial \bar{H}_{\rm odd}}{\partial p_{\phi}}+\bar{H}_{\rm odd}+\frac{A^\text{align}}{H_{\rm even}}\frac{p_{\phi}}{r^2}\left(1+B_{npa}^\text{Kerr\,eq}a_{+}^{2}\right) \right], \\
\dot{p}_{r_*} &= -\frac{M \xi}{H_{\rm{EOB}}}\left(\frac{\partial H_{\rm even}}{\partial r}+p_{\phi}\frac{\partial \bar{H}_{\rm odd}}{\partial r} \right) + \frac{p_{r_*}}{p_\phi} \mF_\phi,
\end{align}
\end{subequations}
where we define $\bar{H}_{\rm odd} \equiv H_{\rm odd}/p_{\phi}$. The derivative of $H_\text{even}$ is given by
\begin{subequations}
\begin{align}
  \frac{\partial H_{\rm even}}{\partial r} &= \frac{1}{2H_{\rm even}} \left(K_{0} p_{\phi}^{2} + K_{1}\right), \\
  K_{0} &\equiv A^\text{align}\left[-\frac{2}{r^3}\left(1+B_{npa}^\text{Kerr\,eq}a_{+}^{2}\right)+\frac{a_{+}^{2}}{r^2}\frac{dB_{npa}^\text{Kerr\,eq}}{dr}\right]\nonumber\\
  &\quad +\frac{dA^\text{align}}{dr}\left(\frac{1}{r^2}+\frac{a_{+}^2}{r^2}B_{npa}^\text{Kerr\,eq}\right), \\
  K_{1} &\equiv A^\text{align}\left \{\frac{p_{r_*}^{2}}{\xi^{2}}\left[\frac{dB_{np}^\text{align}}{dr}-\frac{2}{\xi}\frac{d\xi}{dr}\left(1+B_{np}^\text{align}\right)\right] + \frac{\partial Q^\text{align}}{\partial r}\right \} \nonumber\\
  &\quad + \frac{dA^\text{align}}{dr}\left [\mu^2+\frac{p_{r_*}^{2}}{\xi^{2}}\left(1+B_{np}^\text{align}\right)+Q^\text{align}\right].
\end{align}
\end{subequations}
When evolving the equations of motion, we use the same quasi-circular adiabatic initial conditions derived in Ref.~\cite{Buonanno:2005xu}, 
then integrate numerically Eqs.~\eqref{eq:eom_prst} to solve for the binary dynamics.

In \texttt{SEOBNRv5}, one can also employ the post-adiabatic (PA) approximation for the inspiral dynamics, which allows speeding up 
the evaluation of the model, especially for very long waveforms~\cite{Damour:2012ky,Nagar:2018gnk}.  
This technique has been used extensively with great success in the \texttt{TEOBResumS} family of models 
(see, e.g., Refs.~\cite{Nagar:2018gnk,Rettegno:2019tzh,Nagar:2020pcj,Riemenschneider:2021ppj,Gamba:2021ydi}), 
and recently also in the \texttt{SEOBNRv4HM\_PA} model~\cite{Mihaylov:2021bpf}. 
To obtain explicit algebraic equations for the momenta, we follow the same procedure 
as described in Refs.~\cite{Nagar:2018gnk,Rettegno:2019tzh}, which results in
the following equations:
\begin{align}
  \label{eq:pa_pr}
  &p_{r_*}  = \frac{\xi}{2\left(1+B_{np}^\text{align}\right)}\left[\mF_\phi{\left(\frac{dp_{\phi}}{dr}\right)}^{-1}\frac{2 H_{\rm{EOB}}H_{\rm even}}{MA^\text{align}}-\xi\frac{\partial Q^\text{align}}{\partial p_{r_*}} \right], \\
  \label{eq:pa_pphi}
  &K_{0}p_{\phi}^{2}+2H_{\rm even}\frac{\partial \bar{H}_{\rm odd}}{\partial r} p_{\phi}  + K_{1} \nonumber\\
  &\qquad +\frac{2 H_{\rm even}H_{\rm{EOB}}}{M\xi}\left(\frac{dp_{r_*}}{dr}\frac{dr}{dt} - \frac{p_{r_*}}{p_{\phi}}\mF_\phi\right) = 0.
\end{align}
Here, the only unknowns are the explicit $p_{r_*}$ in the left-hand side of the first equation, and the explicit $p_{\phi}^{2}$ and $p_{\phi}$ in the second; all the other instances of $p_{r_*}$ and $p_{\phi}$ are obtained from previous orders. We employ the PA approximation at $8^{\rm th}$ order.

\section{The SEOBNR\lowercase{v}5 multipolar waveform}
\label{sec:waveforms}

In this section, we describe the building blocks used in the construction
of the multipolar spinning, nonprecessing waveform modes $h_{\ell m}$ of the \texttt{SEOBNRv5HM} model.
We closely follow the construction of the \texttt{SEOBNRv4HM} model~\cite{Cotesta:2018fcv} and highlight differences when needed.

In general, the complex linear combination of GW polarizations,  $h(t) \equiv h_{+}(t) -ih_{\times}(t)$, can be expanded in the basis 
of $-2$ spin-weighted spherical harmonics~\cite{Pan:2011gk} as follows:
\begin{equation}
	h(t;  \bm{\lambda}, \iota, \varphi_0) = \sum_{\ell \geq 2}\sum_{|m|\leq \ell} {}_{-2} Y_{\ell m} (\iota, \varphi_0) \, h_{\ell m}(t;\bm{\lambda}),
\label{eq:hoft_sphericalH}
\end{equation}
where $\bm{\lambda}$ denotes the intrinsic parameters of the compact binary source, such as masses ($m_{1,2}$) and spins ($\chi_{1,2}$). 
The waveform modes $h_{\ell m}$ depend on only three parameters $(q, \chi_1, \chi_2)$, since the waveform scales trivially with the total mass $M$.
The parameters $(\iota, \varphi_0)$ describe the binary's inclination angle (computed with respect to the 
direction perpendicular to the orbital plane) and the azimuthal direction to the observer, respectively~\footnote{In general, the GW polarizations emitted by a quasi-circular BBH depend on its masses and spins $\boldsymbol{\lambda}=\{m_1, m_2,\bm{\chi}_{1,2}\}$,
the angles $(\iota, \varphi_0)$, the luminosity distance of the binary to the observer $d_L$ and the time of arrival $t_c$. 
Inserting back units, the modes scale as $\sim G M/ (c^2 d_L)$.}.

In the EOB framework, the GW modes defined in Eq.~(\ref{eq:hoft_sphericalH})
are decomposed into inspiral-plunge and merger-ringdown modes.
In \texttt{SEOBNRv5HM}, we model the (2,2) and the largest subdominant modes \cite{Cotesta:2018fcv}: (3,3), (2,1), (4,4), (3,2), (5,5) and (4,3).
For aligned-spin binaries $h_{\ell m}=(-1)^{\ell} h_{\ell-m}^*$, therefore we restrict the discussion to $(\ell,m)$ modes with $m > 0$.
We have: 
\begin{equation}
	\label{eq:h_match}
	h_{\ell m}(t)= \begin{cases}h_{\ell m}^{\text {insp-plunge }}(t), & t<t_{\text {match }}^{\ell m} \\ h_{\ell m}^{\text {merger-RD }}(t), & t>t_{\text {match }}^ {\ell m }\end{cases},
\end{equation}
where we define $t_{\text {match }}^{\ell m}$ as
\begin{equation}
	\label{eq:t_match}
	t_{\text {match }}^{\ell m}=
	\begin{cases}
	t_{\text {peak }}^{22}, &(\ell, m) = (2,2),(3,3),(2,1), \\
	& \qquad\quad\,\; (4,4),(3,2),(4,3) \\
	t_{\text {peak }}^{22}-10 M,  &(\ell, m) =(5,5),\end{cases}
\end{equation}
where $t_{\text {peak }}^{22}$ is the peak of the $(2,2)$-mode amplitude. The choice of a different attachment point for the $(5,5)$ mode is motivated,
as in Ref.~\cite{Cotesta:2018fcv}, by the fact that at late times the error in some of the NR waveforms 
used to calibrate the model is too large to accurately extract the quantities that are needed to build the full inspiral-merger-ringdown waveforms (see below).
For the same reason, since typically $t_{\text {peak }}^{\ell m}-t_{\text {peak }}^{22}>0$ \citep{Pan:2011gk,Barausse:2011kb}, we emphasize that the merger-ringdown attachment 
for all other modes is done at the peak of the $(2,2)$ mode, rather than at each mode's peak time as in other EOB models \citep{Pan:2011gk,Nagar:2020pcj}. 

\subsection{Inspiral-plunge $h_{\ell m}$ modes}
\label{subsec:hlm_inspiral_plunge}

The inspiral-plunge EOB waveform modes can be written as
\begin{equation}
	\label{eq:hellm_insp}
	h_{\ell m}^{\text {insp-plunge }}=h_{\ell m}^{\mathrm{F}} N_{\ell m}\,,
\end{equation}
where $h_{\ell m}^{\mathrm{F}}$ is a factorized, resummed form of the PN-expanded GW modes for aligned-spins in circular orbits 
\cite{Damour:2007xr,Damour:2008gu,Pan:2010hz},
while $N_{\ell m}$ is the nonquasi-circular (NQC) correction, aimed at incorporating relevant radial effects during the plunge, toward the merger.

The factorized inspiral modes are written as
\begin{equation}
\label{hlmFactorized}
	h_{\ell m}^{\mathrm{F}}=h_{\ell m}^{\text{N}} \hat{S}_{\mathrm{eff}} T_{\ell m} f_{\ell m} e^{i \delta_{t m}}.
\end{equation}
The first factor, $h_{\ell m}^{(N, \epsilon_{\ell m})}$ is the leading (Newtonian) order waveform and its explicit expression is~\cite{Damour:2008gu,Pan:2010hz}
\begin{equation}\label{eq:hnewt}
	h_{\ell m}^{\rm N} =\frac{\smr M}{d_L} n_{\ell m} c_{\ell+\epsilon_{\ell m}}(\smr)v_\phi^{\ell+\epsilon_{\ell m}}Y_{\ell-\epsilon_{\ell m},-m}\left(\frac{\pi}{2},\phi\right).
\end{equation}
Here $d_L$ is the luminosity distance of the binary to the observer, $Y_{\ell m}$ is the scalar spherical harmonic,
$\epsilon_{\ell m}$ is the parity of the mode, such that
\begin{equation}
	\epsilon_{\ell m} = \left\{\begin{aligned}
		&0, &&\text{$\ell+m$ is even}\\
		&1, &&\text{$\ell+m$ is odd}
	\end{aligned} \right.,
\end{equation}
and the functions $n_{\ell m}$ and $c_{k}(\smr)$ are given by
\begin{equation}
	n_{\ell m} = \left\{\begin{aligned}
		& \frac{8\pi (i m)^\ell}{(2\ell+1)!!}\sqrt{\tfrac{(\ell+1)(\ell+2)}{\ell(\ell-1)}}, &&\text{$\ell+m$ is even}\\
		& \frac{-16i \pi (i m)^\ell}{(2\ell+1)!!}\sqrt{\tfrac{(2\ell+1)(\ell+2)(\ell^2-m^2)}{(2\ell-1)(\ell+1)\ell(\ell-1)}}, &&\text{$\ell+m$ is odd},
	\end{aligned} \right.
\end{equation}
and
\begin{equation}
	c_{k}(\smr) = \left(\frac{1-\sqrt{1-4\smr}}{2} \right)^{k-1}+(-1)^k \left(\frac{1+\sqrt{1-4\smr}}{2} \right)^{k-1}.
\end{equation}
Finally, $v_\phi$ in Eq.~\eqref{eq:hnewt} is given by
\begin{equation}
	v_\phi = M \Omega r_\Omega,
\end{equation}
where $\Omega$ is the orbital frequency and
\begin{equation}
	r_\Omega = \left.\left(\frac{\partial H_{\rm EOB}}{\partial p_\phi} \right)^{-2/3}\right|_{p_r=0}.
\end{equation}
The (dimensionless) effective source term $\hat{S}_\text{eff}$ is given by either the effective energy $E_\text{eff}$
or the orbital angular momentum $p_\phi$, both expressed as functions of $v_{\Omega}\equiv\left(M \Omega\right)^{1/3}$, such that
\begin{equation}
\hat{S}_\text{eff} = \left\{
        \begin{array}{ll}
            \frac{E_\text{eff}(v_{\Omega})}{\mu}, & \quad \ell + m \text{ even}, \\\\
          v_{\Omega} \frac{p_\phi(v_{\Omega})}{M \mu}, & \quad \ell + m \text{ odd},
        \end{array}
    \right. 
\end{equation}
where $E_\text{eff}$ is related to the total energy  $E$ via the EOB energy map $E = M \sqrt{1+2\nu \left(E_\text{eff}/\mu - 1\right)}$.

The factor $T_{\ell m}$ in Eq.~\eqref{hlmFactorized} resums an infinite number of leading logarithms entering the tail contributions~\cite{Blanchet:1997jj}, and is given by
\begin{equation}
T_{\ell m} = \frac{\Gamma\left(\ell + 1 - 2 i \hat{k}\right)}{\Gamma (\ell + 1)} e^{\pi \hat{k}} e^{2i \hat{k} \ln (2m\Omega r_0)},
\end{equation}
where $\Gamma(...)$ is the Euler gamma function, $\hat{k}\equiv m \Omega E$ and the constant $r_0$ takes the value $2M/\sqrt{e}$ to give agreement with waveforms computed in the test-body limit~\cite{Pan:2010hz}.

The remaining part of the factorized modes (\ref{hlmFactorized}) is expressed as an amplitude $f_{\ell m}$ and a phase $\delta_{\ell m}$, which are computed such that the expansion of $h_{\ell m}^\text{F}$ agrees with the PN-expanded modes.
For nonspinning binaries, $f_{\ell m}$ is further resummed as~\cite{Damour:2008gu} $f_{\ell m} = (\rho_{\ell m})^\ell$ to reduce the magnitude of the 1PN coefficient, which grows linearly with $\ell$.
Following Refs.~\cite{Pan:2010hz,Taracchini:2012ig,Taracchini:2013rva}, for spinning binaries we separate the nonspinning and spin contributions for the odd $m$ modes, such that
\begin{align}
\label{frholm}
f_{\ell m} = \left\{
        \begin{array}{ll}
           \rho_{\ell m}^\ell, & \quad m \text{ even}, \\\\
           (\rho_{\ell m}^\text{NS})^\ell + f_{\ell m}^\text{S}, & \quad m \text{ odd},
        \end{array}
    \right. 
\end{align}
where $\rho_{\ell m}^\text{NS}$ is the nonspinning part of $\rho_{\ell m}$, while $f_{\ell m}^\text{S}$ is the spin part of $f_{\ell m}$.

The explicit expressions for $\rho_{\ell m}$, $f_{\ell m}$ and $\delta_{\ell m}$ that are used in the \texttt{SEOBNRv5HM} model are provided in Appendix~\ref{app:modes}, and are mostly similar to those in \texttt{SEOBNRv4HM} as derived in Refs.~\cite{Cotesta:2018fcv,Bohe:2016gbl,Pan:2010hz,Taracchini:2012ig}.
The main differences are as follows:
\begin{itemize}
\item We correct the $\mathcal{O}(v^5 \delta \chi_A \nu)$ coefficient in $\rho_{22}$, whose value is $19/42$, but was mistakenly replaced in the \texttt{SEOBNRv4} code by $196/42$.

\item We add in $\rho_{22}$ the NLO spin-squared contribution at 3PN and the LO spin-cubed part at 3.5PN, which are given by Eq. (4.11a) of Ref.~\cite{Henry:2022dzx}.

\item We add all the known spin terms in the (3,2) and (4,3) amplitudes (Eqs. (B2a) and (B5b) from Ref.~\cite{Henry:2022dzx}).

\item We correct the expressions for the (2,1) mode.
As pointed out in Ref.~\cite{Henry:2022dzx}, the $\mathcal{O}(v^6 \chi^2 \nu^2)$ terms in the $(2,1)$ mode in the \texttt{SEOBNRv4HM} model~\cite{Cotesta:2018fcv} are not correct, 
as well as the $\mathcal{O}(\nu v^5)$ nonspinning part of $\delta_{21}$, whose coefficient had the value $-493/42$ \cite{Damour:2008gu,Pan:2011gk} instead of $-25/2$, due to an error in the (2,1) mode in Ref.~\cite{Blanchet:2008je}, which was later corrected in an erratum.

\item We consistently include the high-order PN terms from Appendix A of Ref.~\cite{Cotesta:2018fcv} in the RR force, and not just in the waveform modes.
\end{itemize}

The new terms we add in the modes were derived in Ref.~\cite{Henry:2022dzx}, which was made public when the model was close to being finalized; 
hence, we only added the terms we considered most important, and we will add in a future update of the model all the 3.5PN contributions to the waveform modes, as derived in Refs.~\cite{Henry:2022dzx,Henry:2022ccf}. 
We remark that adding additional PN information in the waveform modes (except for the phases) modifies the energy flux (i.e., the RR force), and would require a recalibration of the EOB dynamics to NR simulations.

As discussed in the \texttt{SEOBNRv4HM} model of Ref.~\cite{Cotesta:2018fcv}, the presence of minima, close to merger, in the amplitude 
of some modes, leads to the introduction of additional calibration parameters before applying the 
NQC corrections. The modes for which this is needed are the $(2,1)$, $(5,5)$ and $(4,3)$. The minima occur for $q \sim 1$ and large $|\chi_A|$,
and can lead to unphysical features in the amplitude after applying the NQC corrections if they occur close to the attachment point
$t\sim t_{\text{match}}$. For the $(2,1)$ mode, this behavior is also found in NR simulations, while for the $(5,5)$ and $(4,3)$ we do not observe it 
in the NR waveforms at our disposal, and is likely an artifact of the PN-expanded modes \cite{Cotesta:2018fcv}. Calibration terms in the modes 
take the form $c_{\ell m} v_\Omega^{\beta_{\ell m}}$, and are added in $f_{\ell m}$, 
with $\beta_{\ell m}$ being the lowest PN order not already included.
The calibration parameter $c_{\ell m}$ is determined by imposing the following condition:
\begin{equation}\
	\label{eq:cal_par}
	\begin{aligned}
	\left|h_{\ell m}^{\text{F}}\left(t_{\text {match }}^{\ell m}\right)\right| &\equiv\left|h_{\ell m}^{\text{N}} \hat{S}_{\mathrm{eff}} T_{\ell m} e^{\mathrm{i} \delta_{\ell m}} f_{\ell m}\left(c_{\ell m}\right)\right|_{t=t_{\text {match }}^{\ell m}}, \\
	&=\left|h_{\ell m}^{\mathrm{NR}}\left(t_{\text {match }}^{\ell m}\right)\right|, \quad \text { for }(\ell, m)=(2,1),(5,5),(4,3),
	\end{aligned}
\end{equation}
where $|h_{\ell m}^{\mathrm{NR}}\left(t_{\text {match }}^{\ell m}\right)|$ is the amplitude of the NR modes at the matching point,
given by fits in parameter space in Appendix~\ref{appx:NQC_fits}.

The remaining $N_{\ell m}$ factor in the inspiral-plunge modes (\ref{eq:hellm_insp}) is the NQC correction and reads
\begin{equation}
	\begin{aligned}
	N_{\ell m} &=\left[1+\frac{\hat{p}_{r_*}^2}{(r \Omega)^2}\left(a_1^{h_{\ell m}}+\frac{a_2^{h_{\ell m}}}{\hat{r}}+\frac{a_3^{h_{\ell m}}}{\hat{r}^{3 / 2}}\right)\right] \\
	&\quad \times \exp \left[i\left(b_1^{h_{\ell m}} \frac{\hat{p}_{r_*}}{r \Omega}+b_2^{h_{\ell m}} \frac{\hat{p}_{r_*}^3}{r \Omega}\right)\right],
	\end{aligned}
\end{equation}
where $\hat{r}\equiv r/M$ and $\hat{p}_{r_*}\equiv p_{r_*}/\mu$.
The use of the NQC corrections guarantees that the modes' amplitude and frequency agree with NR input values (see below), given in Appendix~\ref{appx:NQC_fits},
at the matching point $t^{\ell m}_{\text{match}}$. In particular, one fixes the 5 constants $(a_1^{h_{\ell m}}$, $a_2^{h_{\ell m}}$, $a_3^{h_{\ell m}}$, $b_1^{h_{\ell m}}$, $b_2^{h_{\ell m}})$ by requiring the following ~\cite{Taracchini:2013rva,Bohe:2016gbl,Cotesta:2018fcv}:

\begin{itemize}
	\item The amplitude of the EOB modes is the same as that of the NR modes at the matching point $t_{\textrm{match}}^{\ell m}$:
	\begin{equation}
	\label{eq:NQC_condition_1}
	\left| h_{\ell m}^{\textrm{insp-plunge}}(t_{\textrm{match}}^{\ell m}) \right| = \left|h_{\ell m}^\textrm{NR}(t_{\textrm{match}}^{\ell m})\right|.
	\end{equation}
	We note that this condition is different from that in Eq.~\eqref{eq:cal_par} because it affects $h_{\ell m}^{\textrm{insp-plunge}}(t_{\textrm{match}}^{\ell m})$ and not $h_{\ell m}^\textrm{F}(t_{\textrm{match}}^{\ell m})$.
	Because of the calibration parameter in Eq.~\eqref{eq:cal_par}, for the modes (2,1), (5,5) and (4,3), this condition becomes simply $|N_{\ell m}| = 1$.
	\item The first derivative of the amplitude of the EOB modes is the same as that of the NR modes at the matching point $t_{\textrm{match}}^{\ell m}$:
	\begin{equation}
	\label{eq:NQC_condition_2}
	\left. \frac{d\left| h_{\ell m}^{\textrm{insp-plunge}}(t) \right|}{dt} \right|_{t =t_{\textrm{match}}^{\ell m}} =
	\left. \frac{d\left| h_{\ell m}^{\textrm{NR}}(t) \right|}{dt} \right|_{t =t_{\textrm{match}}^{\ell m}}.
	\end{equation}
	\item The second derivative of the amplitude of the EOB modes is the same as that of the NR modes at the matching point $t_{\textrm{match}}^{\ell m}$:
	\begin{equation}
	\label{eq:NQC_condition_3}
	\left. \frac{d^2\left| h_{\ell m}^{\textrm{insp-plunge}}(t) \right|}{dt^2} \right|_{t = t_{\textrm{match}}^{\ell m}}
	= \left. \frac{d^2\left| h_{\ell m}^{\textrm{NR}}(t) \right|}{dt^2} \right|_{t = t_{\textrm{match}}^{\ell m}}.
	\end{equation}
	\item The frequency of the EOB modes is the same as that of the NR modes at the matching point $t_{\textrm{match}}^{\ell m}$:
	\begin{equation}
	\label{eq:NQC_condition_4}
	\omega_{\ell m}^{\textrm{insp-plunge}}(t_{\textrm{match}}^{\ell m}) = \omega_{\ell m}^{\textrm{NR}}(t_{\textrm{match}}^{\ell m}).
	\end{equation}
	\item The first derivative of the frequency of the EOB modes is the same as that of the NR modes at the matching point $t_{\textrm{match}}^{\ell m}$:
	\begin{equation}
	\label{eq:NQC_condition_5}
	\left .\frac{d {\omega}_{\ell m}^{\textrm{insp-plunge}}(t)}{dt}\right|_{t = t_{\textrm{match}}^{\ell m}}  =
	\left . \frac{d {\omega}_{\ell m}^{\textrm{NR}}(t)}{dt}\right|_{t = t_{\textrm{match}}^{\ell m}}.
	\end{equation}
\end{itemize}
The RHS of Eqs. \eqref{eq:NQC_condition_1}--\eqref{eq:NQC_condition_5} (usually called \emph{input values}), are given as fitting formulae for every point of the
parameter space $(\nu,\chi_1,\chi_2)$ in Appendix \ref{appx:NQC_fits}. These fits are produced using the NR \texttt{SXS} catalog \cite{SXS:catalog, Boyle:2019kee}, 
and BH-perturbation-theory waveforms described in Sec.~\ref{sec:calibration}. We point out that the NQC corrections and the $c_{\ell m}$ calibration coefficients are not included in the 
\texttt{SEOBNRv5HM} radiation-reaction force. 

In the \texttt{SEOBNRv5} model, the input values are enforced at $t = t^{\ell m}_{\text{match}}$, given in Eq.~(\ref{eq:t_match}) as a function of $t_{\text{peak}}^{22}$. We take
\begin{equation}
	\label{eq:t_attach}
	t_{\text{peak}}^{22} = t_{\rm{ISCO}}+ \Delta t^{22}_{\rm{ISCO}}\,,
\end{equation}
where $t_{\rm{ISCO}}$ is the time at which $r = r_{\rm{ISCO}}$, with $r_{\rm{ISCO}}$ the radius of the geodesic ISCO in Kerr spacetime \cite{Bardeen:1972fi} with the same mass and spin as the remnant, 
computed with NR fitting formulas ~\cite{Jimenez-Forteza:2016oae, Hofmann:2016yih}, 
and $\Delta t^{22}_{\rm{ISCO}}$ is a calibration parameter, to be determined by comparing against NR simulations. In the \texttt{SEOBNRv4} model, the merger time was given by
\begin{equation}
	\label{eq:t_attach_v4}
	t_{\text{peak}}^{22} = t_{\text{peak}}^{\Omega} + \Delta t^{22}_{\text{peak}}\,,
\end{equation}
with $t_{\text{peak}}^{\Omega}$ being the peak of the orbital frequency.
The purpose of $\Delta t^{22}_{\text{peak}}$ is still to introduce a time delay between the peak of the orbital frequency and the peak of the (2,2) mode, as observed in the
test-body limit \cite{Barausse:2011kb,Taracchini:2014zpa,Price:2016ywk}.
However, we find the new definition to be more robust, since it is independent of features in the late dynamics,
like the existence of a peak in the orbital frequency, which is not necessarily present for all
BBH parameters when the Hamiltonian and modes are not the same as the ones used in the \texttt{SEOBNRv4} model. More specifically, in the latter 
the $A$-potential was designed ($\log$-resummation)~\cite{Barausse:2009xi,Barausse:2011ys} in such a way always to guarantee the presence of the light ring (the peak in the 
orbital frequency) for aligned-spin binaries. This is no longer the case when the Pad\'e resummation of the $A$-potential is employed, as done in \texttt{SEOBNRv5} (see below). 

Another notable improvement in the \texttt{SEOBNRv5HM} waveforms is the addition of 2GSF calibration coefficients in the nonspinning modes 
and RR force from Ref.~\cite{VandeMeentv5}. In that work, one defines
\begin{equation}
	\rho_{\ell m}=\rho_{\ell m}^{(0)}+\nu \rho_{\ell m}^{(1)}+\mathcal{O}\left(\nu^2\right),
\end{equation}
and augments the $\rho_{\ell m}^{(1),\rm EOB}$ by adding an extra polynomial $\Delta\rho_{\ell m}^{(1)}$ in $v^2_\Omega$
starting at the lowest order in  $v^2_\Omega$ not already included.
The $\Delta\rho_{\ell m}^{(1)}$ are determined by fitting to the numerical $\rho_{\ell m}^{(1),\rm GSF}$ results, leading to the following expressions:
\begin{subequations}
\begin{align}
	\Delta\rho_{22}^{(1)} &=
	21.2 v^8_\Omega
	-411v^{10}_\Omega,
	\\
	\Delta\rho_{21}^{(1)} &=
	1.65  v^6_\Omega
	+26.5 v^8_\Omega
	+80   v^{10}_\Omega,
	\\
	\Delta\rho_{33}^{(1)} &=
	12 		v^8_\Omega
	-215 	v^{10}_\Omega,
	\\
	\Delta\rho_{32}^{(1)} &=
	0.333  v^6_\Omega
	-6.5 v^8_\Omega
	+98   v^{10}_\Omega,
	\\
	\Delta\rho_{44}^{(1)} &=
	-3.56  v^6_\Omega
	+15.6 v^8_\Omega
	-216   v^{10}_\Omega,
	\\
	\Delta\rho_{43}^{(1)} &=
	-0.654  v^4_\Omega
	-3.69 v^6_\Omega
	+18.5   v^8_\Omega,
	\\
	\Delta\rho_{55}^{(1)} &=
	-2.61  v^4_\Omega
	+1.25 v^6_\Omega
	-35.7   v^8_\Omega.
\end{align}
\end{subequations}
In the 2GSF calibration, terms $\Delta\rho_{\ell m}^{(1)}$ are then added directly to the full (not $\nu$-expanded) $\rho_{\ell m}$ coefficients.
In Ref.~\cite{VandeMeentv5}, it is also found beneficial to include additional terms in the (3,2) and (4,3) modes
obtained by matching to the PN expansions of the test-mass limit (TML) GW energy flux. Thus, we add the following terms:
\begin{subequations}
	\begin{align}
	\label{eq:rho32tml}
	\Delta\rho_{32}^{(0), {\rm{TML}}} &=
	\left( - \frac{1312549797426453052}{176264081083715625} \nonumber \right. \\
	&\qquad+ \left. \frac{18778864}{12629925} \text { eulerlog}(2,v_{\Omega} ) \right) v_{\Omega}^{10},
	\\
	\label{eq:rho43tml}
	\Delta\rho_{43}^{(0), {\rm{TML}}} &=
	\left( - \frac{2465107182496333}{460490801971200} \nonumber \right. \\
	&\qquad+ \left. \frac{174381}{67760} \text { eulerlog}(3,v_{\Omega} ) \right) v_{\Omega}^{8},
\end{align}
\end{subequations}
where we define
\begin{equation}
\label{eulerlog}
	\text { eulerlog }\left(m, v_{\Omega}\right) \equiv \gamma_E+\log \left(2 m v_{\Omega}\right),
\end{equation}
in which $\gamma_E$ is the Euler constant.

\subsection{Merger-ringdown $h_{\ell m}$ modes}

The merger-ringdown modes are constructed with a phenomenological ansatz,
using information from NR simulations and TML waveforms.
The ansatz we employ for the modes $(2,2)$, $(3,3)$, $(2,1)$, $(4,4)$, $(5,5)$, which show monotonic amplitude and frequency evolution,
is the same as the one implemented in Refs.~\cite{Bohe:2016gbl, Cotesta:2018fcv} and reads:
\begin{equation}
h_{\ell m}^{\text {merger-RD }}(t)=\nu \tilde{A}_{\ell m}(t) e^{i \tilde{\phi}_{\ell m}(t)} e^{-i \sigma_{\ell m 0}\left(t-t_{\text {match }}^{\ell m}\right)},
\label{eq:ansatz_hlm}
\end{equation}
where $\sigma_{\ell m 0} = \sigma_{\ell m}^\mathrm{R} - i \sigma_{\ell m}^\mathrm{I} $ is the complex frequency of the least-damped quasi-normal mode (QNM) of the remnant BH.
The QNM frequencies are obtained for each $(\ell, m)$ mode as a function of the BH's final mass and spin using the \texttt{qnm} Python package \cite{Stein:2019mop}.
The BH's mass and spin are in turn computed using the fitting formulas of Refs.~\cite{Jimenez-Forteza:2016oae} and \cite{Hofmann:2016yih}, respectively.
The ansätze for the two functions $\tilde{A}_{\ell m}$ and $\tilde{\phi}_{\ell m}$ in Eq.~(\ref{eq:ansatz_hlm}) are the following \cite{Bohe:2016gbl, Cotesta:2018fcv}
\begin{equation}
\tilde{A}_{\ell m}(t)=c_{1, c}^{\ell m} \tanh \left[c_{1, f}^{\ell m}\left(t-t_{\text {match }}^{\ell m}\right)+c_{2, f}^{\ell m}\right]+c_{2, c}^{\ell m},
\label{eq:ansatz_a}
\end{equation}
\begin{equation}
\tilde{\phi}_{\ell m}(t)=
\phi_{\text {match }}^{\ell m}-d_{1, c}^{\ell m} \log \left[\frac{1+d_{2, f}^{\ell m} e^{-d_{1, f}^{\ell  m}\left(t-t_{\text {match}}^{\ell_{m}}\right)}}{1+d_{2, f}^{\ell m}}\right],
\label{eq:ansatz_phi}
\end{equation}
where $\phi_{\text {match }}^{\ell m}$ is the phase of the inspiral-plunge mode $(\ell, m)$ at $t=t_{\text {match }}^{\ell m}$.
The coefficients $d_{1, c}^{\ell m}$ and $c_{i, c}^{\ell m}$ ($i=1,2$) are constrained by the requirement that 
the amplitude and phase of $h_{\ell m}(t)$ in Eq.~\eqref{eq:h_match}
are continuously differentiable at $t=t_{\text {match}}^{\ell m}$, 
and can  be written in terms of $c_{1, f}^{\ell m}$, $c_{2, f}^{\ell m}$, $\sigma_{\ell m}^{\mathrm{R}}$,
$\left|h_{\ell m}^{\text {insp-plunge }}\left(t_{\text {match }}^{\ell m}\right)\right|$,
$\partial_{t}\left|h_{\ell m}^{\text {insp-plunge }}\left(t_{\text {match }}^{\ell m}\right)\right|$, as follows
\begin{equation}
	\begin{aligned}
	c_{1, c}^{\ell m} &=\frac{1}{c_{1, f}^{\ell m} \nu}\left[\partial_{t}\left|h_{\ell m}^{\text {insp-plunge }}\left(t_{\text {match }}^{\ell m}\right)\right|\right.\\
	&\quad \left.-\sigma_{\ell m}^{\mathrm{R}}\left|h_{\ell m}^{\text {insp-plunge }}\left(t_{\text {match }}^{\ell m}\right)\right|\right] \cosh ^{2}\left(c_{2, f}^{\ell m}\right),
	\end{aligned}
	\label{eq:c1c}
\end{equation}
\begin{equation}
	\begin{aligned}
	c_{2, c}^{\ell m} =&\frac{\left|h_{\ell m}^{\text {insp-plunge }}\left(t_{\text {match }}^{\ell m}\right)\right|}{\nu}
	-\frac{1}{c_{1, f}^{\ell m} \nu}\left[\partial_{t}\left|h_{\ell m}^{\text {insp-plunge }}\left(t_{\text {match }}^{\ell m}\right)\right|\right.\\
	&\left.-\sigma_{\ell m}^{\mathrm{R}}\left|h_{\ell m}^{\text {insp-plunge }}\left(t_{\text {match }}^{\ell m}\right)\right|\right]
	\cosh \left(c_{2, f}^{\ell m}\right) \sinh \left(c_{2, f}^{\ell m}\right),
	\end{aligned}
	\label{eq:c2c}
\end{equation}
or in terms of $d_{1, f}^{\ell m}, d_{2, f}^{\ell m}, \sigma_{\ell m}^{\mathrm{I}}, \omega_{\ell m}^{\text {insp-plunge }}\left(t_{\text {match }}^{\ell m}\right)$ for $d_{1,c}^{\ell m}$
\begin{equation}
d_{1, c}^{\ell m}=
\left[\omega_{\ell m}^{\text {insp-plunge }}\left(t_{\text {match }}^{\ell m}\right)-\sigma_{\ell m}^{\mathrm{I}}\right]
\frac{1+d_{2, f}^{\ell m}}{d_{1, f}^{\ell m} d_{2, f}^{\ell m}}.
\label{eq:d1c}
\end{equation}

The remaining parameters in Eqs.~(\ref{eq:ansatz_a}) and (\ref{eq:ansatz_phi})
are the free coefficients $c^{\ell m}_{i,f}$ and $d^{\ell m}_{i,f}$, $i=1,2$.

The NQC corrections ensure that the waveform's amplitude and frequency 
coincide with the NR input values at $t=t_{\text {match }}^{\ell m}$, and make the merger-ringdown
modes independent of the EOB inspiral modes, allowing for an independent calibration of the two.
To obtain $c^{\ell m}_{i,f}$ and $d^{\ell m}_{i,f}$, we first extract them from each NR and TML waveform by least-square fits, 
and then interpolate the values obtained across the parameter space using polynomial fits in $\nu$ and $\chi$.
While in Ref.~\citep{Cotesta:2018fcv}
the same polynomial was used for most of the free coefficients, in this work we use a recursive-feature-elimination (RFE) \cite{Guyon:2002}
algorithm with polynomial features of third and fourth order, depending on the quantity to fit.
Applying a $\log$ transformation to some of the coefficients is also beneficial,
both to improve the quality of the fits and to ensure the positivity of those quantities when extrapolating
outside of the region where NR data is available.
Finally, we apply a similar RFE strategy to most of the fits for the input values,
the only exceptions being the fits of the amplitude of the odd-m modes and their derivatives.
The odd $m$ modes vanish in the equal-mass and equal-spin limit, since they need to satisfy the symmetry under rotation
$ \varphi_0 \rightarrow \varphi_0 +\pi $, therefore, the corresponding amplitudes are better captured by ad-hoc
nonlinear ansätze that enforce this limit by construction (see also Appendix \ref{appx:coeff_fits}).

\subsection{Mode mixing in the (3,2) and (4,3) merger-ringdown $h_{\ell m}$  modes}

The merger-ringdown $(3,2)$ and $(4,3)$ modes show post-merger oscillations~\cite{Buonanno:2006ui, Kelly:2012nd}, mostly
related to the mismatch between the \textit{spherical} harmonic basis used for extraction in NR simulations, and
the \textit{spheroidal} harmonics adapted to the perturbation theory of Kerr BHs. Because of this, it is not
possible to use the same ansatz of Eqs.~(\ref{eq:ansatz_hlm}),~(\ref{eq:ansatz_a}) and (\ref{eq:ansatz_phi}) straightforwardly.

Equation~(\ref{eq:hoft_sphericalH}) can be formulated in terms of $-2$ spin-weighted \textit{spheroidal} harmonics as:
\begin{equation}
	h(t;  \bm{\lambda}, \iota, \varphi_0) = \sum_{\ell^{\prime}\geq 2} \sum_{|m|\leq \ell^{\prime}}
	\sum_{n\geq 0} {}_{-2} S_{\ell^{\prime}mn} (\iota, \varphi_0) \; {}^{S}h_{\ell mn}(t,\bm{\lambda}),
\label{eq:hoft_spheroidalH}
\end{equation}
where $S_{\ell mn}\equiv S_{\ell m}(a_f \sigma_{\ell m n})$ are the $-2$ spin-weighted spheroidal harmonics associated with the QNM frequencies
$\sigma_{\ell  m n}$, and with $a_f M_f$ being the spin angular momentum of the final BH of mass $M_f$~\cite{Berti:2005gp}.
The superscript $S$ denotes that the ${}^{S}h_{\ell mn}$ modes are expanded in the spheroidal harmonics basis.

One can switch from the spherical harmonic basis to the spheroidal harmonic basis via:
\begin{equation}
	{}_{-2}S_{\ell^{\prime}mn} = \sum_{\ell \geq |m|} \mu^{*}_{m\ell \ell^{\prime}n}  {}_{-2}Y_{\ell m}\,,
\label{eq: basis_tansform}
\end{equation}
where $\mu_{m\ell \ell^{\prime}n}$ are mode mixing coefficients, which we compute using fits provided in Ref.~~\cite{Berti:2014fga}
(more complex fits can be found in Ref.~\cite{London:2018nxs}), and the star denotes the usual complex conjugation.
Inserting Eq.~(\ref{eq: basis_tansform}) in Eq.~(\ref{eq:hoft_spheroidalH}) for the spheroidal harmonics we get,
\begin{equation}
	h(t; \iota, \varphi_0) =\sum_{\ell^{\prime}\geq 2} \sum_{|m|\leq \ell^{\prime}}  \sum_{n\geq 0} \sum_{\ell \geq |m|}
{}_{-2}Y_{\ell m} (\iota, \varphi_0) \, {}^{S}h_{\ell mn}(t)\,	\mu^{*}_{m\ell \ell^{\prime}n},
\label{eq:spheroidal_to_spherical}
\end{equation}
where we have suppressed the $\bm{\lambda}$ parameter from the expression to ease the notation.
Comparing Eq.~(\ref{eq:spheroidal_to_spherical}) with Eq.~(\ref{eq:hoft_sphericalH}), we obtain the following relation between spherical and spheroidal modes,
\begin{equation}
	h_{\ell m} (t) =  \sum_{\ell^{\prime} \geq |m|}   \sum_{n\geq 0}  {}^{S}h_{\ell^{\prime} mn}(t) \, \mu^{*}_{m\ell \ell^{\prime}n}.
\label{eq:modes_basis_change}
\end{equation}

Starting from Eq.~(\ref{eq:modes_basis_change}), we can model the mode-mixing behavior \citep{KumarMehta:2019izs,Estelles:2020twz} 
to obtain monotonic functions that can be fitted by the ansatz already used for the other modes.
Practically, it is not feasible to sum over all the spheroidal modes to get each spherical mode, so we make a few reasonable approximations. 
First, we neglect the overtone ($n>0$) contributions in the right-hand side of Eq.~(\ref{eq:modes_basis_change}), because their decay times
are $\gtrsim 3$ times smaller than the dominant overtone $n=0$. Second, for a given $(\ell, m)$ mode, we neglect the contributions
from the spheroidal modes with $\ell^{\prime}> \ell $ since their amplitudes are subdominant compared to the $(\ell, m, 0)$ mode. 
With these approximations, we can rewrite Eq.~(\ref{eq:modes_basis_change}) as
\begin{equation}
h_{\ell m} (t) \simeq  \sum_{\ell^{\prime}\leq \ell}  {}^{S}h_{\ell^{\prime} m0}(t) \,\mu^{*}_{m\ell \ell^{\prime}0}.
\label{eq:modes_basis_change_final}
\end{equation}
Writing it explicitly for the modes of interest,
\begin{subequations}
\begin{align}
h_{22}(t) &\simeq \mu_{2220}^{*}\, {}^{S}h_{220}(t),\\
h_{33}(t)&\simeq \mu_{3330}^{*}\, {}^{S}h_{330}(t),\\
h_{32}(t)& \simeq \mu_{2320}^{*}\, {}^{S}h_{220} (t)+ \mu_{2330}^{*}\, {}^{S}h_{320}(t), \label{eq:mixed_modes_1}\\
h_{43}(t)& \simeq \mu_{3430}^{*}\, {}^{S}h_{330} (t)+ \mu_{3440}^{*}\, {}^{S}h_{430}(t) \label{eq:mixed_modes_2}.
\end{align}
\end{subequations}
From these equations, we can solve for the ${}^{S}h_{\ell m0}$ modes to obtain
\begin{subequations}
\begin{align}
{}^{S}h_{320}(t)&\simeq \dfrac{ h_{32}(t)\mu_{2220}^{*} - h_{22}(t) \mu_{2320}^{*}}{\mu_{2330}^{*}\mu_{2220}^{*}},
\label{eq:unmixed_modes_1} \\
{}^{S}h_{430}(t)&\simeq \dfrac{h_{43}(t)\mu_{3440}^{*} - h_{33}(t)\mu_{3430}^{*}}{\mu_{3330}^{*}\mu_{3440}^{*}}.
\label{eq:unmixed_modes_2}
\end{align}
\end{subequations}

\begin{figure*}[htb!]
	\includegraphics[width=1.0\columnwidth]{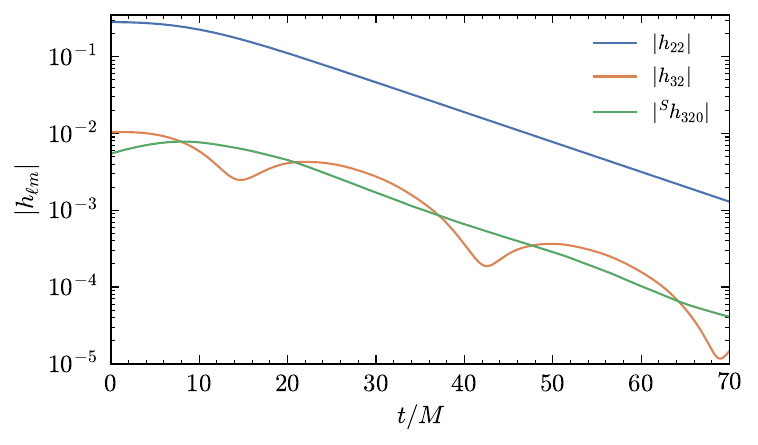}
	\includegraphics[width=1.0\columnwidth]{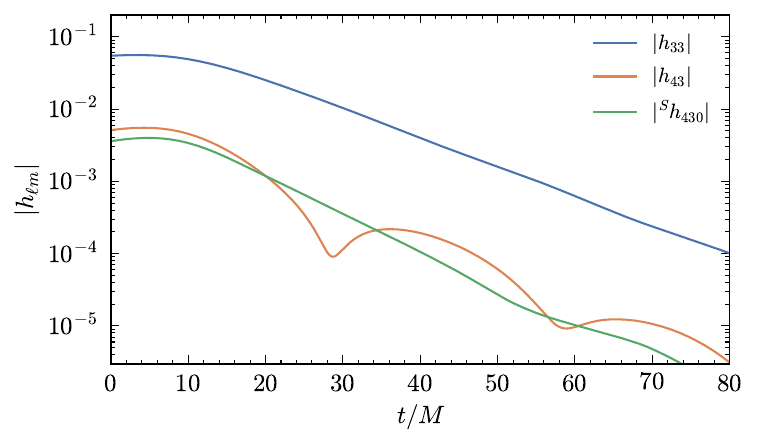}
	\includegraphics[width=1.0\columnwidth]{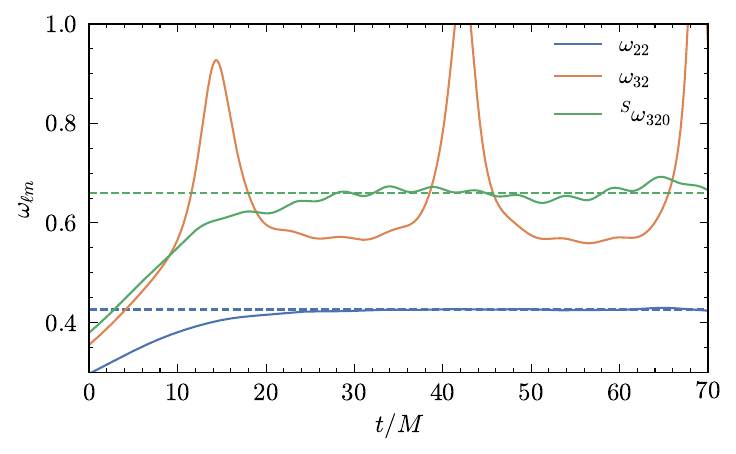}
	\includegraphics[width=1.0\columnwidth]{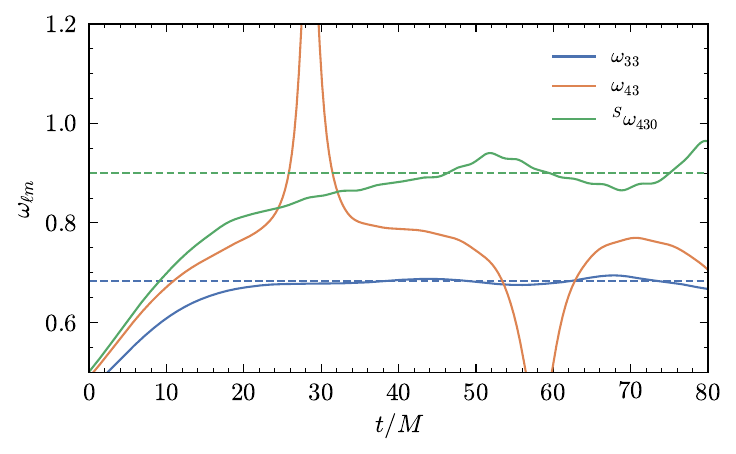}
	\caption{
	Mode-mixing in the NR simulation \texttt{SXS:BBH:2138} ($q=3.0, \chi_{1}=-0.6, \chi_{2}=0.4$).
	\textit{Upper panel}: Amplitude of the modes $|h_{\ell m}|$ and of $|{}^S h_{\ell m0}|$, after the mode-mixing removal
	(Eqs.~(\ref{eq:unmixed_modes_1}) and (\ref{eq:unmixed_modes_2})). We denote with $t=0$ the time of the peak of the $(2,2)$-mode amplitude.
	\textit{Lower panel}: Frequencies of the modes $h_{\ell m}$ and of ${}^S h_{\ell m0}$.
	The ringdown frequencies of the ${}^{S} h_{320}$ and ${}^{S} h_{430}$ modes are well approximated by the (3,2,0) and (4,3,0) QNM frequencies
	(dashed horizontal lines) after the mode-mixing removal.}
	\label{fig:mode_mixing_remov}
\end{figure*}

We show in Fig.~\ref{fig:mode_mixing_remov} the characteristics of the ${}^{S}h_{\ell m 0}$ mode obtained from the spherical mode $h_{\ell m}$ via Eqs.~(\ref{eq:unmixed_modes_1})
and~(\ref{eq:unmixed_modes_2}) for the NR waveform \texttt{SXS:BBH:2138}. The $h_{32}$ mode shows oscillations in its amplitude and frequency,
while the ${}^{S}h_{320}$ mode obtained from Eq.~(\ref{eq:unmixed_modes_1}) has a nearly monotonic behavior.
Most importantly, the frequency of the ${}^{S}h_{320}$ mode oscillates around the QNM frequency predicted in BH perturbation theory for the spheroidal (3,2,0) mode.

Thus, we model the spheroidal ${}^{S}h_{lm0}$ modes using the ansatz of Eq.~(\ref{eq:ansatz_hlm}),
where in Eq.~(\ref{eq:ansatz_phi}) $\phi^{\rm{match}}_{\ell m}$ is replaced by ${}^{S}\phi^{\rm{match}}_{\ell m 0}$, which is
the phase of ${}^{S}h_{lm0}$ at $t=t^{\rm{match}}_{\ell m}$. In Eqs.~(\ref{eq:c1c}) and (\ref{eq:c2c}) we replace $h_{\ell m}$ 
by ${}^{S}h_{lm0}$, and in Eq.~(\ref{eq:d1c}) we replace $\omega_{\ell m}$ by ${}^{S}\omega_{\ell m 0}$. Once we have a model for ${}^{S}h_{320}$ and ${}^{S}h_{430}$,
it is straightforward to obtain the (3,2) and (4,3) modes by combining them with the (2,2) and (3,3) ones previously obtained by inverting Eqs.~(\ref{eq:unmixed_modes_1}) and (\ref{eq:unmixed_modes_2}).

The NQC corrections for the inspiral-plunge $h_{\ell m}$ modes require the values for the spherical NR modes $h_{\ell m}^{\rm{NR}} (t_{\ell m}^{\rm{match}})$, 
and those are the quantities that we fit and interpolate across the parameter space.
However, we need the input values for ${}^{S}h_{lm0}^{\rm{match}} \equiv {}^{S}h_{lm0}(t_{\ell m}^{\rm{match}})$ and its derivative in order to fix 
the coefficients $c_{i,c}^{\ell m}$ and $d_{i, c}^{\ell m}$. They can be derived from Eqs.~(\ref{eq:unmixed_modes_1}) and~(\ref{eq:unmixed_modes_2}) starting from the $h_{\ell m}$ input values.

First, we introduce the following quantities: 
\begin{subequations}
\begin{align}
\rho &= |\mu_{m \ell \ell^{\prime} 0}| \dfrac{|h_{\ell^{\prime} m}^{\rm{match}}|}{|\mu_{m \ell^{\prime} \ell^{\prime} 0}||h_{\ell m}^{\rm{match}}|}, \\
\delta \phi &= \phi^{\ell^{\prime} m}_{\rm{match}} - \phi^{\ell m}_{\rm{match}}  - \arg(\mu_{m \ell \ell^{\prime} 0}) + \arg(\mu_{m \ell^{\prime} \ell^{\prime} 0}),\\
F &= \sqrt{  (1-\rho \cos(\delta \phi))^{2} + \rho^2\sin^{2}(\delta \phi) },\\
\dot{\rho} &= |\mu_{m \ell \ell^{\prime} 0}| \Bigg( \dfrac{ \partial_{t}|h_{\ell^{\prime} m}^{\rm{match}} |} {|h_{\ell m}^{\rm{match}}| }
- \dfrac{|h_{\ell^{\prime} m}^{\rm{match}}| }{|h_{\ell m}^{\rm{match}}|^2 }\partial_{t}|h_{\ell m}^{\rm{match}}| \Bigg), \\
\delta\dot{\phi} &= \partial_{t}\phi^{\ell^{\prime} m}_{\rm{match}} - \partial_{t} \phi^{\ell m}_{\rm{match}},\\
\dot{F} &= \frac{(\rho \dot{\rho} + \rho\sin(\delta \phi) \delta\dot{\phi} - \dot{\rho}\cos(\delta \phi) )}{F}, 
\end{align}
\end{subequations}
where $|h_{\ell m}^{\rm{match}}|\equiv \left|h_{\ell m}^{\text {insp-plunge }}\left(t_{\text {match }}^{\ell m}\right)\right|$. 
Then,
\begin{subequations}
\begin{align}
|^{S}h_{\ell m 0}^{\rm{match}}| &= \dfrac{|h_{\ell m}^{\rm{match}}| F}{|\mu_{m\ell \ell 0}|}, \\
^{S}\phi_{\rm{match}}^{\ell m 0} &= \phi_{\rm{match}}^{\ell m} + \arg(\mu_{m \ell \ell 0}) + \arctan\Bigg( \dfrac{-\rho \sin(\delta \phi)}{1-\rho\cos(\delta \phi)}  \Bigg), \\
\partial_{t} |^{S}h_{\ell m 0}^{\rm{match}}|  &= \dfrac{ (\partial_{t}|h_{\ell m}^{\rm{match}}| F + |h_{\ell m}^{\rm{match}}| \dot{F}) }{|\mu_{m\ell \ell 0}|}, \\
^{S}\omega_{\ell m 0}^{\rm{match}} &= \omega_{\ell m }^{\rm{match}} + \frac{(\rho^2 \delta \dot{\phi} - \rho \cos(\delta \phi)\delta \dot{\phi} - \dot{\rho}\sin(\phi))}{F^2},
\end{align}
\end{subequations}
where for the (3,2) mode $m=2, \ell=3, \ell^{\prime} =2$, and for the (4,3) mode $m=3, \ell=4, \ell^{\prime} =3$.

\section{Calibration to numerical-relativity waveforms}
\label{sec:calibration}

The inspiral-plunge modes described in Sec. \ref{sec:performance} are functions of the binary parameters $(q, \chi_1, \chi_2)$, the initial orbital frequency
$\Omega_0$ at which the evolution is started, and a set of calibration parameters, which are determined as a function of $(q, \chi_1, \chi_2)$ such that we maximize the agreement between the waveform model and
NR simulations of BBHs. In the \texttt{SEOBNRv5} model we employ the following calibration parameters:
\begin{itemize}
	\item $\bm{a_6}$: a 5PN, linear in $\nu$, parameter  that enters the nonspinning $A_\text{noS}(u)$ potential of Eq.~(\ref{eq:Ataypm}). 
 	\item $\bm{d_{\rm{SO}}}$: a 4.5PN spin-orbit parameter, that enters the odd-in-spin part of the effective Hamiltonian (see Eqs.~(\ref{HeffAnzAlign}) and (\ref{eq:SO_calib})).
    \item $\bm{\Delta t^{22}_{\rm{ISCO}}}$: a parameter that determines the time shift between the Kerr ISCO, computed
	from the final mass and spin of the remnant~\cite{Jimenez-Forteza:2016oae, Hofmann:2016yih}, and the peak of the (2,2)-mode amplitude, as given by Eq.~(\ref{eq:t_attach}).
	We remark that this quantity is different from $\Delta t^{22}_{\rm{peak}}$ used in the \texttt{SEOBNRv4} model, where it corresponded
	to the time difference between the peak of the orbital frequency (light ring) and the peak of the (2,2)-mode amplitude.
\end{itemize}

The resummation of the analytical information that enters the EOB potentials is critical in determining the model's flexibility to reduce differences with NR waveforms. 
In the \texttt{SEOBNRv5} model we perform a (1,5) Pad\'e resummation of the Taylor-expanded potential $A_\text{noS}^\text{Tay}(u)$, given by Eq.~(\ref{eq:Ataypm}), while treating $\ln u$ as a constant, i.e., we use
\begin{equation}
\label{eq:ApmPade}
A_\text{noS}(u) = P^1_5[A_\text{noS}^\text{Tay}(u)].
\end{equation}
The Pad\'e resummation of $A_\text{noS}$ was originally introduced in Ref.~\cite{Damour:2000we} to guarantee the presence of 
an ISCO in the EOB dynamics at 3PN order for any mass ratio. 
It was then adopted in nonspinning and initial spinning \texttt{EOBNR} models (e.g., see 
Refs.~\cite{Buonanno:2007pf,Pan:2011gk,Pan:2011gk,Pan:2009wj}), and in all \texttt{TEOBResumS} models 
(e.g., see Refs.~\cite{Damour:2007yf,Damour:2008qf,Nagar:2018zoe,Nagar:2020pcj,Gamba:2021ydi}).
For $\bar{D}_\text{noS}(u)$ we perform a (2,3) Pad\'e resummation of the 5PN Talyor-expanded $\bar{D}_\text{noS}^\text{Tay}(u)$ given by Eq.~\eqref{DpmTay} in Appendix~\ref{app:HamCoeffs}, such that
\begin{equation}
\label{DbpmPade}
\bar{D}_\text{noS}(u) = P^2_3[\bar{D}_\text{noS}^\text{Tay}(u)].
\end{equation}
This resummation of $\bar{D}_\text{noS}(u)$ was recently explored in Ref.~\citep{Nagar:2021xnh}, although combined with different choices for 
$A_\text{noS}(u)$ and $Q_\text{noS}(u)$ than the ones used in \texttt{SEOBNRv5}. 
{\tt TEOBResumS} includes information through 3PN order in $\bar{D}_\text{noS}(u)$, which is Taylor expanded ($D_\text{noS}(u) \equiv 1/\bar{D}_\text{noS}(u)$ is inverse-Taylor resummed)~\cite{Nagar:2020pcj,Riemenschneider:2021ppj}.

The \texttt{SEOBNRv4} model adopted a $\log$-resummation for these potentials, which was designed to guarantee the presence of a light ring 
(a peak in the orbital frequency) for aligned-spin binaries. The light ring was needed to determine the point at which 
to attach the merger-ringdown waveforms, based on $\Delta t^{22}_{\rm{peak}}$.
The use of $\Delta t^{22}_{\rm{ISCO}}$ as reference for the attachment of the merger-ringdown in the \texttt{SEOBNRv5} model eliminates the dependence on the existence of a peak in the orbital frequency.
This enables us to use resummed potentials that may not necessarily exhibit a light ring, but lead to a better agreement with NR simulations compared to the $\log$-resummed ones in \texttt{SEOBNRv4}.
\footnote{An updated NR calibration of the \texttt{SEOBNRv4} nonspinning Hamiltonian, using the \texttt{SEOBNRv5} RR force and gravitational modes, 
is presented in Appendix A of Ref.~\cite{VandeMeentv5}, confirming that the improvements observed can be predominantly attributed to the updated Hamiltonian.}

Similarly, the different resummation of the generic-spin Hamiltonian in \texttt{SEOBNRv5}, based on that of a test mass in a deformed Kerr background \cite{Damour:2001tu,Balmelli:2015zsa,Khalil:2020mmr,Khalilv5}, 
instead of on the one of a test spin \cite{Barausse:2009aa,Barausse:2011ys,Vines:2016unv} in as in \texttt{SEOBNRv4}, is a crucial factor in achieving high faithfulness compared to NR simulations.
Notably, this change allows us to reach higher accuracy with just one spin-dependent calibration parameter in the Hamiltonian ($d_{\rm{SO}}$), surpassing what could be obtained by tuning three such parameters in \texttt{SEOBNRv4}. 

We calibrate \texttt{SEOBNRv5HM} to 442 NR waveforms, all produced with the pseudo-Spectral Einstein code (\texttt{SpEC}) 
of the Simulating eXtreme Spacetimes (\texttt{SXS}) collaboration 
\cite{SXS:catalog,Boyle:2019kee,Chu:2015kft,Blackman:2017dfb,Hemberger:2013hsa,Scheel:2014ina,Lovelace:2014twa,Abbott:2016apu,Bohe:2016gbl,Blackman:2015pia,Lovelace:2016uwp,Varma:2018mmi,Abbott:2016nmj,Varma:2019csw,Kumar:2015tha,Mroue:2013xna,Yoo:2022erv}, 
except for a simulation with mass ratio and dimensionless spins $q = 8, ~\chi_1 = 0.85, ~\chi_2 = 0.85$ produced with the \texttt{Einstein Toolkit} code \cite{EinsteinToolkit:2022_11, Cotesta:2018fcv}.
We also incorporate information from 13 waveforms computed by solving the Teukolsky equation in the framework of BH 
perturbation theory~\cite{Barausse:2011kb,Taracchini:2014zpa}, with mass ratio $q = 10^{3}$ and dimensionless spins values in the range $-0.99 \leq \chi \leq 0.99$.
~\footnote{The full list of simulations is provided as an ancillary file in \url{https://arxiv.org/src/2303.18039v1/anc/NR_simulations.json}. 
For each simulation we list the mass-ratio $q$, the dimensionless spins $\chi_{1,2}$, the initial orbital frequency 
$\Omega_{0}$, the initial eccentricity $e_0$ and the number of orbits $N_{\mathrm{orb}}$ up to the merger. }

\begin{figure}
	\includegraphics[width=\linewidth]{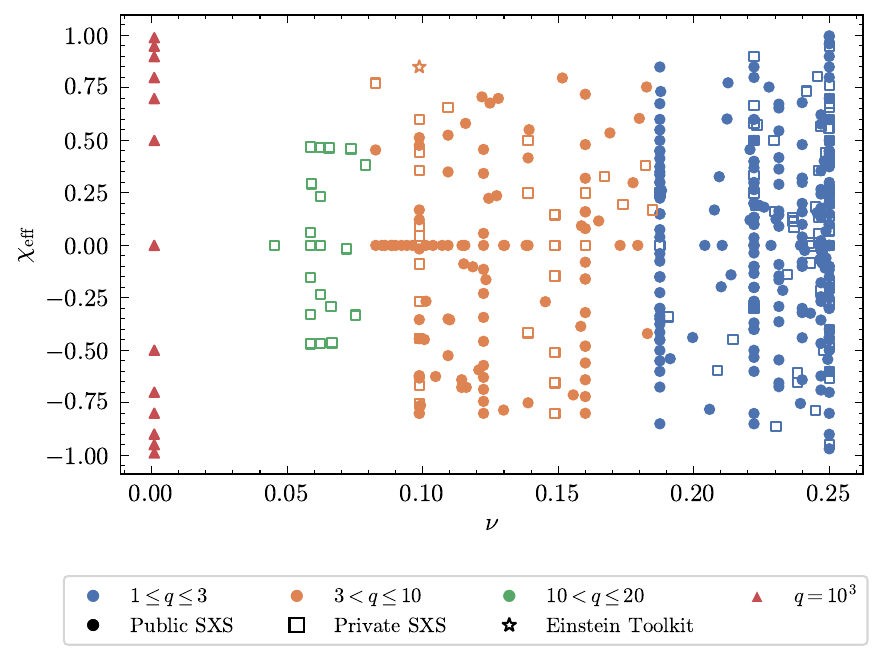}
	\caption{
	NR and BH-perturbation-theory waveforms used to calibrate \texttt{SEOBNRv5HM}, projected on the binary's parameters $\nu$ and $\chi_{\mathrm{eff}}=(\chi_1 m_1 + \chi_2 m_2)/M$.
	We highlight four regions as explained in the text, and use different markers to distinguish between 327 simulations from the public \texttt{SXS} catalog~\cite{Boyle:2019kee}, 
	114 private \texttt{SXS} waveforms, one \texttt{Einstein Toolkit} simulation and 13 Teukolsky-code waveforms.
	We refer to private waveforms as all those which can not be downloaded from the \texttt{SXS} website~\cite{SXS:catalog} at the time of this publication. }
	\label{fig:params_NR}
\end{figure}
In Fig.~\ref{fig:params_NR} we show the coverage of NR and BH-perturbation-theory waveforms projected
on the binary's parameters $\nu$ and $\chi_{\mathrm{eff}}=(\chi_1 m_1 + \chi_2 m_2)/M$, separated in different regions.
In the first region $1 \leq q \leq 3$ there is a large number of configurations
with both BHs carrying spin. The spins' magnitude reach $\chi_{1,2} = 0.998$ in the equal-mass limit,
while they are limited to $\chi_{1,2} = 0.85$ for $q = 3$. The NR coverage in this region is mostly comparable to \texttt{SEOBNRv4HM}.
The second region is between $3 < q \leq 10$. This region includes a significant number of configurations,
with primary spins $-0.9 \leq \chi_1 \leq 0.85$, and is much more densely populated than for \texttt{SEOBNRv4HM}.
The third region is between $10 < q \leq 20$ and it includes simulations with spins only on the heavier BH,
with spin magnitudes only up to $\chi_1 = 0.5$, or nonspinning waveforms. \texttt{SEOBNRv4HM} was not calibrated to any
NR simulation in this region.
Finally, the fourth region covers the 13 Teukolsky-code waveforms, with $q = 10^{3}$ and dimensionless spins values in the range $-0.99 \leq \chi \leq 0.99$.

The rest of this section explains how we determine the calibration parameters by comparing the \texttt{SEOBNRv5} waveform model to NR waveforms. 
We closely follow the procedure adopted in Ref.~\cite{Bohe:2016gbl} and highlight differences when needed.

\subsection{Calibration requirements}

In order to calibrate the waveform model to NR we first need to establish when two waveforms are close to each other.  
Given two waveforms $h_1(t)$ and $h_2(t)$, we introduce the match, which is defined as the noise-weighted inner product \cite{Finn:1992xs,Sathyaprakash:1991mt}
\begin{equation}
	\label{eq:overlap}
	\left(h_1 \mid h_2\right) \equiv 4 \operatorname{Re} \int_{f_l}^{f_h} \frac{\tilde{h}_1(f) \tilde{h}_2^*(f)}{S_n(f)} \mathrm{d} f ,
\end{equation}
where $\tilde h_1(f)$ and $\tilde h_2(f)$ indicate Fourier transforms, and $S_n(f)$ is the one-sided power
spectral density of the detector noise, which we assume to be the design zero-detuned high-power
noise PSD of Advanced LIGO \cite{Barsotti:2018}.
The faithfulness is then defined as the overlap between the normalized waveforms, maximized over
the relative time and phase shift, that is 
\begin{equation}
	\label{eq:def_match}
	\left\langle h_1 \mid h_2\right\rangle=\max _{\phi_c, t_c} \frac{\left(h_1\left(\phi_c, t_c\right) \mid h_2\right)}{\sqrt{\left(h_1 \mid h_1\right)\left(h_2 \mid h_2\right)}}.
\end{equation}
In Eq.~(\ref{eq:overlap}), we fix $f_h = 2048 ~\text{Hz}$ and choose $f_l$  to be $f_l = 1.35 f_{\rm{start}}$, where we  
identify the start of the NR simulation $f_{\rm{start}}$ as the peak of the NR waveform in the frequency domain. 
The choice of a buffer factor of $1.35$ is needed to exclude features caused by the Fourier transform, which would spoil the match.  
This is particularly important when comparing a time-domain signal and a frequency-domain approximant, as will be done in following sections.%
\footnote{If $f_l<10~\text{Hz}$, or when comparing different waveform models between each other, 
we instead take $f_l=10~\text{Hz}$.} 
We fix $f_h = 2048 ~\text{Hz}$. We taper the time-domain waveforms using a Planck window function~\cite{McKechan:2010kp}, before transforming them in frequency domain.

Given the binary parameters:
\begin{equation}
	\boldsymbol{\Lambda} \equiv \{q, \chi_1, \chi_2\},
\end{equation}
and calibration parameters
\begin{equation}
	\boldsymbol{\theta} \equiv \{a_6, d_{\rm{SO}}, \Delta t^{22}_{\rm{ISCO}}\},
\end{equation}
we define the \emph{unfaithfulness} (or \emph{mismatch}) of $h_{\mathrm{EOB}}$ to $h_{\mathrm{NR}}$, for the same physical parameters $\boldsymbol{\Lambda}$, and as a function of the calibration parameters $\boldsymbol{\theta}$, as
\begin{equation}
	\mathcal{M}(\boldsymbol{\theta})=1 - \left\langle h_{\mathrm{EOB}}(\boldsymbol{\Lambda} ; \boldsymbol{\theta}) \mid h_{\mathrm{NR}}(\boldsymbol{\Lambda})\right\rangle.
\end{equation}
The goal that we set for the calibration of the \texttt{SEOBNRv5}
model is to find values of the calibration parameters
$\boldsymbol{\theta}(\boldsymbol{\Lambda})$ such that the (2,2) mode
matches with the NR (2,2) mode above $99.9 \%$ (for the \texttt{SEOBNRv4} model 
the goal was set to $99 \%$).  The $10^{-3}$ requirement as maximum mismatch is
challenging, but still reasonable, considering that other 
state-of-the-art aligned-spin approximants~\cite{Pratten:2020fqn,Estelles:2020osj,Nagar:2018zoe} 
can reach mismatches of $10^{-3}$ or smaller against most of NR configurations. More importantly, we need to 
push the accuracy of the \texttt{SEOBNR} models in view of more sensitive 
runs with current facilities and new detectors on the ground and in space{~\cite{Purrer:2019jcp}.  A $10^{-4}$ goal would be
extremely challenging, and would demand a more sophisticated
calibration with additional parameters, as well as a careful treatment
of NR errors, which are often of this order of magnitude (as estimated, for example, 
by comparing different resolutions or extrapolation orders of the same
simulation).  We also require, as in the \texttt{SEOBNRv4} model, that the
difference in merger time $\delta t_{\rm{merger}}$ (defined as the 
peak of the (2,2)-mode amplitude) after a low-frequency phase
alignment is smaller than $5M$, as the mismatch alone is not very
sensitive to such differences.

\subsection{Nested-sampling analysis}
\label{subsec:nested_sampling}

Given the dimensionality of the problem and the large number of NR simulations at our disposal, it is
especially important to devise a computationally efficient and flexible calibration procedure.
For this work, we improve on the strategy adopted in the \texttt{SEOBNRv4} model, which consisted in a
Markov-chain Monte Carlo (MCMC) analysis to obtain a posterior distribution for the calibration parameters for each NR simulation.
MCMC methods allow to easily explore high-dimensional parameter spaces, and have the advantage of providing information on the structure
of the calibration space, particularly on the correlations between calibration parameters.
For our problem, we find the best computational performance with nested samplig~\cite{Skilling:2006gxv},
using the sampler \texttt{nessai}~\cite{Williams:2021qyt} through \texttt{Bilby}~\cite{Ashton:2018jfp}.
We compare our result to other samplers available in \texttt{Bilby} and to the \texttt{emcee} \cite{Foreman-Mackey:2012any} MCMC sampler
used to calibrate \texttt{SEOBNRv4} for a few cases, finding consistent results.

We define the likelihood function to be:
\begin{equation}
	\label{eq:calib_likelihood}
	P(h^\mathrm{NR} | \boldsymbol{\theta}) \propto \exp \left[-\frac{1}{2}\left(\frac{{\mathcal{M}}_{\max }(\boldsymbol{\theta})}{\sigma_{\mathcal{M}}}\right)^2-\frac{1}{2}\left(\frac{\delta t_{\text {merger}}(\boldsymbol{\theta})}{\sigma_t}\right)^2\right],
\end{equation}
where  ${\mathcal{M}}_{\max}(\boldsymbol{\theta})$ is the maximum unfaithfulness between EOB and NR waveforms over the total mass range $10 M \leq M_{\odot} \leq 200 M$,
$\sigma_{\mathcal{M}}$ is chosen to be $10^{-3}$, and $\sigma_{t}$ is chosen to be $5 M$, to impose our calibration requirements.
We carry out the calibration for 441 \texttt{SXS} NR waveforms plus 1 \texttt{Einstein Toolkit} NR waveform, as summarized above.
We take uniform priors for all calibration parameters, specifically $a_6 \in [-500, 500]$,~$\Delta t^{22}_{\rm{ISCO}} \in [-100, 40]$,~$d_{\rm{SO}} \in [-500, 500]$.

For each NR simulation we obtain a posterior distribution $P(\boldsymbol{\theta} | h^\mathrm{NR})$ whose mean 
and variance (and mutual correlations between the parameters) relate to the calibration requirements.
The next step in the calibration procedure is to compute a fit for the calibration parameters as functions of the binary parameters $\boldsymbol{\theta} (\boldsymbol{\Lambda})$,
starting from the set of calibration posteriors. 
In some cases, the correlations between the parameters lead to a secondary mode. To 
obtain a more regular fit, we select only one mode of each calibration posterior, based on continuity considerations. 
After this step, we discard samples that do not satisfy the calibration requirements for each posterior.
If this would discard more than $50 \%$ of the points, we instead keep half of the original samples of the selected mode with the best likelihood values. 
We do this since, for a few of the most challenging NR simulations, like \texttt{SXS:BBH:1124} with $q = 1, \chi_1 = \chi_2 = 0.998 $, we do not find values of the
calibration parameters that satisfy both requirements on $\mathcal{M}_{\max}$ and $\delta t_{\rm{merger}}$. 
In Fig.~\ref{fig:corner_plot} we show an example of a calibration posterior for the NR simulation \texttt{SXS:BBH:2420}.

As done for the \texttt{SEOBNRv4} model, we find it convenient to perform the calibration hierarchically, starting from nonspinning (noS) and then moving to aligned-spin waveforms.
First, we sample over 18 nonspinning configurations (the remaining 21 nonspinning simulations are only used for validation) using as calibration parameters
\begin{equation}
	\boldsymbol{\theta}_{\rm{noS}} \equiv \{a_6,\Delta t^{22}_{\rm{ISCO, noS}}\}.
\end{equation}
We then fix $a_6(\nu)$, $\Delta t^{22}_{\rm{ISCO, noS}}(\nu)$ by the respective fits, as described in the next section, and sample over the remaining 403 aligned-spin configurations using as calibration parameters
\begin{equation}
	\boldsymbol{\theta}_{\rm{S}} \equiv \{d_{\rm{SO}}, \Delta t^{22}_{\rm{ISCO, S}}\},
\end{equation}
where
\begin{equation}
	\label{eq:dt_fit}
	\Delta t^{22}_{\rm{ISCO}} = \Delta t^{22}_{\rm{ISCO, noS}} + \Delta t^{22}_{\rm{ISCO, S}},
\end{equation}
and $\Delta t^{22}_{\rm{ISCO, S}}$ is assumed to vanish in the nonspinning limit.
\begin{figure}
	\includegraphics[width=\linewidth]{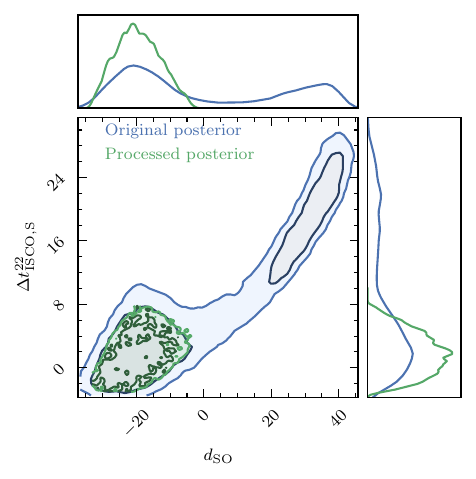}
	\caption{Posterior for the calibration parameters $\{d_{\rm{SO}}, \Delta t^{22}_{\rm{ISCO, S}}\}$,
	obtained by comparing to the NR simulation \texttt{SXS:BBH:2420} ($q = 1.0, \chi_1 = 0.2, \chi_2 = 0.2$).
	The blue posterior is the result of the nested-sampling analysis described in Sec.~\ref{subsec:nested_sampling},
	and shows values mostly clustered around two distinct regions or \emph{modes}.
	The green posterior is what we obtain after removing one of the two modes and keeping only the points with $\mathcal{M}_{\max}<10^{-3}$ and $\delta t_{\rm{merger}}<5 M$.
	We use these \emph{processed posteriors} to obtain fits for the calibration parameters across parameter space.}
	\label{fig:corner_plot}
\end{figure}
We investigate the possibility of adding a spin dependence to $a_6$, or adding a spin-spin calibration parameter $d_{\rm{SS}}$ at 5PN order similar to the one used in the \texttt{SEOBNRv4} model, but we find no significant improvements --- for example by comparing the mismatch and time to merger against NR taking the maximum likelihood points of the calibration posteriors.
On the other hand, limiting the sampling to two dimensions makes it faster, and produces more Gaussian-like posteriors which are significantly simpler to fit.

\subsection{Calibration-parameter fits and extrapolation}

We now discuss how we obtain fits for the calibration parameters $\boldsymbol{\theta} = \{a_6, \Delta t^{22}_{\rm{ISCO}}, d_{\rm{SO}}\}$
as functions of the binary parameters $\boldsymbol{\Lambda} = \{q, \chi_1, \chi_2\}$, given the calibration posteriors.
To help with the extrapolation, we also use some knowledge of the conservative dynamics in the $\nu \rightarrow 0$ limit. For $a_6$  we employ 
Eq.~(\ref{eq:a6_TPL}), which is obtained by requiring that the ISCO shift predicted by the \texttt{SEOBNRv5} Hamiltonian agrees with the 
1GSF ISCO shift, as explained is Sec.~(\ref{sec:zerospin_H}). 
For $\Delta t^{22}_{\rm{ISCO}}$ we estimate the test-mass values, for different spin magnitudes, 
using results of Ref.~\cite{Taracchini:2014zpa}. We do so by imposing that the difference between the peak of the (2,2) mode and the peak 
of the orbital frequency in the EOB test-mass--limit waveforms matches the one measured in the Teukolsky-code waveforms (see, e.g., Fig.~13 of Ref.~\cite{Taracchini:2014zpa}). 
We then convert the corresponding value to the difference between the ISCO and the peak of the (2,2)-mode amplitude.
Since the Teukolsky-code waveforms were produced using a different EOB dynamics, we prefer to relate those quantities closer to merger,
and not directly match the difference between the ISCO and the peak of the (2,2) mode of Teukolsky-code and EOB waveforms. 
Nevertheless, we find that the difference is not be very large. 

In the nonspinning limit, the data for $\boldsymbol{\theta}_{\rm{noS}} = \{a_6, \Delta t^{22}_{\rm{ISCO, noS}}\}$ are simple enough to allow for an independent direct fit of the maximum-likelihood
point of the calibration posteriors and TML values, using least square fits.
For $a_6$ we use a quartic polynomial in $\nu$, while for $\Delta t^{22}_{\rm{ISCO, noS}}$, that is an ansatz of the form 
\begin{equation}
	\Delta t^{22}_{\rm{ISCO, noS}} = (a_0 + a_1 \,\nu + a_2 \,\nu^2 + a_3 \,\nu^3 )\,\nu^{-1/5 + a_4\, \nu},
\end{equation}
where the $\nu^{-1/5}$ factor ensures the expected test-mass scaling for $(t^{\text{22}}_{\text{peak}} - t_{\text{ISCO}})$~\cite{Buonanno:2000ef}, and provides a better extrapolation of the fit in the $\nu \rightarrow 0$ limit.
Figure~\ref{fig:fit_nonspinning} shows the $\{a_6, \Delta t^{22}_{\rm{ISCO, noS}}\}$ data and the resulting fits.

\begin{figure}
	\includegraphics[width=\linewidth]{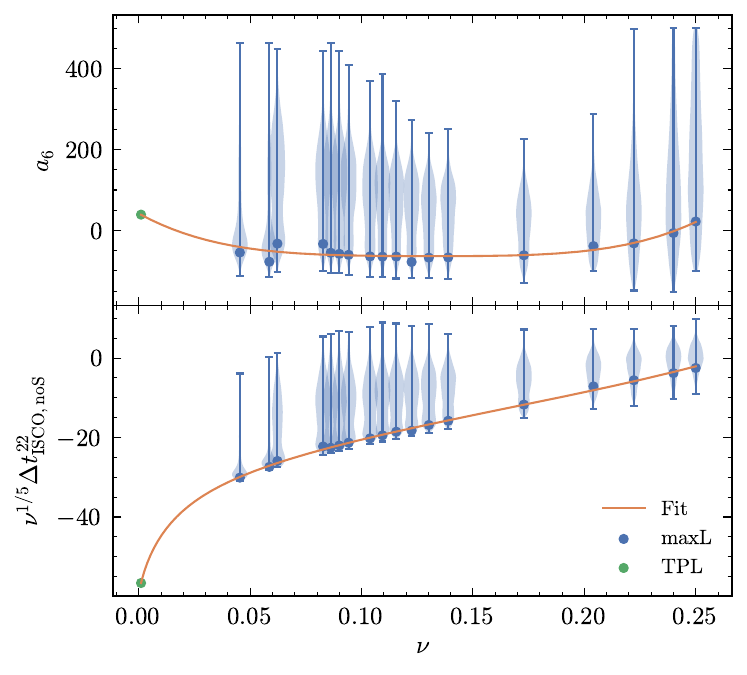}
	\caption{Fits for the nonspinning calibration parameters $\boldsymbol{\theta}_{\rm{noS}} = \{a_6, \Delta t^{22}_{\rm{ISCO, noS}}\}$.
	The parameters are obtained by least-square fits of the maximum likelihood points (blue dots) of the calibration 
	posteriors (shaded violins), for a set of NR simulations with different mass-ratios $\nu$, together with estimates 
	of the test-mass limit values (green dots). We rescale $ \Delta t^{22}_{\rm{ISCO, noS}}$ by 
	$\nu^{1/5}$ to improve its extrapolation in the $\nu \rightarrow 0$ limit. 
	No processing is needed for the nonspinning calibration posteriors, as the maximum likelihood 
	point lies in the same mode for all configurations. 
    }
	\label{fig:fit_nonspinning}
\end{figure}

For the aligned-spin fit of $\boldsymbol{\theta}_{\rm{S}} = \{d_{\rm{SO}}, \Delta t^{22}_{\rm{ISCO, S}}\}$, we use a similar approach as in the \texttt{SEOBNRv4} model~\cite{Bohe:2016gbl}, with a few 
important differences. We fit the median of the calibration posteriors, instead of the mean, as this provides better unfaithfulness when comparing against NR. In principle, fitting the maximum-likelihood also for aligned-spin cases would give the best result, 
but does not turn out to be a viable option due to the lack of regularity in the data. We use three variables in the fit $(\nu,a_{+},a_{-})$, instead of just $(\nu,\chi)$, where $\chi=\chi_S+\chi_A \delta / (1-2 \nu)$, as this provides a better result, also
when using a subset of NR simulations for the fit (see also Appendix~\ref{appx:robustness}), or when comparing to independent sets of \texttt{NRHybSur3dq8}~\cite{Varma:2018mmi} waveforms not used in the calibration. 
We rescale $\Delta t^{22}_{\rm{ISCO, S}}$ by $\nu^{1/5}$ to ensure the correct test-mass scaling.

More specifically, after removing secondary modes and discarding samples that don't meet the calibration requirements, 
and after rescaling $\Delta t^{22}_{\rm{ISCO, S}}$ by $\nu^{1/5}$, 
we consider the medians $\langle\boldsymbol{\theta}_{\mathrm{S}}\rangle_{(n)}$ and 
covariance matrices ${C_{\rm{S}}}_{(n)}$ of the calibration posteriors, with $n$ labeling each of the 442 NR simulations.
We parametrize $d_{\rm{SO}}$ by a cubic polynomial in $(\nu,a_{+},a_{-})$ and $\Delta t^{22}_{\rm{ISCO, S}} \nu^{1/5}$ by a cubic polynomials in $(\nu,a_{+},a_{-})$
with an additional $a_+^4$ feature. We determine the coefficients of these polynomials by minimizing the following function, using a Sequential Least Squares Programming (SLSQP) minimization algorithm~\cite{Bohe:2016gbl}
\begin{equation}
	\chi_{\mathrm{s}}^2 \equiv \sum_{n \in \mathcal{S}_{\mathrm{s}}} \frac{w}{2}\left(\boldsymbol{\theta}_{\mathrm{S}}-\left\langle\boldsymbol{\theta}_{\mathrm{S}}\right\rangle_{(n)}\right)\left(C_{\mathrm{S}}^{-1}\right)_{(n)}\left(\boldsymbol{\theta}_{\mathrm{S}}-\left\langle\boldsymbol{\theta}_{\mathrm{S}}\right\rangle_{(n)}\right)^{\mathrm{T}} +\chi_{\mathrm{TML}}^2,
\end{equation}
where $\chi_{\mathrm{TML}}^2$ is a term that penalizes deviations from the test-mass limit of $\Delta t^{22}_{\rm{ISCO, S}}$ and takes the form
\begin{equation}
	\chi_{\mathrm{TML}}^2 = \sum_{\chi_{i} \neq 0} \frac{\Big( \Delta t^{22}_{\rm{ISCO, S}} - \Delta t^{22, \mathrm{TML}}_{\rm{ISCO},\chi_i} \Big)^2}{\sigma^2_{\mathrm{TML}}},
\end{equation}
in which $\Delta t^{22, \mathrm{TML}}_{\rm{ISCO},\chi_i}$ are the estimated test-mass values of $\Delta t^{22}_{\rm{ISCO},S}$, for different spin magnitudes $\chi_i$ for which Teukolsky waveforms are available,
and we take $\sigma_{\mathrm{TML}} = 5 M$.
As for the \texttt{SEOBNRv4} model, the function $w$ is a weighting function of the form
\begin{equation}
	w \equiv \chi_1^2+\chi_2^2+\frac{|\chi|}{2 \nu}\,,
\end{equation}
which accounts for the inhomogeneous distribution of NR simulations in the BBH parameter space.

We finally list the calibration-parameter fits:
\begin{widetext}
\begin{equation}
	\label{eq:a6_fit}
	a_6 = 329523.262 \nu^{4} - 169019.14 \nu^{3} + 33414.4394 \nu^{2} - 3021.93382 \nu + 41.787788,
\end{equation}
\begin{equation}
	\label{eq:dt_fit_ns}
	\Delta t^{22}_{\rm{ISCO, noS}} = \nu^{- 1/5 + 10.051322 \nu } \left(55565.2392 \nu^{3} - 9793.17619 \nu^{2} - 1056.87385 \nu - 59.62318\right),
\end{equation}
\begin{align}
	\label{eq:dt_fit_s}
	\Delta t^{22}_{\rm{ISCO, S}} = &\nu^{-1/5}\left(- 6.789139 a_{+}^{4} + 5.399623 a_{+}^{3} + 6.389756 a_{+}^{2} a_{-} - 132.224951 a_{+}^{2} \nu + 49.801644 a_{+}^{2} \right.\nonumber \\
	&+ 8.392389 a_{+} a_{-}^{2} + 179.569825 a_{+} a_{-} \nu - 40.606365 a_{+} a_{-} + 384.201019 a_{+} \nu^{2} - 141.253182 a_{+} \nu \nonumber \\
	&\left.+ 17.571013 a_{+} - 16.905686 a_{-}^{2} \nu + 7.234106 a_{-}^{2} + 144.253396 a_{-} \nu^{2} - 90.192914 a_{-} \nu + 14.22031 a_{-}\right),
\end{align}
\begin{align}
	\label{eq:dSO_fit}
	d_{\rm{SO}} = &- 7.584581 a_{+}^{3} - 10.522544 a_{+}^{2} a_{-} - 42.760113 a_{+}^{2} \nu + 18.178344 a_{+}^{2} - 17.229468 a_{+} a_{-}^{2} \nonumber \\
	&+ 362.767393 a_{+} a_{-} \nu - 85.803634 a_{+} a_{-} - 201.905934 a_{+} \nu^{2} - 90.579008 a_{+} \nu + 49.629918 a_{+} \nonumber \\
	&- 7.712512 a_{-}^{3} - 238.430383 a_{-}^{2} \nu + 69.546167 a_{-}^{2} - 1254.668459 a_{-} \nu^{2} + 472.431938 a_{-} \nu \nonumber \\
	&- 39.742317 a_{-} + 478.546231 \nu^{3} + 679.52177 \nu^{2} - 177.334832 \nu - 37.689778.
\end{align}

\end{widetext}

To ensure a robust behavior of the fits between the last calibration points and 
extreme-mass-ratio limit, we perform exhaustive checks of the sanity of the 
waveform model across a broad range of the parameter space ($q\in[1,100]$, covering the full spin range). 
The tests include visual inspections of the waveforms, 
assessing stability in response to perturbations of the intrinsic parameters, 
verifying the monotonicity of the amplitude and frequency of the $(2,2)$ mode up its peak
and confirming that the higher modes consistently maintain amplitudes smaller than the $(2,2)$ mode, up to the merger.

\section{Performance of the \texttt{SEOBNRv5HM} model against numerical-relativity simulations}
\label{sec:performance}

To assess the impact of the improvements introduced in the \texttt{SEOBNRv5HM} waveform model, we compare it to 
the set of NR simulations described in Sec.~\ref{sec:calibration}, and to other state-of-the-art aligned-spin approximants. 
We do so by performing unfaithfulness computations, as well as comparisons of angular-momentum flux and binding energy against NR.
Finally, we assess the computational efficiency of the model for GW data-analysis purposes, providing benchmarks.


\subsection{Faithfulness for multipolar waveforms}
\label{subsec:faithfulness}

The GW signal emitted by a quasi-circular aligned-spin BBH system depends on 11 parameters: the masses and spins $\boldsymbol{\lambda}=\{m_{1,2},\chi_{1,2}\}$, 
the direction of the observer from the source described by $(\iota, \varphi_0)$, the luminosity distance $d_L$, the polarization angle $\psi$, 
the location in the sky of the detector $(\theta, \phi)$, and the time of arrival $t_c$. 
The strain in the detector caused by a passing GW can be  expressed as
\begin{align}
\label{eq:det_strain}
h(t) \equiv & F_+(\theta,\phi,\psi) \ h_+(t;\iota,\varphi_0, d_L, \bm{\lambda},t_{\mathrm{c}}) \nonumber \\
&+ F_\times(\theta,\phi,\psi)\ h_\times(t;\iota,\varphi_0, d_L, \bm{\lambda},t_\mathrm{c})\,,
\end{align}
where $F_{+,\times}$ are the antenna pattern functions \cite{Sathyaprakash:1991mt,Finn:1992xs}. 
The strain in Eq.~(\ref{eq:det_strain}) can be expressed in terms of an effective polarization angle $\kappa(\theta,\phi,\psi)$ as
\begin{align}
\label{eq:strainKappa}
h(t) =\mathcal{A}(\theta, \phi) (h_+ \cos \kappa + h_\times \sin \kappa),
\end{align}
where the dependences of $\kappa$, $h_+$ and $h_\times$ have been removed to ease the notation, 
and the definition of the coefficient $\mathcal{A}(\theta,\phi)$ can be found in Refs.~\cite{Cotesta:2018fcv, Ossokine:2020kjp}.

To assess the agreement between two waveforms 
with higher-order multipoles~\cite{Cotesta:2018fcv,Ossokine:2020kjp,Garcia-Quiros:2020qpx}, which we denote as the signal, $h_s$ and the template, $h_t$, 
observed by a detector, we define the faithfulness function \cite{Cotesta:2018fcv,Ossokine:2020kjp},
\begin{equation}
\mathcal{F}(M_{\textrm{s}},\iota_{\textrm{s}},{\varphi_0}_{\textrm{s}},\kappa_{\textrm{s}}) =  \max_{t_c, {\varphi_0}_{t}, \kappa_{t}} \left[\left . \frac{ \langle h_s|h_t \rangle}{\sqrt{  \langle h_s|h_s \rangle  \langle h_t|h_t \rangle}}\right \vert_{\substack{\iota_{\mathrm{s}} = \iota_{t} \\\boldsymbol{\lambda}_\mathrm{s}(t_{\mathrm{s}} = t_{0_\mathrm{s}}) = \boldsymbol{\lambda}_{t}(t_t = t_{0_\mathrm{t}})}} \right ],
\end{equation}
where the inner product is defined in Eq.~(\ref{eq:overlap}). Typically, we set the inclination angle of the template and the signal to be the same, 
while the coalescence time, azimuthal and effective polarization angles of the template, $(t_{0_t},\varphi_{0_t}, \kappa_t)$, 
are adjusted to maximize the faithfulness of the template.  The maximizations over the coalescence time $t_c$, 
and coalescence phase ${\varphi_0}_{t}$ are performed numerically, while the optimization over the effective 
polarization angle $\kappa_{t}$ is done analytically as described in Ref.~\cite{Capano:2013raa}.

To reduce the dimensionality of the faithfulness function it is useful to define the sky-and-polarization-averaged faithfulness \cite{Babak:2016tgq,Ossokine:2020kjp} as
\begin{equation}
	\overline{\mathcal{F}}\left(M_{\mathrm{s}}, \iota_{\mathrm{s}}\right) \equiv \frac{1}{8 \pi^2} \int_0^{2 \pi} d \kappa_{\mathrm{s}} \int_0^{2 \pi} d \varphi_{0 \mathrm{~s}} \mathcal{F}\left(M_{\mathrm{s}}, \iota_{\mathrm{s}}, \varphi_{0 \mathrm{~s}}, \kappa_{\mathrm{s}}\right).
\end{equation}
We also define the sky-and-polarization-averaged, signal-to-noise-ratio (SNR)-weighted faithfulness as \cite{Ossokine:2020kjp,Cotesta:2018fcv}:
\begin{widetext}
\begin{equation}
	\overline{\mathcal{F}}_{\mathrm{SNR}}\left(M_{\mathrm{s}}, \iota_{\mathrm{s}}\right) \equiv \sqrt[3]{\frac{\int_0^{2 \pi} d \kappa_{\mathrm{s}} \int_0^{2 \pi} d \varphi_{0 \mathrm{~s}} \mathcal{F}^3\left(M_{\mathrm{s}}, \iota_{\mathrm{s}}, \varphi_{0 \mathrm{~s}}, \kappa_{\mathrm{s}}\right) \mathrm{SNR}^3\left(\iota_{\mathrm{s}}, \varphi_{0 \mathrm{~s}}, \kappa_{\mathrm{s}}\right)}{\int_0^{2 \pi} d \kappa_{\mathrm{s}} \int_0^{2 \pi} d \varphi_{0 \mathrm{~s}} \operatorname{SNR}^3\left(\iota_{\mathrm{s}}, \varphi_{0 \mathrm{~s}}, \kappa_{\mathrm{s}}\right)}} ,
\end{equation}
\end{widetext}
where the $\operatorname{SNR}\left(\iota_{\mathrm{s}}, \varphi_{0_{\mathrm{s}}}, \theta_{\mathrm{s}}, \phi_{\mathrm{s}}, \kappa_{\mathrm{s}}, d_{L \mathrm{s}}, \boldsymbol{\lambda}_{\mathrm{s}}, t_{c \mathrm{~s}}\right)$ is defined as
\begin{equation}
	\label{eq:mm_SNR_weighted}
	\operatorname{SNR}\left(\iota_{\mathrm{s}}, \varphi_{0_{\mathrm{s}}}, \theta_{\mathrm{s}}, \phi_{\mathrm{s}}, \kappa_{\mathrm{s}}, d_{L \mathrm{s}}, \boldsymbol{\lambda}_{\mathrm{s}}, t_{c \mathrm{~s}}\right) \equiv \sqrt{\left(h_{\mathrm{s}}, h_{\mathrm{s}}\right)}.
\end{equation}
The weighting by the SNR in Eq. \eqref{eq:mm_SNR_weighted} takes into account the
dependence on the phase and effective polarization of the signal at a fixed
distance. Finally, we define the sky-and-polarization-averaged, SNR-weighted unfaithfulness (or mismatch) as
\begin{equation}
	\overline{\mathcal{M}}_{\mathrm{SNR}}=1-\overline{\mathcal{F}}_{\mathrm{SNR}}.
\end{equation}

\begin{figure*}
	\includegraphics[width=\textwidth]{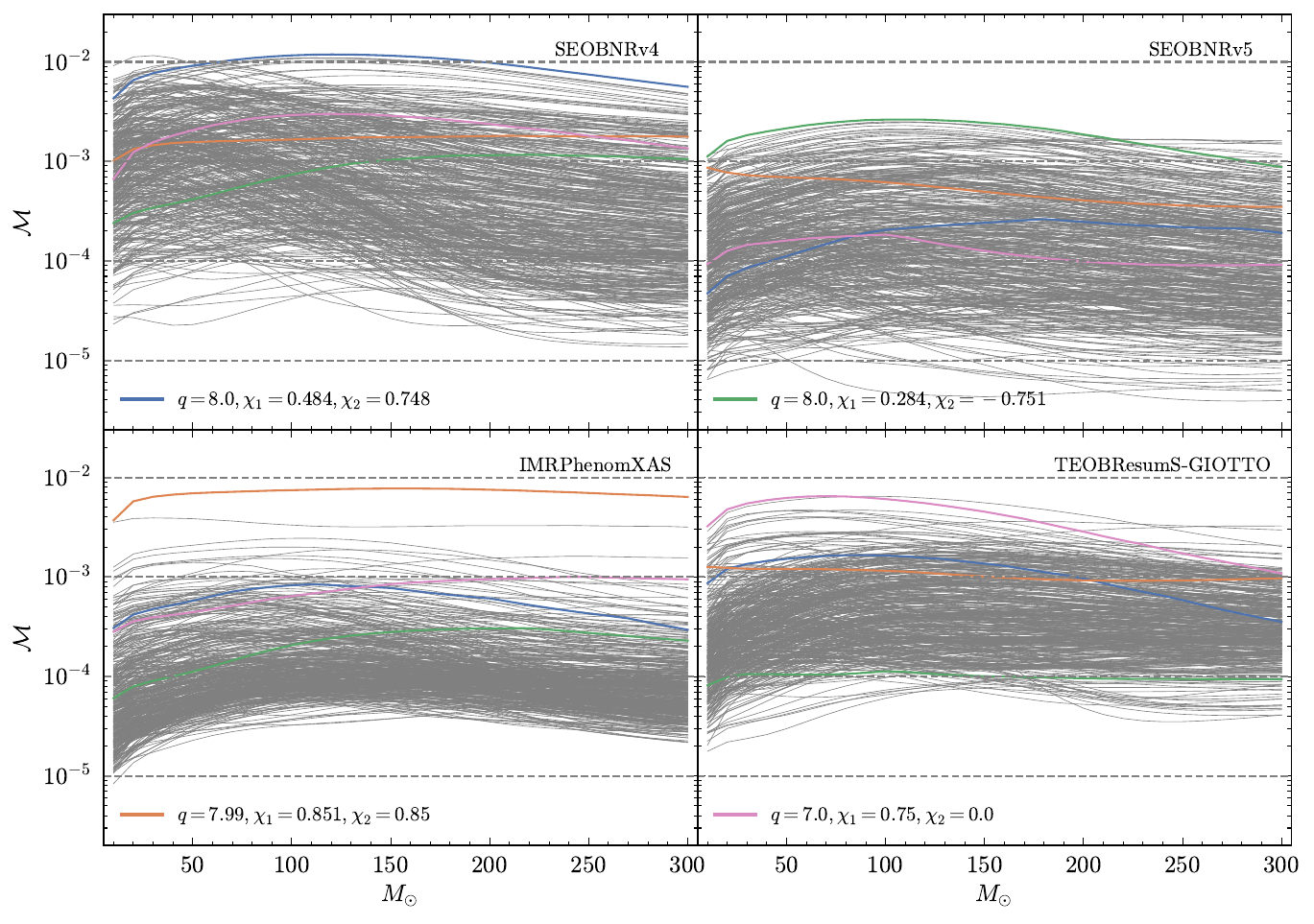}
	\caption{(2,2)-mode mismatch over a range of total masses between 10 and 300 $M_{\odot}$, 
	between different aligned-spin approximants and the 442 NR simulations used in this work.
	The colored lines highlight cases with the worst maximum mismatch for each model.
	Note that \texttt{SEOBNRv5} has no outliers beyond $0.3 \%$ and many more cases at lower unfaithfulness,
	especially compared to \texttt{SEOBNRv4} and \texttt{TEOBResumS-GIOTTO}.}
	\label{fig:mismatch_spaghetti_22}
\end{figure*}
%

\subsection{Accuracy of \texttt{SEOBNRv5} (2,2) mode}

We start by considering (2,2)-mode only mismatches. In this case, the result does not depend on the inclination, 
and the mismatch definition reduces to the one used in Sec.~\ref{sec:calibration}.
Figure~\ref{fig:mismatch_spaghetti_22} shows the (2,2)-mode mismatch over a range of total masses between 10 and 300$M_\odot$  
using the 442 NR simulations summarized in Sec.~\ref{sec:calibration} for different state-of-the-art aligned-spin approximants: 
\texttt{SEOBNRv5}, its predecessor \texttt{SEOBNRv4} \cite{Bohe:2016gbl}, 
the aligned-spin model from the other EOB family \texttt{TEOBResumS} \cite{Riemenschneider:2021ppj,Nagar:2020pcj,Nagar:2019wds,Nagar:2018zoe} and 
\texttt{IMRPhenomXAS} \cite{Pratten:2020fqn}, from the 4th generation of Fourier-domain
phenomenological waveform models.
All approximants are called through \texttt{LALSimulation}, except for \texttt{SEOBNRv5} and for 
\texttt{TEOBResumS}, for which we use the latest available public version \texttt{TEOBResumSv4.1.4-GIOTTO}.
\footnote{This corresponds to the commit fc4595df72b2eff4b36e563f607eab5374e695fe of the public bitbucket
repository \url{https://bitbucket.org/eob_ihes/teobresums}, and it's the latest tagged version at the time of this publication.}

The colored lines highlight cases with the worst maximum mismatch for each model: as expected, 
the most challenging cases have high mass ratio and high spins, 
as all models have been calibrated to few NR simulations in this region of parameter space.
We note that \texttt{SEOBNRv5} has no outliers beyond $0.3 \%$ and many more cases at lower unfaithfulness,
especially compared to \texttt{SEOBNRv4} and \texttt{TEOBResumS-GIOTTO}. 
Comparing the two upper panels of Fig.~\ref{fig:mismatch_spaghetti_22}, we can see in particular that \texttt{SEOBNRv5} yields
unfaithfulnesses almost one order of magnitude smaller than those of its predecessor \texttt{SEOBNRv4} model.
\begin{figure}
	\includegraphics[width=\linewidth]{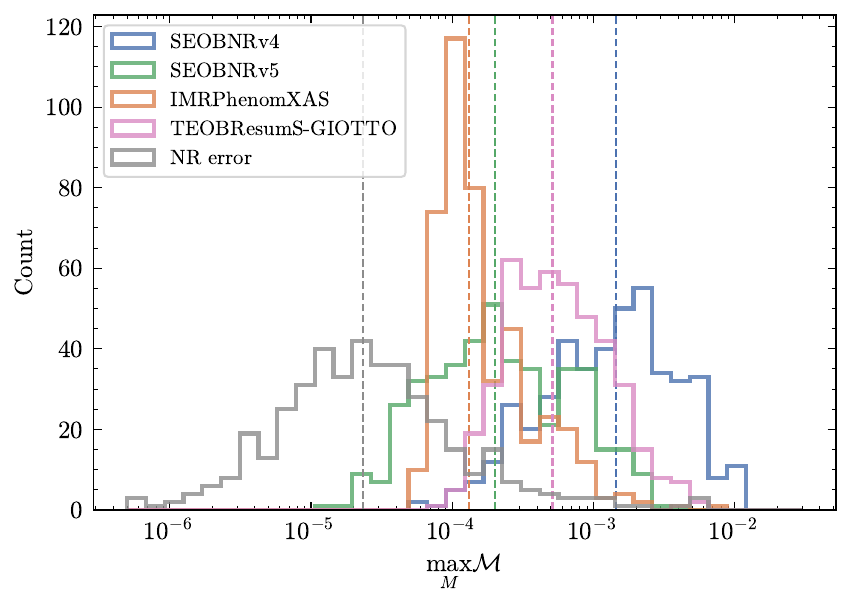}
	\includegraphics[width=\linewidth]{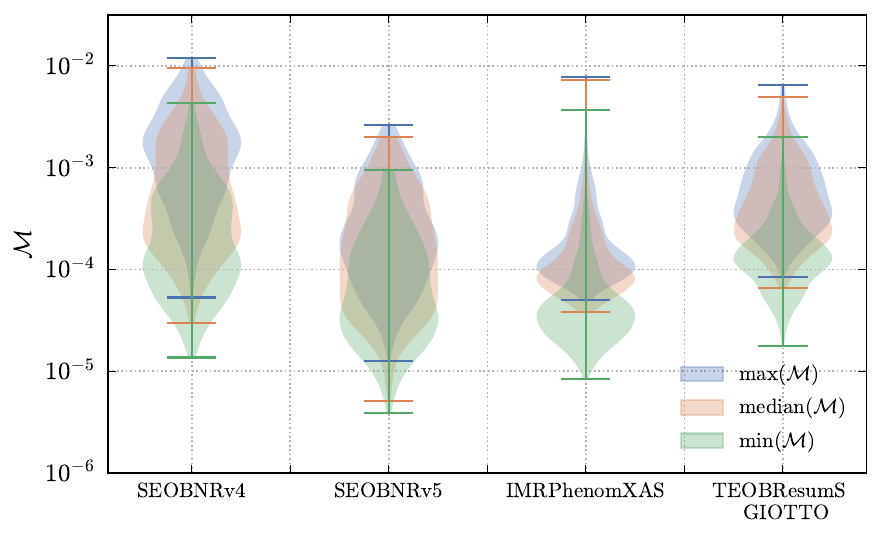}
	\caption{
	\textit{Top panel}: Histogram of the maximum (2,2)-mode mismatch over a range of total masses between 10 and 300 $M_{\odot}$, 
	between different aligned-spin approximants and the 442 NR simulations used in this work.
	The NR error is estimated by computing the mismatch between NR simulations with the highest and second-highest resolutions. 
	The vertical dashed lines show the medians. 
	\textit{Bottom panel:} distribution of the maximum (blue), median (orange) and minimum (green) mismatch over the same range of total masses for the different models.
	}
	\label{fig:mismatch_hist_22}
\end{figure}

The top panel of Fig.~\ref{fig:mismatch_hist_22} shows histograms of the maximum (2,2)-mode mismatch over the same range of total masses. 
We also show an estimate of the NR error computed as the mismatch between NR simulations with the highest and second-highest resolutions, if available. 
The mismatch between NR simulations of the highest resolution and different extrapolation order is typically 
one order of magnitude smaller than the one obtained comparing different resolutions, hence we do not show it in these comparisons.
The vertical dashed lines correspond to the medians of the distributions. Overall \texttt{IMRPhenomXAS} 
achieves the lowest median unfaithfulness ($1.31 \times 10^{-4}$), while still having two outliers above $0.3 \%$, with 
\texttt{SEOBNRv5} closely following with median mismatch $1.99 \times 10^{-4}$, but a larger tail of cases with low unfaithfulness approaching $10^{-5}$.
\texttt{TEOBResumS-GIOTTO} is slightly less accurate with median mismatch $5.12 \times 10^{-4}$, 
while \texttt{SEOBNRv4} is the least faithful model with median value $1.44 \times 10^{-3}$, almost one 
order of magnitude larger than \texttt{SEOBNRv5}. 
These results are summarized in Table~\ref{tab:mm_22}, together with the fraction of cases falling below $10^{-3}$ and $10^{-4}$ for each approximant. 

The NR error is about one order of magnitude smaller than the \texttt{SEOBNRv5} modeling error, 
with median value $\sim 2 \times 10^{-5}$. Still, there are a few cases where the two are comparable, and 
improving the accuracy of the NR simulations used to calibrate the model would be critical to 
reducing the modeling errors by another order of magnitude. 
The bottom panel of Fig.~\ref{fig:mismatch_hist_22} provides a complementary summary of the unfaithfulness calculation, 
by showing the distribution of the maximum (blue), median (orange) and minimum (green) mismatch over the same range of total masses for the different models.

We find that $10 \%$ of the cases are above $0.1 \%$ maximum mismatch for \texttt{SEOBNRv5}: most of those correspond, as expected, to high spins, 
both for large mass-ratios and for $q \simeq 1$ where spin magnitudes can reach values up to $0.998$.
In a future update of the model, the description of these cases could be improved by suitably including 
the full 5PN spin contributions (NNNLO SO and SS, NLO S$^3$ and S$^4$) 
to the conservative dynamics recently obtained in Refs.~\cite{Antonelli:2020aeb,Antonelli:2020ybz,Mandal:2022nty,Kim:2022pou,Mandal:2022ufb,Kim:2022bwv,Levi:2022dqm,Levi:2022rrq,Levi:2019kgk,Levi:2020lfn},
by including all spin-contributions up to 3.5PN to the waveform modes, as derived in Refs.~\cite{Henry:2022dzx,Henry:2022ccf}, 
or by additional spin-dependent calibration coefficients other than $d_{\rm{SO}}$.

Other challenging cases for \texttt{SEOBNRv5} are those with large mass-ratio, small $a_{+}$, 
but large secondary spin, for example \texttt{SXS:BBH:1430}, with parameters $(q, \chi_1, \chi_2) = (8.0, 0.284, -0.751)$.
The calibration term, which has the form $\sim a_{+} d_{\rm{SO}}$, is suppressed, and deviations of the model 
from NR are only partially captured by having $d_{\rm{SO}}$ itself depending also on the spin difference $a_{-}$. 
To understand what could be the error when one has exactly $a_{+}=0$, but $a_{-}$ is large, 
we can compare the model to \texttt{NRHybSur3dq8} waveforms: 
taking $q=8$ and varying $\chi_2$, while fixing $\chi_1$ so that $a_{+}=0$, 
we see at most mismatches around $0.004$ for large negative secondary spin $\chi_2<-0.9$, 
where \texttt{NRHybSur3dq8} is also extrapolating from its training region ($\chi_i \leq 0.8$).
While additional calibration terms with a different spin dependence could improve these cases, 
this shows that for the moment the analytical spin information captures the correct behavior at a level
comparable to other modeling errors. 

\setlength{\extrarowheight}{8pt}
\begin{table*}
    \centering
    \begin{ruledtabular}
    	\begin{tabular}{ l  c  c  c  c  c  }
 Approximant & \makecell[cc]{\texttt{SEOBNRv4}} & \makecell[cc]{\texttt{SEOBNRv5}} & \makecell[cc]{\texttt{IMRPhenomXAS}} & \makecell[cc]{\texttt{TEOBResumS-GIOTTO}} \\
 \hline
 \centering
 $\text{median} \max_{M} \mathcal{M}$ &   $1.44 \times 10^{-3}$ & $1.99 \times 10^{-4}$ & $1.31 \times 10^{-4}$ & $5.12 \times 10^{-4}$ \\
 $ \% \max_{M} \mathcal{M} < 10^{-3}$ &   $38 \%$ & $90 \%$ & $97 \%$ & $76 \%$ \\
 $ \% \max_{M} \mathcal{M} < 10^{-4}$ &   $1 \%$ & $ 27 \%$ & $29 \%$ & $1 \%$ \\
 \end{tabular}
    \end{ruledtabular}
 \caption{Summary of the (2,2)-mode mismatch over a range of total masses between 10 and 300 $M_{\odot}$, 
 between different aligned-spin approximants and the 442 NR simulations used in this work. 
 We display the median of the maximum mismatch across total mass, and the fraction of cases falling below $10^{-3}$ and $10^{-4}$. 
 }
 \label{tab:mm_22}
\end{table*}

\subsection{Accuracy of \texttt{SEOBNRv5HM} modes}

\begin{figure*}
	\includegraphics[width=\textwidth]{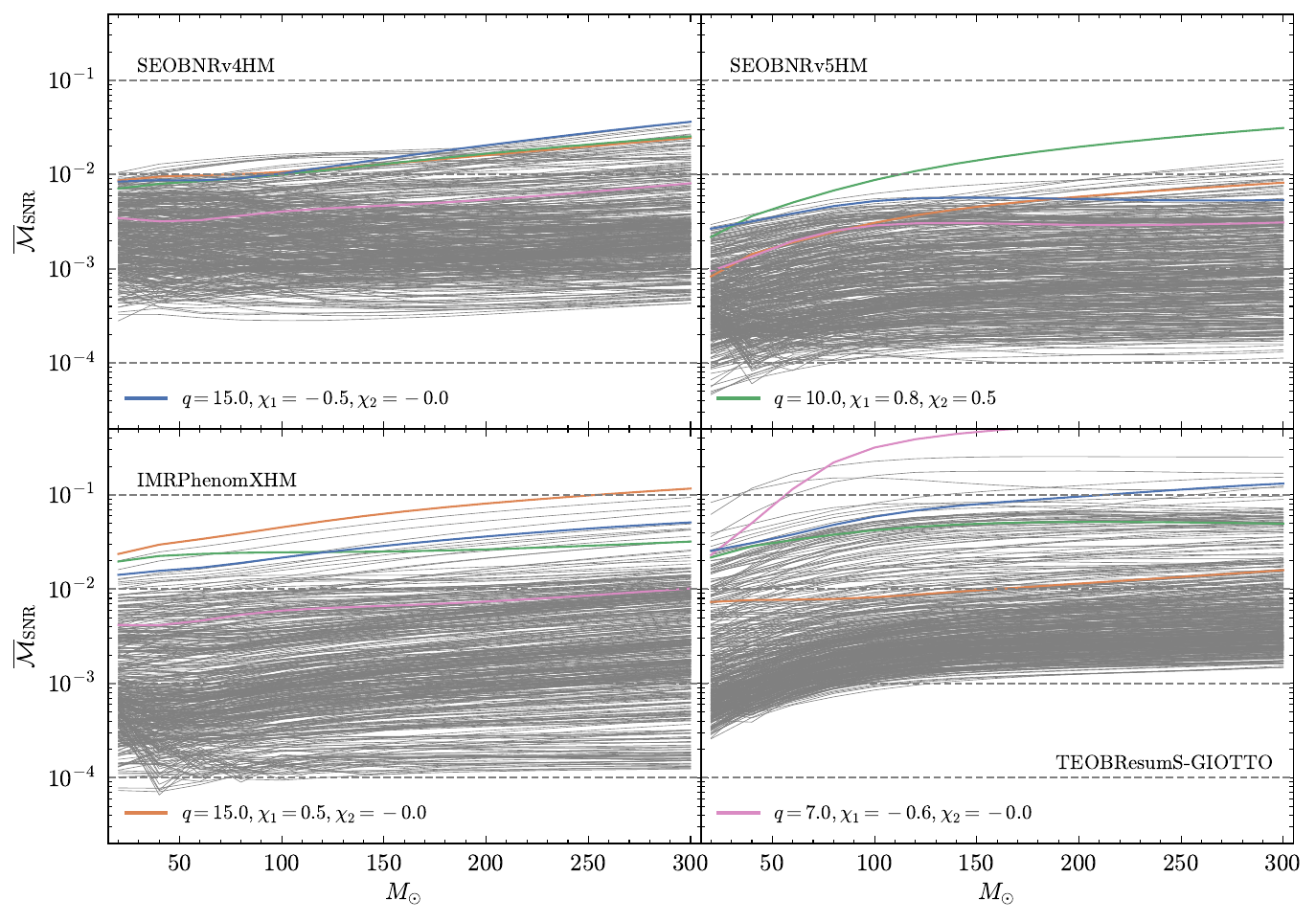}
	\caption{The sky-and-polarization averaged, SNR-weighted mismatch, for inclination $\iota = \pi/3$, over a range of total masses between 20 and 300 $M_\odot$ 
	between different aligned-spin multipolar approximants and the 441 \texttt{SXS} NR simulations used in this work.
	The colored lines highlight cases with the worst maximum mismatch for each model.}
	\label{fig:mismatch_spaghetti_HM}
\end{figure*}

We now turn to mismatches for the full polarizations, including higher-multipoles. 
Figure~\ref{fig:mismatch_spaghetti_HM} shows the sky-and-polarization averaged, SNR-weighted mismatch, 
for inclination $\iota = \pi/3$, over a range of total masses between 20 and 300 $M_\odot$ 
between the 441 \texttt{SXS} NR simulations used in this work and different multipolar aligned-spin approximants:
\texttt{SEOBNRv4HM}~\cite{Cotesta:2018fcv}, \texttt{SEOBNRv5HM}, \texttt{TEOBResumS-GIOTTO} \cite{Riemenschneider:2021ppj,Nagar:2020pcj,Nagar:2019wds,Nagar:2018zoe} 
and \texttt{IMRPhenomXHM}~\cite{Garcia-Quiros:2020qpx}.
For each approximant we include all modes available%
\footnote{For \texttt{TEOBResumS-GIOTTO} we do not include the $(5,5)$ mode, after finding 
that, in the version of the code used for these comparisons, 
it has an unphysically large amplitude close to merger in some corners of the parameter space 
(equal-mass, large opposite spins, as for example \texttt{SXS:BBH:2132}).}, 
while for NR waveforms we use modes up to $\ell_{\rm{max}} = 5$.
The modes included are specifically 
$(\ell, |m|)=(2,2),(2,1),(3,3),(4,4),(5,5)$ for \texttt{SEOBNRv4HM}, 
$(\ell, |m|)=(2,2),(2,1),(3,3),(3,2),(4,4),(4,3),(5,5)$ for \texttt{SEOBNRv5HM}, 
$(\ell, |m|)=(2,2),(2,1),(3,3),(3,2),(4,4)$ for \texttt{IMRPhenomXHM} and 
$(\ell, |m|)=(2,2),(2,1),(3,3),(3,2),(3,1),(4,4),(4,3),(4,2)$ for \texttt{TEOBResumS-GIOTTO}.

In this comparison we omit the \texttt{Einstein Toolkit} simulation, for which we only have the (2,2) mode.
As in the previous results, we highlight with a different color cases with the worst maximum mismatch for each model: 
unsurprisingly the worst cases are at the corners of the NR parameter space, 
and correspond to configurations with very high $q$ and non-zero spins, where the impact of higher-multipoles is substantial, 
also due to the significant inclination $\iota = \pi/3$.

First of all, we note that all models perform worse compared to the (2,2)-mode only case, 
as expected due to the limited alignment freedom with a global phase and time shift, 
but also because the higher modes are available today at lower PN order than the dominant one, and 
their modeling close to merger is complicated by numerical noise in NR simulations.

Focusing on the upper panels, comparing \texttt{SEOBNRv4HM} and \texttt{SEOBNRv5HM}, we see 
an overall improvement, with many more cases between $10^{-4}$ and $10^{-3}$ for \texttt{SEOBNRv5HM}, 
and just a few outliers above $1 \%$ for large values  of the total mass.
The improvement for low total mass, where an accurate inspiral is primarily important, 
is particularly significant, and \texttt{SEOBNRv5HM} is always well below $1 \%$, never exceeding $0.3 \%$.
On the other hand the increase of the mismatch with the total mass for \texttt{SEOBNRv5HM}, absent in the (2,2)-mode 
only comparison, points to limitations in the merger-ringdown modeling of the higher modes, as in other models.
A related limitation is the absence of some of the higher modes in the waveform models, 
which contribute significantly to the ringdown signal for high mass-ratio systems at a high inclination, as we 
quantify below. Focusing on the bottom panels, we see that \texttt{IMRPhenomXHM} also has many cases between $10^{-4}$ and $10^{-3}$, 
but reaches high values of the unfaithfulness when compared to high mass-ratio, spinning configurations, 
in which higher-mode content is more significant. 
Specifically, the unfaithfulness exceeds $10 \%$ for the most challenging configurations with $q=15$. 
We point out that \texttt{IMRPhenomXHM} has not been calibrated to $q=15$ \texttt{SXS} simulations, 
that became only recently available \cite{Yoo:2022erv}, but was calibrated to private $q=18$ \texttt{BAM} waveforms, 
with different spin values, which have not been used for \texttt{SEOBNRv5HM}.
\texttt{TEOBResumS-GIOTTO} achieves unfaithfulness between $10^{-3}$ and $10^{-2}$ for most cases, 
but also has an appreciable number of outliers reaching mismatch $10 \%$, possibly pointing to 
robustness issues in some of the higher modes close to merger. 

\begin{figure}
	\includegraphics[width=\linewidth]{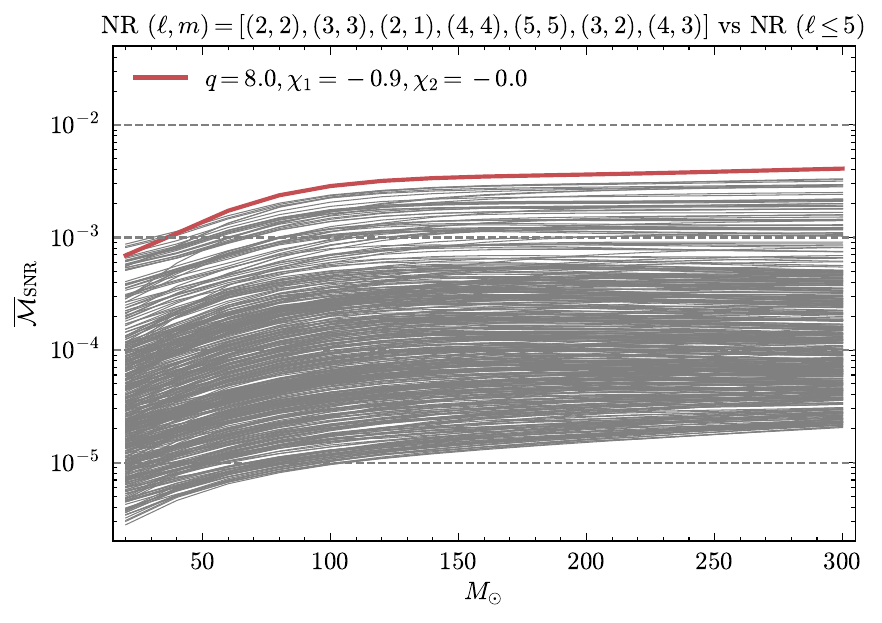}
	\caption{The sky-and-polarization averaged, SNR-weighted mismatch, for inclination $\iota = \pi/3$, 
	over a range of total masses between 20 and 300 $M_{\odot}$, 
	of NR waveforms with the same modes as \texttt{SEOBNRv5HM} $(\ell,m) = (2,2), (3,3), (2,1), (4,4), (5,5), (3,2), (4,3)$
	against NR waveforms with all $(\ell \leq 5)$ modes}.
	\label{fig:mismatch_HM_NR_custom_modes}
\end{figure}

In order to quantify how much the increase of the mismatch with the total mass is related to the missing modes, 
we show in Fig.~\ref{fig:mismatch_HM_NR_custom_modes} 
the sky-and-polarization averaged, SNR-weighted mismatch, for inclination $\iota = \pi/3$, 
over a range of total masses between 20 and 300 $M_\odot$ 
of NR waveforms with the same modes as \texttt{SEOBNRv5HM} $(\ell,m) = (2,2), (3,3), (2,1), (4,4), (5,5), (3,2), (4,3)$
against NR waveforms with all $(\ell \leq 5)$ modes. 
As expected we see an increase of the mismatch with total mass, 
indicating that the error due to neglecting some higher modes is mostly important in the ringdown, 
and we see it can reach more than $0.4\%$ for high $q$ and large spins. 
This tells us that to reach the same accuracy 
of just the (2,2) mode ($<0.3 \%$) for the full polarizations at high $\iota$ one would need to include additional modes in \texttt{SEOBNRv5HM}.

\begin{figure}
	\includegraphics[width=\columnwidth]{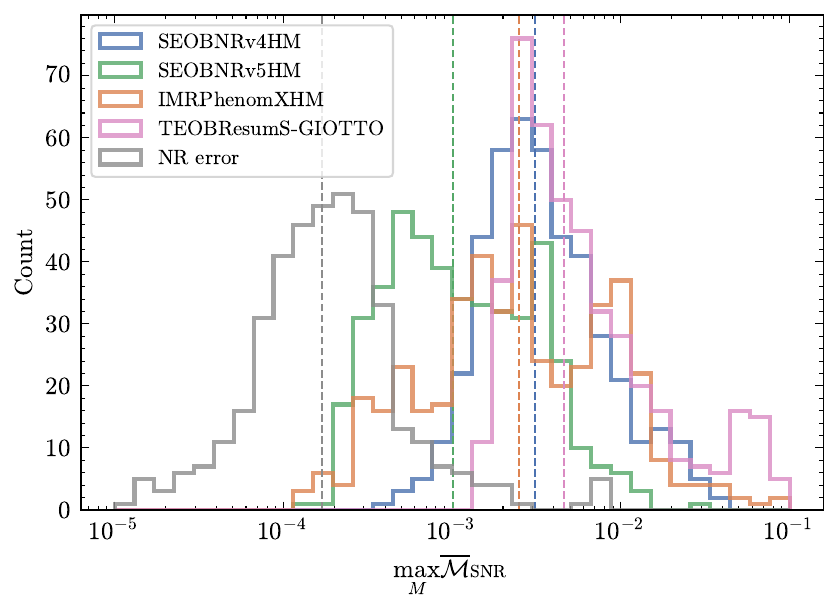}
	\includegraphics[width=\columnwidth]{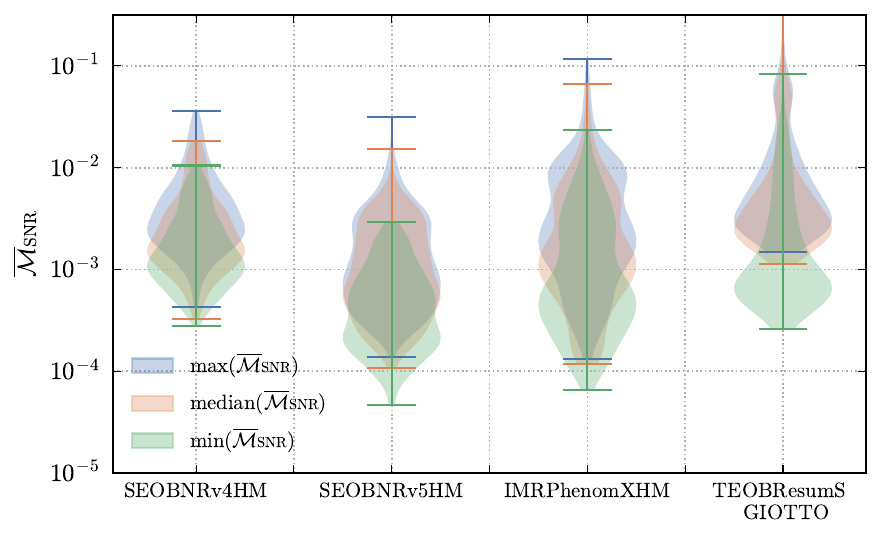}
	\caption{
	\textit{Top panel:} Histogram of the maximum sky-and-polarization averaged, SNR-weighted mismatch, for inclination $\iota = \pi/3$, over a range of total masses between 20 and 300 $M_{\odot}$, 
	between different aligned-spin multipolar approximants and the 441 \texttt{SXS} NR simulations used in this work.
	The NR error is estimated by computing the mismatch between NR simulations with the highest and second-highest resolutions. The vertical dashed lines show the medians.
	\textit{Bottom panel:} distribution of the maximum (blue), median (orange) and minimum (green) mismatch over the same range of total masses for the different models.
	}
	\label{fig:mismatch_hist_HM}
\end{figure}

Figure~\ref{fig:mismatch_hist_HM} summarizes the comparison of Fig.~\ref{fig:mismatch_spaghetti_HM}:
in the top panel we show histograms of the maximum unfaithfulness over the same range of total masses, 
with the vertical lines corresponding to the medians of the distributions,
and an estimate of the NR error computed as the mismatch between NR simulations with different resolutions.
As for the (2,2)-mode only case, the NR error is about one order of magnitude smaller than the \texttt{SEOBNRv5HM} modeling error, 
with median $\sim 1 \times 10^{-4}$.
Overall \texttt{SEOBNRv5HM} achieves a lower unfaithfulness than \texttt{SEOBNRv4HM}, \texttt{IMRPhenomXHM} and \texttt{TEOBResumS-GIOTTO},  
with the median value $1.01 \times 10^{-3}$ and only 7 cases above $1 \%$, as summarized in Table~\ref{tab:mm_hm}.
The violin plots in the bottom panel provide a further comparison by showing the distribution of 
the maximum (blue), median (orange) and minimum (green) mismatch for each model.

We note that in the unfaithfulness computation we include all modes up to $\ell_{\rm{max}} = 5$ in the NR waveforms, while the $(5,5)$ mode is not included in \texttt{IMRPhenomXHM} and \texttt{TEOBResumS-GIOTTO}. To check the impact of neglecting the $(5,5)$ mode in these two models, 
we also repeat the comparison presented in this section using only multipoles up to $\ell_{\rm{max}}=4$, in both the models and the NR waveforms.
We find a result very similar to what is shown above, with all models displaying a slightly better performance, due to fewer missing modes, and the same hierarchy for the accuracy of different approximants.

\setlength{\extrarowheight}{8pt}
\begin{table*}
    \centering
    \begin{ruledtabular}
    	\begin{tabular}{ l  c  c  c  c  }
 Approximant & \makecell[cc]{\texttt{SEOBNRv4HM}} & \makecell[cc]{\texttt{SEOBNRv5HM}} & \makecell[cc]{\texttt{IMRPhenomXHM}} & \makecell[cc]{\texttt{TEOBResumS-GIOTTO}} \\
 \hline
 \centering
 $\text{median} \max_{M} {\overline{\mathcal{M}}}_{\text{SNR}}$ &   $3.11 \times 10^{-3}$ & $1.01 \times 10^{-3}$ & $2.50 \times 10^{-3}$ & $4.59 \times 10^{-3}$ \\
 $ \% \max_{M} {\overline{\mathcal{M}}}_{\text{SNR}} < 10^{-2}$ &   $88 \%$ & $98 \%$ & $86 \%$ & $74 \%$ \\
 $ \% \max_{M} {\overline{\mathcal{M}}}_{\text{SNR}} < 10^{-3}$ &   $5 \%$ & $49 \%$ & $23 \%$ & $0 \%$ \\
 \end{tabular}
    \end{ruledtabular}
 \caption{Summary of the sky-and-polarization averaged, SNR-weighted mismatch, for inclination $\iota = \pi/3$, 
 over a range of total masses between 20 and 300 $M_{\odot}$, 
 between different aligned-spin multipolar approximants and the 441 \texttt{SXS} NR simulations used in this work. 
 We display the median of the maximum mismatch across total mass, and the fraction of cases falling below $10^{-2}$ and $10^{-3}$. 
 } \label{tab:mm_hm}
\end{table*}



\begin{figure}
	\includegraphics[width=\linewidth]{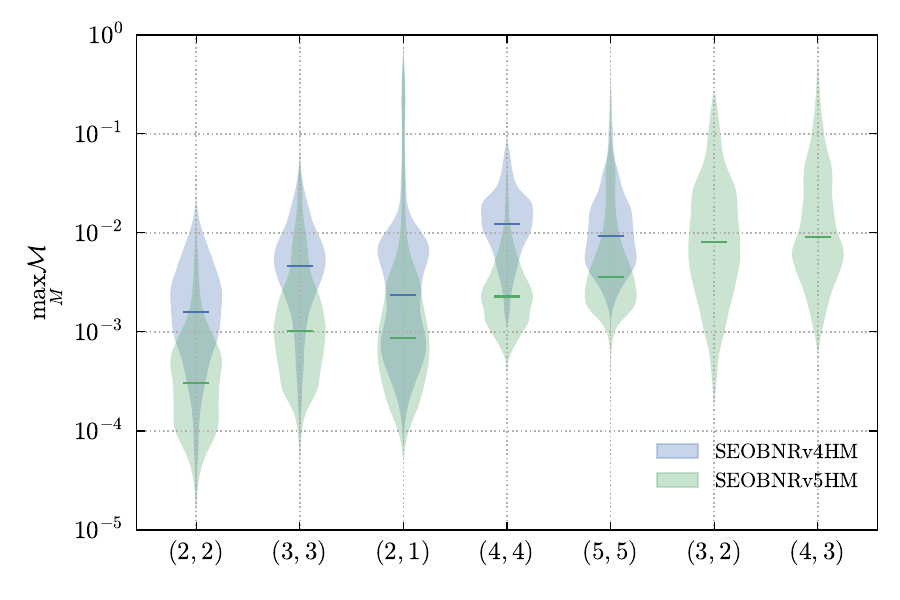}
	\caption{Mode-by-mode mismatches between \texttt{SEOBNRv4HM}, \texttt{SEOBNRv5HM} and \texttt{NRHybSur3dq8}, for 5000 random configurations with $q \in [1,8], |\chi_i|\leq 0.9$.
	For each mode we show the maximum mismatch over a range of total masses between 10 and 300 $M_{\odot}$. The horizontal lines show the medians.}
	\label{fig:mismatch_modes_3dq8}
\end{figure}
\begin{figure}
	\includegraphics[width=\linewidth]{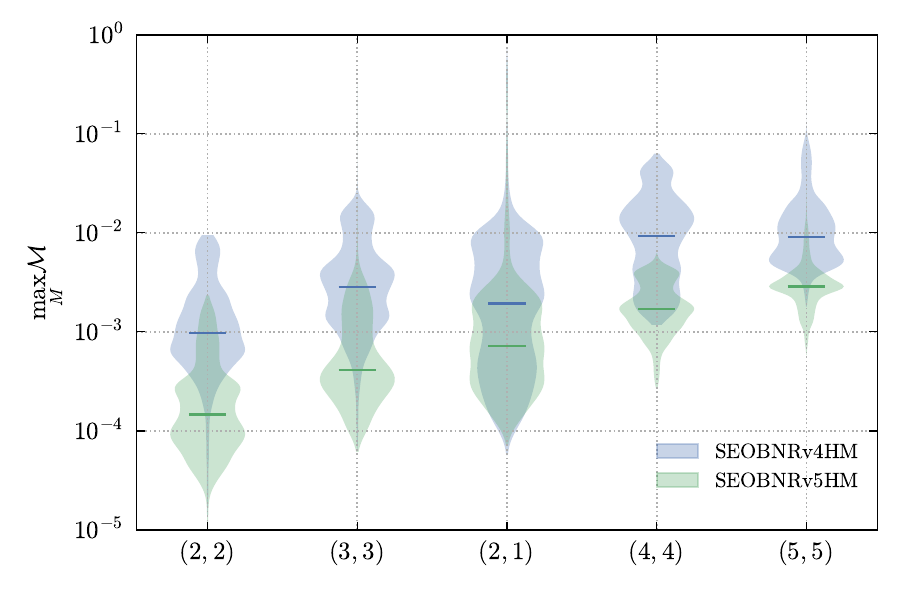}
	\caption{Mode-by-mode mismatches between \texttt{SEOBNRv4HM}, \texttt{SEOBNRv5HM} and \texttt{NRHybSur2dq15}, for 5000 random configurations with $q \in [1,15], |\chi_1|\leq 0.6, \chi_2 = 0$.
	For each mode we show the maximum mismatch over a range of total masses between 10 and 300 $M_{\odot}$. The horizontal lines show the medians.}
	\label{fig:mismatch_modes_2dq15}
\end{figure}

To validate \texttt{SEOBNRv5HM}, we compare it to the multipolar aligned-spin surrogate model 
\texttt{NRHybSur3dq8} \cite{Varma:2018mmi}.
This model was built for binaries with mass-ratios $1-8$ and spin magnitudes up to $0.8$, and 
provides waveforms with errors comparable to the NR accuracy in the region where the model was trained.
\texttt{NRHybSur3dq8} waveforms were not used in the construction of \texttt{SEOBNRv5HM}, so this is an 
important validation check of the NR calibration pipeline. 
We point out that \texttt{NRHybSur3dq8} is trained on NR waveforms hybridized with PN and \texttt{SEOBNRv4} waveforms in the early inspiral.
In the following comparisons, we generate waveforms from an initial geometric frequency of $0.015$, for which the impact of the hybridization should not be large. 

Figure~\ref{fig:mismatch_modes_3dq8} compares \texttt{SEOBNRv4HM} and \texttt{SEOBNRv5HM} against \texttt{NRHybSur3dq8}, 
showing a kernel density estimation of the distribution of the maximum mode-by-mode mismatches between them.
We use 5000 random configurations with $q \in [1,8], |\chi_i|\leq 0.9$, allowing 
some extrapolation outside of the surrogate's training region, as to also test the extrapolation of the \texttt{SEOBNRv5HM} calibration.

First, we notice that the (2,2)-mode median mismatch $\sim 3 \times 10^{-4}$ is comparable to the one against NR, 
only slightly higher because of the larger number of challenging cases with high $q$ and high spin in this comparison. 
The maximum unfaithfulness for the (2,2) mode, which is reached, as expected, for large mass ratios and positive spins,
remains below $0.01$, if we limit the comparison to the region $q \in [1,8], |\chi_i|\leq 0.8$ where the surrogate was trained, and can be only slightly above $0.01$ if going up to $|\chi_i| = 0.9$ in the surrogate's extrapolation region. This 
confirms a good extrapolation of the \texttt{SEOBNRv5HM} fits. 
Comparing to \texttt{SEOBNRv4HM}, we have as expected fewer cases above $0.01$, and much lower median unfaithfulness.

Going to the higher multipoles, we see larger errors for the smaller higher modes, as for most other state-of-the-art models.
The subdominant higher modes in NR simulations are noisier, and more difficult to model (both for EOB models and for \texttt{NRHybSur3dq8}).
Some of the higher modes also include considerably less analytical information compared to the (2,2) mode (see Appendix~\ref{app:modes}), 
and adding the full 3.5PN contributions from Refs.~\cite{Henry:2022dzx, Henry:2022ccf} 
would likely bring a significant improvement to some of them. 
Nonetheless, we see a consistent improvement comparing \texttt{SEOBNRv5HM} to \texttt{SEOBNRv4HM}, mostly due to the enhanced calibration and merger-ringdown description.

The (2,1) mode shows a tail of cases with large mismatches for both \texttt{SEOBNRv5HM} and \texttt{SEOBNRv4HM}:
as also discussed in Ref.~\cite{Cotesta:2018fcv} those are cases with a minimum in the amplitude close to merger, 
which can be especially difficult to model given that the current merger-ringdown ansatz assumes a monotonic post-merger amplitude evolution.
Nonetheless, these are configurations where the (2,1) mode is highly suppressed, and would not impact significantly in the full polarizations.
We also compare the (3,2) and (4,3) modes of \texttt{SEOBNRv5HM} against \texttt{NRHybSur3dq8}
(these modes are not included in \texttt{SEOBNRv4HM}). We see that these modes show the largest modeling errors, 
which is expected considering they are among the smallest modes for most configurations, and also 
keeping in mind that the mode-mixing modeling in the ringdown is approximated.

Figure~\ref{fig:mismatch_modes_2dq15} shows a similar comparison against \texttt{NRHybSur2dq15} \cite{Yoo:2022erv}, 
limited to the modes modeled by the surrogate. This model was built for binaries with mass-ratios $1-15$, 
primary spin up to $0.5$ and no secondary spin. 
We consider 5000 random configurations with $q \in [1,15], |\chi_1|\leq 0.6, \chi_2 = 0$, allowing again 
some extrapolation outside of the surrogate's training region, as to also test the extrapolation of the \texttt{SEOBNRv5HM} calibration fits.
We see a similarly large improvement for all the modes comparing \texttt{SEOBNRv5HM} to \texttt{SEOBNRv4HM}, and 
the (2,2) mode result, with maximum value $2.3 \times 10^{-3}$ and median $1.5 \times 10^{-4}$, confirms the 
robustness of the calibration procedure.

\begin{figure*}
	\includegraphics[width=\textwidth]{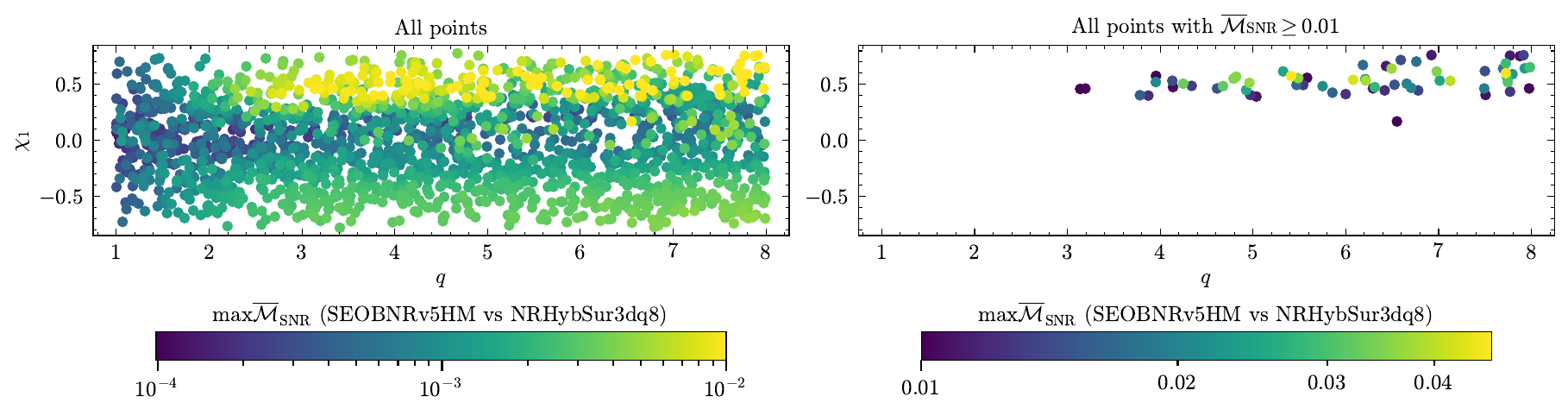}
	\caption{Sky-and-polarization averaged, SNR-weighted mismatch, for inclination $\iota = \pi/3$, 
	between \texttt{SEOBNRv5HM} and \texttt{NRHybSur3dq8}, for 2000 random configurations with $q \in [1,8], |\chi_i|\leq 0.8$.
	We show maximum mismatch over a range of total masses between 20 and 300 $M_{\odot}$ as a function of the mass-ratio $q$ and the primary spin $\chi_1$.}
	\label{fig:scatter_3dq8_ipi3}
\end{figure*}

In Fig.~\ref{fig:scatter_3dq8_ipi3} we show the sky-and-polarization averaged, SNR-weighted mismatch, for inclination $\iota = \pi/3$, 
between \texttt{SEOBNRv5HM} and \texttt{NRHybSur3dq8}, for 2000 random configurations with $q \in [1,8], |\chi_i|\leq 0.8$. 
In particular, we plot the maximum mismatch as a function of the mass-ratio $q$ and the primary spin $\chi_1$.
The unfaithfulness grows with mass ratio and spin, with the highest unfaithfulness reaching $0.04$. This effect also is enhanced by
the fact that we start all the waveforms at the same frequency and for higher mass ratios, the number of cycles in band grows as $\sim 1/\nu$.

\begin{figure*}
	\includegraphics[width=\textwidth]{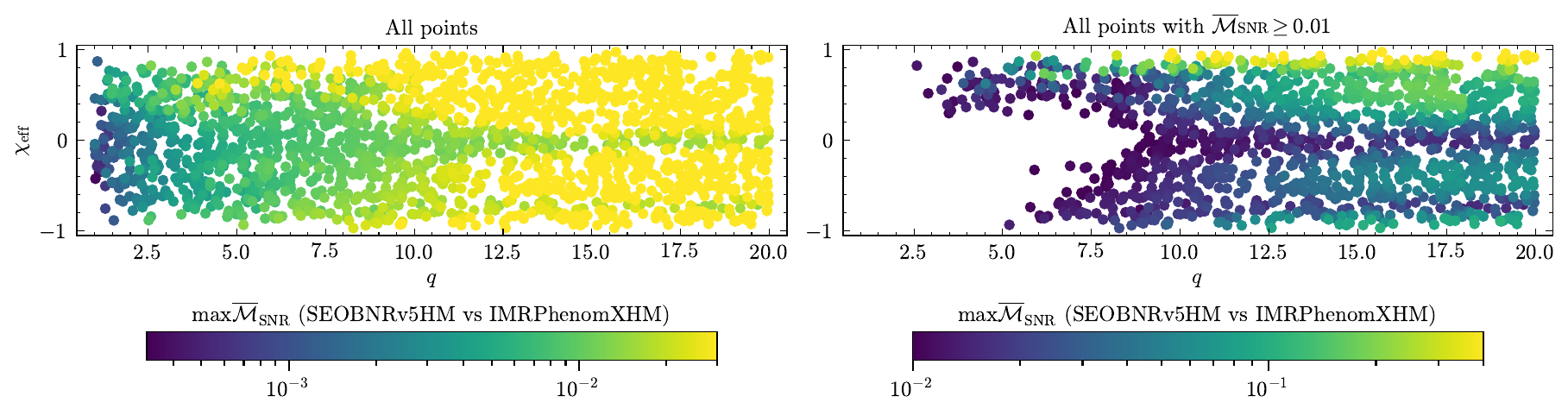}
	\caption{Sky-and-polarization averaged, SNR-weighted mismatch, for inclination $\iota = \pi/3$, 
	between \texttt{SEOBNRv5HM} and \texttt{IMRPhenomXHM}, for 2000 random configurations with $q \in [1,20], |\chi_i|\leq 0.99$.
	We show maximum mismatch over a range of total masses between 20 and 300 $M_{\odot}$ as a function of the mass-ratio $q$ and the effective spin $\chi_{\text{eff}}$.}
	\label{fig:scatter_xhm_ipi3}
\end{figure*}

We plot in Fig.~\ref{fig:scatter_xhm_ipi3} a similar comparison between \texttt{SEOBNRv5HM} and \texttt{IMRPhenomXHM}, 
for 2000 random configurations with $q \in [1,20], |\chi_i|\leq 0.99$ 
in order to examine the behavior of the models outside of the region in which they were calibrated to NR. 
As in the previous comparsion, the unfaithfulness grows with mass-ratio and spin, and can reach very large values for $q\simeq20$ and high $\chi_{\text{eff}}$. 
This confirms that waveform systematics are important, even for aligned-spin systems observed by current detectors, in the region where waveform models are not calibrated to NR simulations.



\subsection{Accuracy of \texttt{SEOBNRv5} angular-momentum flux and binding energy}

The performance of waveform models is typically assessed by computing the unfaithfulness between the waveforms produced by the model and 
NR waveforms with corresponding parameters, as the waveform itself is the relevant quantity used in data analysis. 
In EOB models, however, the knowledge of the binary's dynamics allows us to complement the waveform comparison with other dynamical quantities. 
Since the calibration of the model to NR is based on the waveforms, seeing an improvement in different dynamical quantities is a powerful check of the physical 
robustness of the model. In particular, we examine the angular-momentum flux radiated at infinity \cite{Boyle:2008ge, Albertini:2021tbt}, 
and the binding energy \cite{Damour:2011fu,Nagar:2015xqa,Ossokine:2017dge}.

We compute the NR angular-momentum flux at infinity from the waveform modes using 
\begin{equation}
	\dot{J} =-\frac{1}{8 \pi} \sum_{\ell=2}^{\ell_{\max }} \sum_{m=-\ell}^{\ell} m \Im\left(\dot{h}_{\ell m} h_{\ell m}^*\right),
\end{equation}
where we assume $\ell_{\rm{max}} = 8$. For clarity, we normalize the flux by the leading (Newtonian) one for circular orbits, 
\begin{equation}
	\dot{J}_{N}=\frac{32}{5} \nu^2\left(M \Omega \right)^{7 / 3},
\end{equation}
where we estimate the NR orbital frequency $\Omega_{\mathrm{NR}}$ from the NR (2,2)-mode frequency as 
\begin{equation}
	\Omega_{\mathrm{NR}} \equiv \frac{\omega_{22}^{\mathrm{NR}}}{2}.
\end{equation}
We denote the normalized flux as
\begin{equation}
	\dot{\hat{J}} = \frac{\dot J}{\dot J_N}.
\end{equation}
We note again that the \texttt{SEOBNRv5} flux does not include NQC corrections, and we practically compute it from the dynamics as 
$\dot{J} = \dot p_{\phi}$. In the following, we always consider it as a function of $\Omega_{\rm{EOB}}$, which is read from the orbital dynamics.

As an example, in Fig.~\ref{fig:flux_comparison} we compare the \texttt{SEOBNRv4} and \texttt{SEOBNRv5} angular-momentum fluxes 
against the one extracted from the NR simulation \texttt{BFI:q2-3d-95:001} with parameters $(q, \chi_1, \chi_2) = (1.0, -0.95, -0.949)$. 
We plot the fluxes as function of $v = ( M \Omega)^{1/3}$, where it is intended that $v = (M \Omega_{\rm{NR}})^{1/3}$ for NR, and $v = (M \Omega_{\rm{EOB}})^{1/3}$ for the EOB models,
and we highlight with the triangle, square and diamond where 3, 1 and 0 GW cycles before merger (taken as the peak of $|h_{22}|$) are.
The \texttt{SEOBNRv5} flux shows a better agreement, thanks to the additional PN information summarized in Sec.~\ref{subsec:hlm_inspiral_plunge}
and the calibration to 2GSF. As highlighted in Ref.~\cite{VandeMeentv5}, the latter seems to be the most significant source of improvement.

\begin{figure}
	\includegraphics[width=\linewidth]{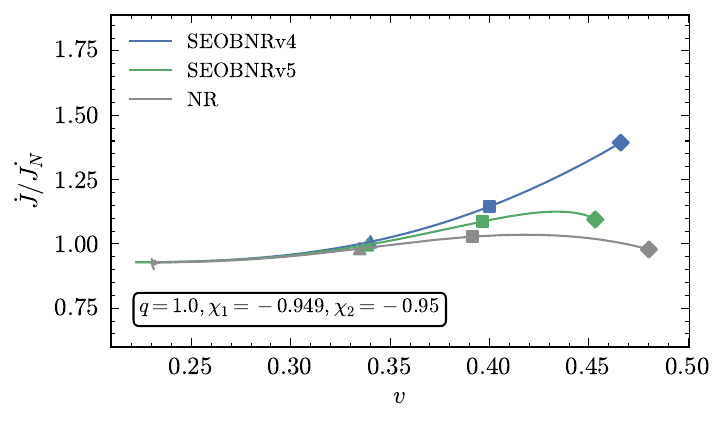}
	\caption{Comparison of the Newtonian-normalized angular-momentum flux between \texttt{SEOBNRv4}, \texttt{SEOBNRv5}, and
	the NR simulation \texttt{SXS:BBH:0156}. The triangle, square and diamond correspond, respectively, to 
	3, 1 and 0 GW cycles before merger, which is taken as the peak of the (2,2)-mode amplitude for each model.}
	\label{fig:flux_comparison}
\end{figure}

To quantify the improvement of the \texttt{SEOBNRv5} model with respect to \texttt{SEOBNRv4} across parameter space, we show
in Fig.~\ref{fig:flux_scatter} the fractional difference between of the Newtonian-normalized angular-momentum flux $\dot{\hat{J}}$ of \texttt{SEOBNRv4} and \texttt{SEOBNRv5}, and the one
obtained from the NR simulations described in Sec.~\ref{sec:calibration}, evaluated two cycles before merger.
The median fractional difference goes from $4.83 \%$ to $1.15 \%$, and while the difference can be as high as $18 \%$ for the \texttt{SEOBNRv4} model, 
it is always below $9 \%$ for the \texttt{SEOBNRv5} model.

\begin{figure}
	\includegraphics[width=\linewidth]{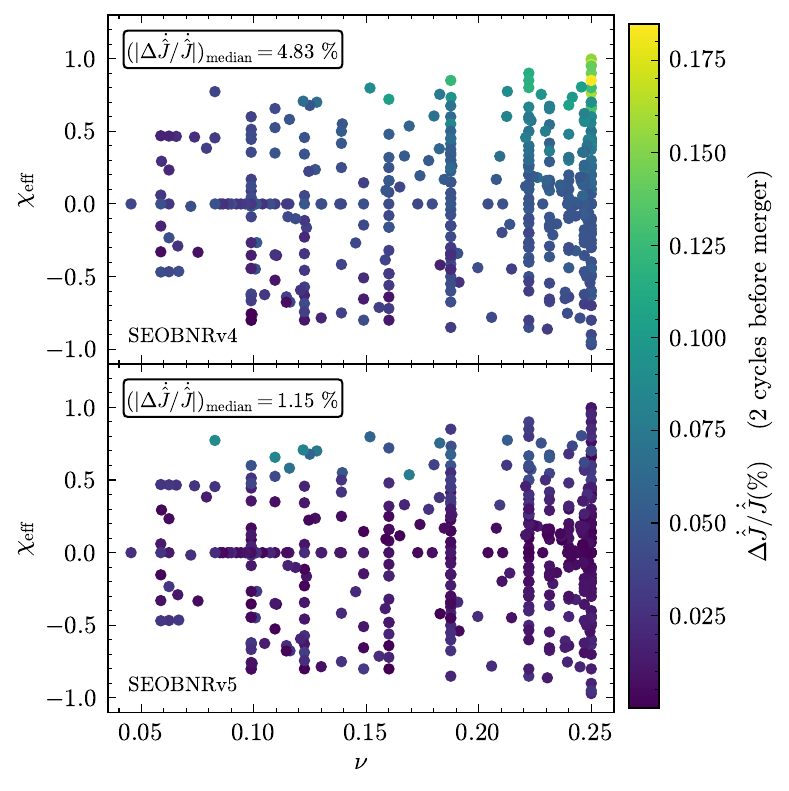}
	\caption{Fractional difference between of the Newtonian-normalized angular-momentum flux $\dot{\hat{J}}$ of \texttt{SEOBNRv4} and \texttt{SEOBNRv5}, and the one
	obtained from the NR simulations used in this work, evaluated two cycles before merger.}
	\label{fig:flux_scatter}
\end{figure}

The other comparison we consider is of the binding energy \cite{Damour:2011fu, Nagar:2015xqa,Ossokine:2017dge}.
The NR binding energy data used here was obtained in Ref.~\citep{Ossokine:2017dge}, 
while the EOB binding energy is simply computed by evaluating 
\begin{equation}
	E^{\mathrm{bind}}_{\mathrm{EOB}}=H_{\mathrm{EOB}}-M,
\end{equation}
along the EOB dynamics. Henceforth, to ease the notation, we will refer to $E_{\mathrm{EOB}}$ instead of 
$E^{\mathrm{bind}}_{\mathrm{EOB}}$.
The EOB orbital frequency is obtained from $\Omega_{\rm{EOB}}={\partial H_{\mathrm{EOB}}}/{\partial p_\phi}$,
 to be consistent with the gauge-invariant definition used for NR in Ref.~\cite{Ossokine:2017dge}.

In Fig.~\ref{fig:binding_nonspinning} we show the fractional difference between the NR binding energy for nonspinning configurations, and the one
of \texttt{SEOBNRv4} and \texttt{SEOBNRv5}, for different mass-ratios. 
The gray region is an estimate of the NR error obtained from the $q=1$ data.
Both EOB models show minor errors during most of the inspiral, and stay within the NR uncertainty until 
around 3 GW cycles before merger. The \texttt{SEOBNRv5} model shows, however, a much better agreement in the late-inspiral, between 
3 and 1 cycles before merger, and remains within the error until $v \simeq 0.45$ for all mass-ratios.
As highlighted in Ref.~\cite{VandeMeentv5}, this improvement is mostly a consequence of the calibration to 2GSF results.

\begin{figure}
	\includegraphics[width=\linewidth]{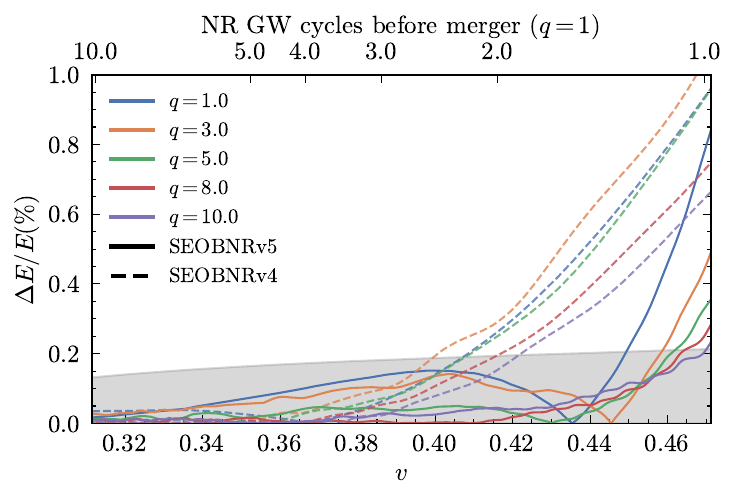}
	\caption{Fractional difference between the EOB and NR nonspinning binding energy as a function of $v$, for \texttt{SEOBNRv5} and \texttt{SEOBNRv4}.
	The gray region represents an estimate of the NR error. Notice the improvement in the agreement of 
	\texttt{SEOBNRv5} compared to \texttt{SEOBNRv4}, especially between 3 and 1 cycles before merger.}
	\label{fig:binding_nonspinning}
\end{figure}

We now turn to aligned-spin cases, and as starting point we compare different spin contributions to the binding energy, 
which can be extracted by combining results for various spin combinations as in Refs.~\cite{Ossokine:2017dge,Dietrich:2016lyp,Khalil:2020mmr}
\begin{subequations}
\begin{align}
	\label{eq:e_SO}
	E_{\mathrm{SO}}=&-\frac{1}{6}(-0.6,0)+\frac{8}{3}(0.3,0)-2(0,0)-\frac{1}{2}(0.6,0), \\
	E_{\mathrm{S^2}}=&\frac{3}{2}(-0.6,0)-2(0,0)+\frac{3}{2}(0.6,0)-(0.6,-0.6), \\
	E_{\mathrm{S^3}}=&-\frac{5}{6}(-0.6,0)-\frac{8}{3}(0.3,0)+3(0,0)-\frac{1}{2}(0.6,0) \nonumber \\
	&+\frac{1}{2}(0.6,-0.6)+\frac{1}{2}(0.6,0.6),
\end{align}
\end{subequations}
where the numbers in brackets correspond to the dimensionless spins $(\chi_1, \chi_2)$ of the BHs.
The spin-squared contributions to the binding energy $E_{\mathrm{S^2}}$ refer to
both $\mathrm{S_i^2}$ and $\mathrm{S_1 S_2}$ interactions, and similarly the spin-cubic contributions 
$E_{\mathrm{S^3}}$ refer to both $\mathrm{S_i^3}$ and $\mathrm{S_i^2 S_j}$. Among these contributions 
the spin-orbit term dominates throughout the inspiral, while the quadratic
and cubic-in-spin terms have comparable magnitudes,
with the quadratic terms growing larger close to merger.

\begin{figure}
	\includegraphics[width=\linewidth]{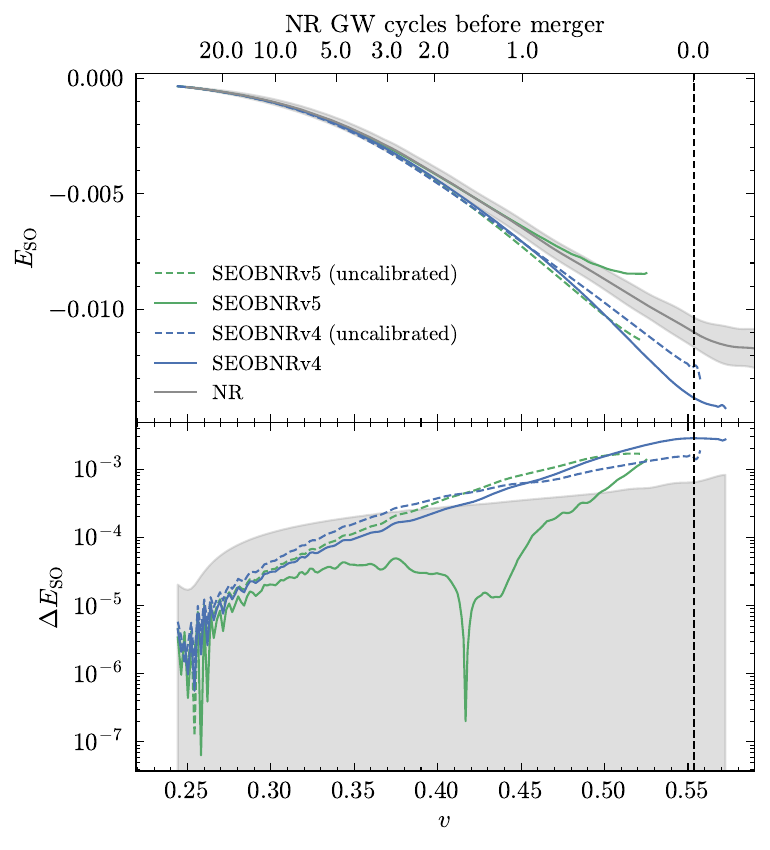}
	\caption{
	Spin-orbit contribution to the binding energy as a function of $v$ 
	for \texttt{SEOBNRv4} (blue), \texttt{SEOBNRv5} (green) and NR (gray).
	The uncalibrated models are obtained by setting to zero the calibration parameters entering the Hamiltonian. 
	The dashed vertical line represents the merger of the NR configuration in Eq.~(\ref{eq:e_SO}) that merges at the lowest frequency, 
	and the numbers of cycles also refer to the same simulation, while the EOB curves terminate at EOB merger. The shaded regions represent the NR error. 
	\texttt{SEOBNRv5} has a better agreement with NR compared to \texttt{SEOBNRv4}, and remains within the NR error almost until merger.}
	\label{fig:plot_SO}
\end{figure}

We begin by considering the spin-orbit effects. In Fig.~\ref{fig:plot_SO} we compare the NR data to \texttt{SEOBNRv4} and \texttt{SEOBNRv5}.
In both cases, we consider calibrated and uncalibrated models, where by uncalibrated we mean that we set to zero all calibration parameters entering the Hamiltonian
(the values of ${\Delta t^{22}_{\rm{ISCO}}}$ or ${\Delta t^{22}_{\rm{peak}}}$, on the other hand, do not affect these comparisons, as they only determine the time at which
the merger-ringdown waveform modes are attached).
\texttt{SEOBNRv5} has a better agreement with NR compared to \texttt{SEOBNRv4}, and remains within the NR error almost until merger.
Moreover, the calibrated \texttt{SEOBNRv5} model performs better than the uncalibrated model during the entire inspiral, 
whereas in \texttt{SEOBNRv4} the calibration degrades the agreement after $v \simeq 0.45$. 

\begin{figure*}
	\includegraphics[width=1.0\columnwidth]{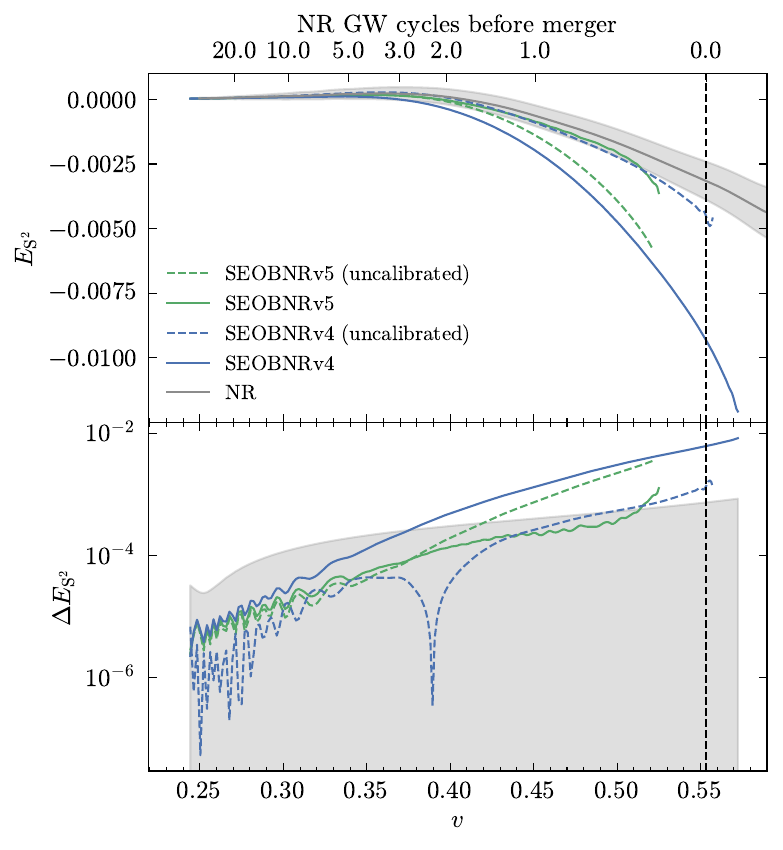}
	\includegraphics[width=1.0\columnwidth]{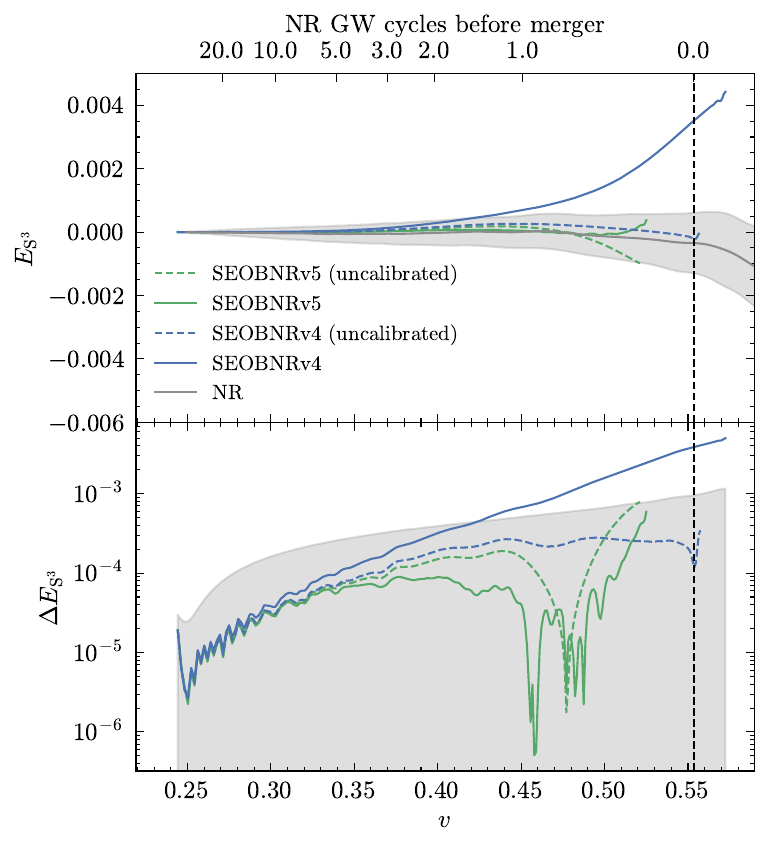}
	\caption{Same as Fig~\ref{fig:plot_SO}, but for spin-spin contributions (left panel) and for cubic-in-spin contributions (right panel).}
	\label{fig:plot_S2_S3}
\end{figure*}

The results for the spin-spin term are shown in the left panel of Fig.~\ref{fig:plot_S2_S3}: again, 
\texttt{SEOBNRv5} clearly outperforms \texttt{SEOBNRv4}, and has differences 
compatible with the NR uncertainty almost up to merger. 
An interesting difference is that, while uncalibrated \texttt{SEOBNRv4} 
has a smaller difference with NR compared to the calibrated model, the same trend is 
not present in \texttt{SEOBNRv5}.
This shows that the calibration of the model, which focuses on producing accurate waveforms, 
is not guaranteed to provide a better description of the conservative dynamics in the strong-field regime. 
A possible reason for this difference might be the additional presence of a spin-spin calibration parameter $d_{\rm{SS}}$ in \texttt{SEOBNRv4}, 
breaking the symmetry underlying the extraction of the terms used here. 
It is also possible that, due to degeneracies between changes in the dissipative and conservative dynamics, 
the less accurate flux of \texttt{SEOBNRv4} is compensated by the calibration of the Hamiltonian, and results in an overall worse agreement of the conservative dynamics with NR.

We consider cubic-in-spin contributions to the binding energy in the right panel of Fig.~\ref{fig:plot_S2_S3}.
These effects are minor, and contribute little to the overall disagreement, however one can see similarly 
to the spin-squared contributions that for \texttt{SEOBNRv4} the calibration worsens the agreement with NR,
making it the only model that does not stay within the NR error.

\begin{figure}
	\includegraphics[width=\linewidth]{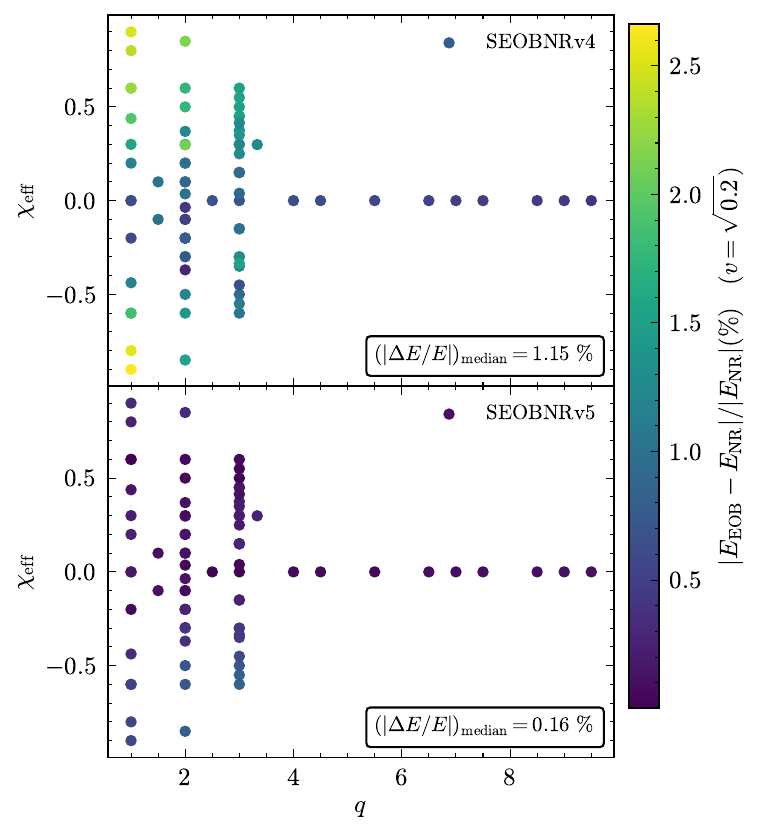}
	\caption{Fractional difference between the EOB and NR binding energy, for \texttt{SEOBNRv5} and \texttt{SEOBNRv4}, at $v = \sqrt{0.2}\simeq 0.45$.}
	\label{fig:summary_scatter}
\end{figure}

We finally quantify the improvement across parameter space by computing the fractional energy difference in the binding energy $\left|E_{\mathrm{EOB}} - E_{\mathrm{NR}}\right| / E_{\mathrm{NR}}$ 
at a fixed frequency $v = \sqrt{0.2} \simeq 0.45$ 
for several configurations. Constructing the binding energy curves is not a straightforward process, as one needs to take into account 
a shift of the curves due to the presence of junk radiation in NR waveforms, therefore we only focus on the simulations examined in Ref.~\cite{Ossokine:2017dge}.
In Fig.~\ref{fig:summary_scatter} we show such a comparison for the \texttt{SEOBNRv4} and \texttt{SEOBNRv5} models. In the first case 
the difference in the binding energy can reach more than $2.5 \%$, especially for large values of the effective spin $\chi_{\rm{eff}}$, 
while for \texttt{SEOBNRv5} we always find deviations from the NR binding energy at the sub-percent level.
The median relative difference is also considerably smaller, going from $1.15 \%$ for \texttt{SEOBNRv4} to only $0.16 \%$ for the \texttt{SEOBNRv5} model.

As highlighted in Ref.~\cite{VandeMeentv5}, an improved modeling of the binding energy and angular momentum flux
doesn't necessarily correspond to a higher faithfulness of the waveforms in the regime where they are calibrated to NR, 
due to a significant degeneracy between the calibration terms in the EOB Hamiltonian and changes in the RR force. 
Nevertheless, achieving a more accurate representation of both the conservative and dissipative dynamics improves the overall consistency and naturalness of the model.
This reduces the model's reliance on NR calibration and provides greater confidence that \texttt{SEOBNRv5} will maintain a certain faithfulness to NR when 
extrapolated beyond its calibration region, in particular for higher mass ratios.

\subsection{Computational performance}
\label{sec:computational_performance}

The fifth generation of \texttt{SEOBNR} models, starting from \texttt{SEOBNRv5HM}, is implemented in \pySEOBNR, a Python package for developing and using waveform models within the \texttt{SEOBNR} framework. 
As described in Ref.~\cite{Mihaylovv5}, \pySEOBNR~offers a simple, object-oriented interface for building, calibrating, deploying, and profiling waveform models in both time and frequency domain.
The \pySEOBNR~package moves the development core of the \texttt{SEOBNR} framework from the previously used C-based LALSuite \cite{lalsuite} to a much more flexible, modern and widely used Python infrastructure, 
setting a new standard for developing waveform models for current and future GW detectors. 
The user interface is implemented in pure Python, to facilitate ease of use and quick adoption by other researchers. 
The backend of the package relies on well-known, regularly maintained packages under open-software licenses, including 
\texttt{Cython}~\cite{behnel2011cython} and \texttt{Numba}~\cite{Numba-LLVM} for fast Hamiltonian evaluation and waveform generation, and \texttt{NumExpr}~\cite{numexpr} for efficient \texttt{numpy}~\cite{harris2020array} vectorized operations. 

In this section we discuss the computational performance of the \texttt{SEOBNRv5HM} implementation in \pySEOBNR, in terms of walltime for generating a waveform, 
and compare the model to other time-domain aligned-spin approximants that include higher modes, \texttt{SEOBNRv4HM}, with and without PA approximation, 
\texttt{TEOBResumS-GIOTTO}, which also employs the PA approximation, and \texttt{IMRPhenomTHM}.

Figure~\ref{fig:benchamrks} shows the walltime for generating a waveform in the time domain, including interpolation on a constant time step, 
for total masses between $10$ and $100 M_{\odot}$, at starting frequency of $10 ~\text{Hz}$, for three values of the mass ratio $q = {1, 3, 10}$ and spins $\chi_{1} = 0.8,~ \chi_{2} = 0.3$. 
For all approximants we include all modes up to $\ell = 4$, and keep all other settings as default.
We choose the sampling rate such that the Nyquist criterion is satisfied for the $\ell = 4$ multipoles.
\footnote{All benchmarks were performed on the \texttt{Hypatia} computer cluster 
at the Max Planck Institute for Gravitational Physics in Potsdam, on a compute node equipped with a dual-socket 64-core AMD EPYC (Rome) 7742 CPU.}

Comparing the \texttt{SEOBNRv5HM} and \texttt{SEOBNRv4HM} models without the use of the PA approximation (dashed lines), 
we find a major performance improvement across all values of the total mass $M$. The speedup
is most significant for lower total mass $(\sim 50 \times)$, and decreases for higher total mass to $\sim 10 \times$. 
The difference between \texttt{SEOBNRv5HM} and \texttt{SEOBNRv4HM\_PA}, with the PA approximation being used in both cases (plotted in solid lines),
is less drastic. Nonetheless, \texttt{SEOBNRv5HM} is consistently faster, despite including two additional modes. The speed-up is up to $\sim 70 \%$ for low total-mass binaries.
When using the PA approximation, a significant improvement in \texttt{SEOBNRv5HM} is the use of analytic equations for the momenta (see Eqs.~(\ref{eq:pa_pr}) and~\ref{eq:pa_pphi}), 
whereas these quantities are determined numerically in \texttt{SEOBNRv4HM}. 
We note that the difference between \texttt{SEOBNRv4HM} with and without the PA approximation is not limited to the use of the 
PA approximation, since \texttt{SEOBNRv4HM\_PA} features several optimizations, such as the use 
of analytic derivatives of the Hamiltonian, which have also been implemented in the \texttt{SEOBNRv5HM} model 
independently of the use of the PA approximation.
This is one of the reasons why the difference between \texttt{SEOBNRv5HM} with and without PA is not as large as in the previous generation of \texttt{SEOBNR} models. 
It can reach up to $\sim 2 \times$ for low total mass systems, while it is between $10 \mbox{--} 40 \%$ for $M\sim 100 M_{\odot}$, for cases where the cost of integrating the dynamics is less high.
Comparing \texttt{SEOBNRv5HM} to a different EOB model, \texttt{TEOBResumS-GIOTTO}, employing in both cases the PA approximation, we see that 
\texttt{TEOBResumS-GIOTTO} is faster for high total-mass binaries, with a difference ranging from $\sim 3 \times$ for $q=1$ to $\sim 1.5 \times$ for $q=10$,
while the two are comparable for low total masses.
The time-domain phenomenological model \texttt{IMRPhenomTHM} outperforms all EOB models, for large total-mass systems, by over an order of magnitude. 
This is due to its use of fast closed-form expressions, rather than ODE integration. 
The gap between the models narrows as the total mass decreases, as the mode interpolation on a constant time-step needed for the 
Fast-Fourier-Transform becomes a major cost for long inspirals (excluding \texttt{SEOBNRv4HM} without PA approximation, where ODE integration remains by far the main cost factor). 

\begin{figure}
	\includegraphics[width=\linewidth]{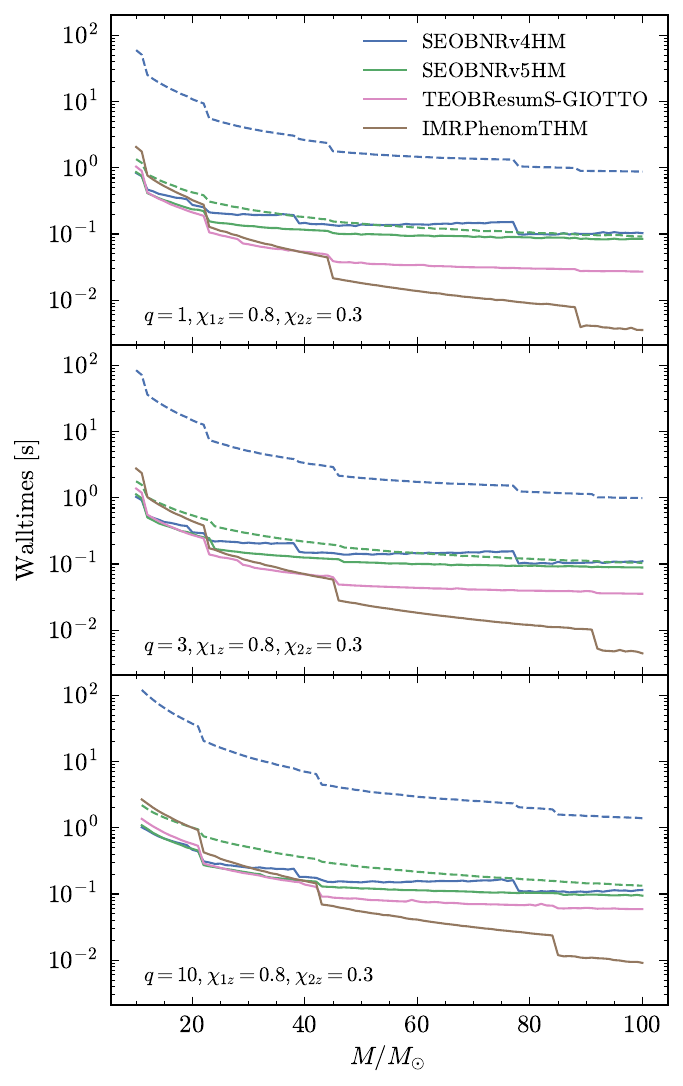}
	\caption{
	Walltimes for \texttt{SEOBNRv5HM} and \texttt{SEOBNRv4HM}, with PA approximation (solid lines) and without (dashed lines), 
	\texttt{TEOBResumS-GIOTTO} and \texttt{IMRPhenomTHM}, starting from $f_{\rm{start}}=10~\rm{Hz}$, as a function of the total mass $M$.
	\texttt{SEOBNRv5HM} outperforms \texttt{SEOBNRv4HM}, particularly for low total mass systems, 
	both with and without the PA approximation, and shows walltimes close to \texttt{TEOBResumS-GIOTTO}. 
	\texttt{IMRPhenomTHM} is the fastest model for low total masses 
	due to its use of closed-form expressions, with the gap narrowing for lower total masses. 
	The analytic PA approximation and several optimizations, such as the use of analytic derivatives of the Hamiltonian, 
	play a crucial role in the \texttt{SEOBNRv5HM} performance. 
	}
	\label{fig:benchamrks}
\end{figure}

\section{Parameter-estimation study}
\label{sec:PE}

One of the most relevant applications of waveform models is to perform parameter inference for GW signals. Current parameter-estimation codes for inferring the properties of compact-binary coalescences are based on Bayesian inference, 
where the posterior probability distribution $P(\boldsymbol{\lambda} | d|)$ for the parameters $\boldsymbol{\lambda}$, given a signal $d$, is given by the Bayes theorem~\cite{Thrane:2018qnx}
\begin{equation}
P(\boldsymbol{\lambda} | d) = \dfrac{\pi(\boldsymbol{\lambda}) \, \mathcal{L}(d | \boldsymbol{\lambda})}{Z},
\end{equation}
where $\mathcal{L}(d | \boldsymbol{\lambda})$ is the likelihood of reproducing the data given a set of parameter values and a model for the signal, 
$\pi(\boldsymbol{\lambda})$ is the prior probability and $Z=\int d\boldsymbol{\lambda} \, \pi(\boldsymbol{\lambda}) \, \mathcal{L}(d | \boldsymbol{\lambda})$ 
is the evidence of the model reproducing the data. The posterior distribution is stochastically sampled across the model parameter space, 
typically using nested sampling~\cite{Skilling:2006gxv} or MCMC methods, which require from millions to hundred of millions of waveform evaluations 
(see e.g. Refs.~\cite{Ashton:2021anp, Williams:2021qyt, Williams:2023ppp}). 
Therefore, besides requiring that the waveform models accurately reproduce the data, it is also important that they are computationally efficient, 
to perform parameter estimation with reasonable resources and in a reasonable time. 
In this section we study the performance of \texttt{SEOBNRv5HM} for the recovery of parameters with a synthetic signal and three GW events observed during O1, O2 and O3. 

\subsection{Inference with a numerical-relativity  synthetic signal}

\setlength{\extrarowheight}{8pt}
\begin{table}[h!]
    \centering
    \begin{ruledtabular}
    \begin{tabular}{ l  c  c  c  }
 Parameter & \makecell[cc]{Injected \\ value} & \makecell[cc]{IMRPhenomXHM \\ recovery} & \makecell[cc]{SEOBNRv5HM \\ recovery} \\
 \hline
 \centering
 $M/M_\odot$ &  162.0 & $139.6^{+9.55}_{-10.93}$ & $160.58^{+11.57}_{-12.91}$ \\
 $\mathcal{M}/M_\odot$ & 29.53 & $29.65^{+1.46}_{-0.94}$ & $29.7^{+1.07}_{-0.9}$ \\
 $m_1/M_\odot$ & 151.88 & $128.09^{+9.95}_{-11.81}$ & $150.27^{+12.12}_{-13.64}$ \\
 $m_2/M_\odot$ & 10.13  & $11.54^{+1.26}_{-0.86}$ & $10.32^{+0.98}_{-0.78}$ \\
 $1/q$ & 0.067  & $0.09^{+0.02}_{-0.01}$ & $0.07^{+0.01}_{-0.01}$ \\
 $\chi_{\text{eff}}$ & 0.469 & $0.37^{+0.06}_{-0.07}$ & $0.47^{+0.05}_{-0.06}$ \\
 $\chi_{1z}$ &  0.50 & $0.4^{+0.07}_{-0.07}$ & $0.5^{+0.05}_{-0.06}$ \\
 $\chi_{2z}$ &  0.0 & $0.02^{+0.56}_{-0.49}$ & $0.03^{+0.59}_{-0.51}$ \\
 $\iota/\mathrm{rad}$ & 1.047  & $1.08^{+0.2}_{-0.23}$ & $0.98^{+0.2}_{-0.2}$ \\
 $d_{L}/\mathrm{Mpc}$ & 700.0 & $792.04^{+262.38}_{-222.3}$ & $798.97^{+198.04}_{-180.23}$ \\
 $\phi_{\text{ref}}/\mathrm{rad}$ & 0.80 & $3.57^{+1.98}_{-2.1}$ & $3.05^{+2.92}_{-2.73}$ \\
 $\psi/\mathrm{rad}$ & 2.17 & $2.29^{+0.3}_{-0.28}$ & $2.33^{+0.22}_{-0.23}$ \\
 $\alpha/\mathrm{rad}$   & 3.81 & $3.84^{+0.09}_{-0.09}$ & $3.84^{+0.07}_{-0.07}$ \\
 $\delta/\mathrm{rad}$  & 0.63 & $0.6^{+0.09}_{-0.11}$ & $0.59^{+0.06}_{-0.09}$ \\
 $\rho^{\mathrm{H1}}_{\mathrm{mf}}$  & 8.42 & $8.05^{+0.08}_{-0.15}$ & $8.26^{+0.07}_{-0.14}$ \\
 $\rho^{\mathrm{L1}}_{\mathrm{mf}}$  & 9.98 & $9.54^{+0.09}_{-0.17}$ & $9.79^{+0.08}_{-0.17}$ \\
 $\rho^{\mathrm{V1}}_{\mathrm{mf}}$  & 10.18 & $9.67^{+0.08}_{-0.16}$ & $9.98^{+0.08}_{-0.16}$ \\
 $ \log \mathcal{BF} $  &  & $91.26 \pm 0.20 $ & $97.53 \pm  0.21$ \\
    \end{tabular}
    \end{ruledtabular}
 \caption{
 Injected and median values of the posterior distributions for the synthetic NR injection, 
 corresponding to the NR simulation {\tt SXS:BBH:q15Sur002} from the SXS Collaboration, recovered  
 with \texttt{IMRPhenomXHM} and \texttt{SEOBNRv5HM}. The binary parameters correspond to the total mass $M$, 
 chirp mass $\mathcal{M}$, individual masses $m_{1,2}$, inverse mass ratio $1/q$, effective spin parameter  $\chi_{\text{eff}}$, 
 individual spin components $\chi_{1z,2z}$, inclination angle $\iota$, 
 luminosity distance $d_{L}$, coalescence phase $\phi_{\text{ref}}$,  polarization angle $\psi$, 
 right ascension $\alpha$, declination $\delta$, matched-filtered SNR for LIGO-Hanford/Livingston and Virgo detectors 
 $\rho^{\mathrm{H1},\mathrm{L1},\mathrm{V1}}_{\mathrm{mf}}$ and signal-versus-noise log Bayes factor $ \log \mathcal{BF} $.}
 \label{tab:injection_settings}
\end{table}

We begin by examining the parameter recovery on a synthetic signal injected in a network of three detectors, at the locations of LIGO Hanford, 
LIGO Livingston and Virgo, with a zero-noise configuration, to decouple the impact of the model's accuracy from any particular noise realization. 
We inject the NR waveform \texttt{SXS:BBH:2464} from the \texttt{SXS} Collaboration with intrinsic parameters 
$1/q=m_2/m_1=0.067$, $\chi_1=0.5$ and $\chi_2=0$, choosing a detector-frame total mass of $162M_{\odot}$, 
inclination $\iota=\pi/3$ in order to emphasize the higher harmonics of the signal, and a luminosity distance of $700~\text{Mpc}$ to give a network SNR of $\sim16.6$. 
These and the selected injected values for the phase and the sky-location parameters are listed in the left column of Table \ref{tab:injection_settings}. 

We employ the \texttt{Bilby} parameter-estimation code \cite{Ashton:2018jfp}, with version $2.0.0$ and the nested sampler \texttt{dynesty} \cite{Speagle:2019ivv} using the \texttt{acceptance-walk} method, which is well-suited for executing on a multicore machine, in particular, we run on 1 node of 64 CPUs. For the sampler settings for the recovery, 
we employ a number of accepted jumps during each MCMC chain $\mathrm{naccept}=20$ and a total number of live points $\mathrm{nlive}=1000$. 
We employ the sky parameterization option H1L1, which enables us to sample the sky position in azimuth and zenith, converted in post-processing to right ascension and declination, 
since this typically improves the convergence of the sampler, and we enable distance marginalization, to further improve convergence. We leave the rest of the sampler parameters with their default values. 

The prior distributions are uniform for most of the parameters, except for the individual dimensionless spin values, which follow a distribution implied by the 
isotropic spin prior commonly employed in GW parameter estimation. 
Though a non-uniform prior could shift the posterior from the true values for moderate SNR in a zero-noise setup, 
we decide to employ this spin prior as it is commonly employed in actual analyses
~\cite{Callister:2021gxf,LIGOScientific:2018mvr,LIGOScientific:2020ibl,LIGOScientific:2021usb,LIGOScientific:2021djp}.

\begin{figure*}[htpb!]
\includegraphics[width=0.95\columnwidth]{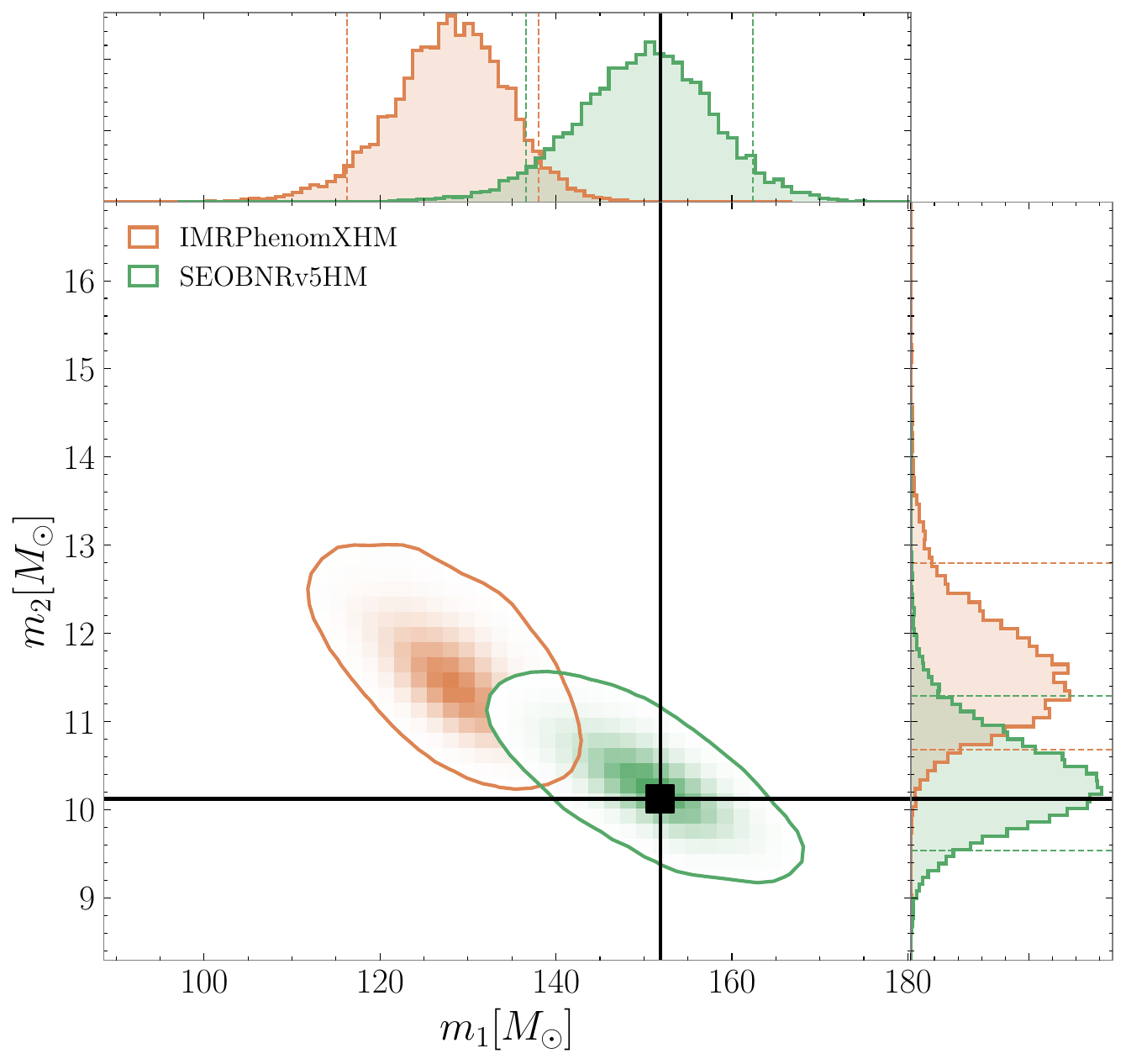}
\includegraphics[width=0.95\columnwidth]{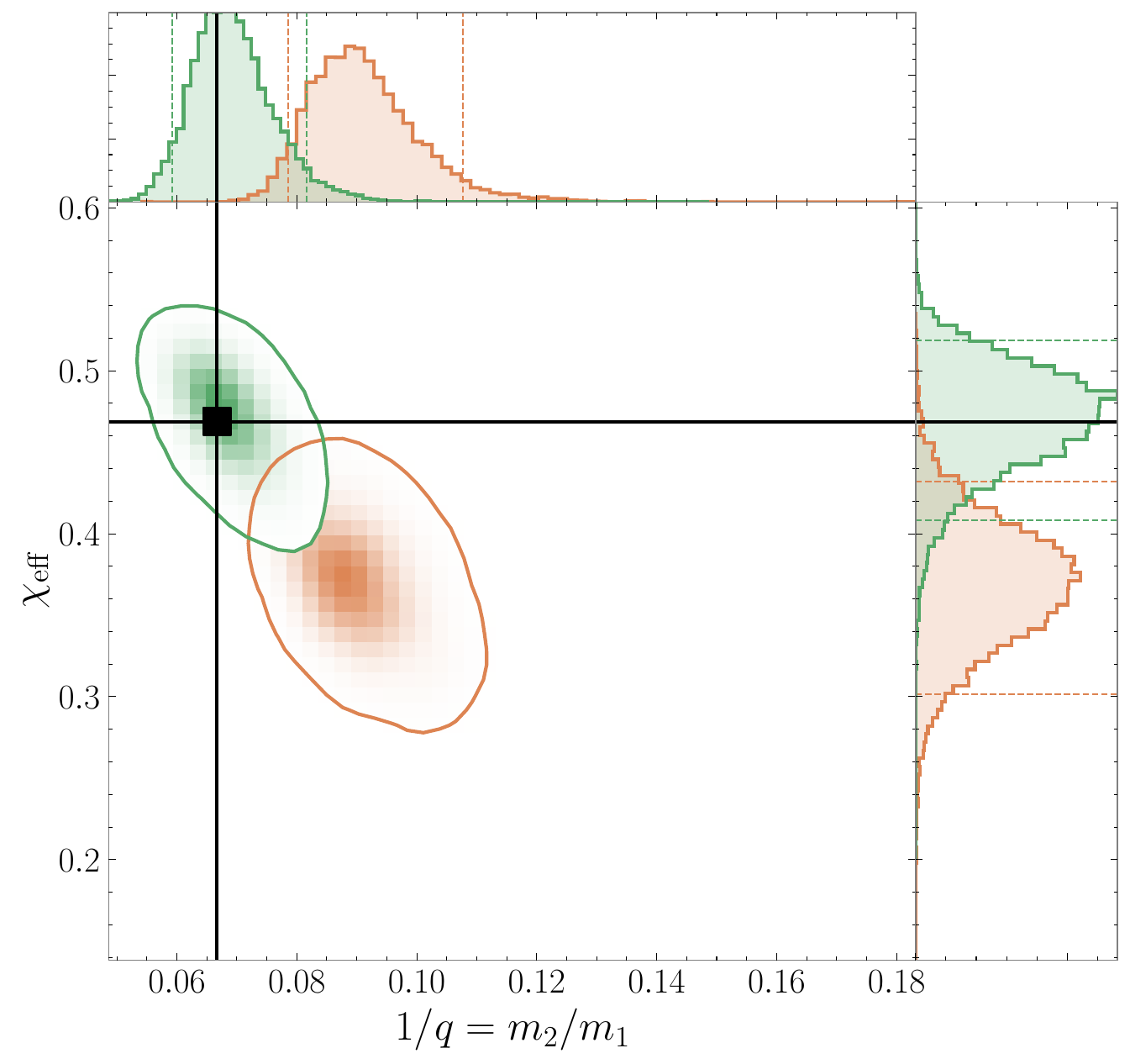}\\
\includegraphics[width=0.95\columnwidth]{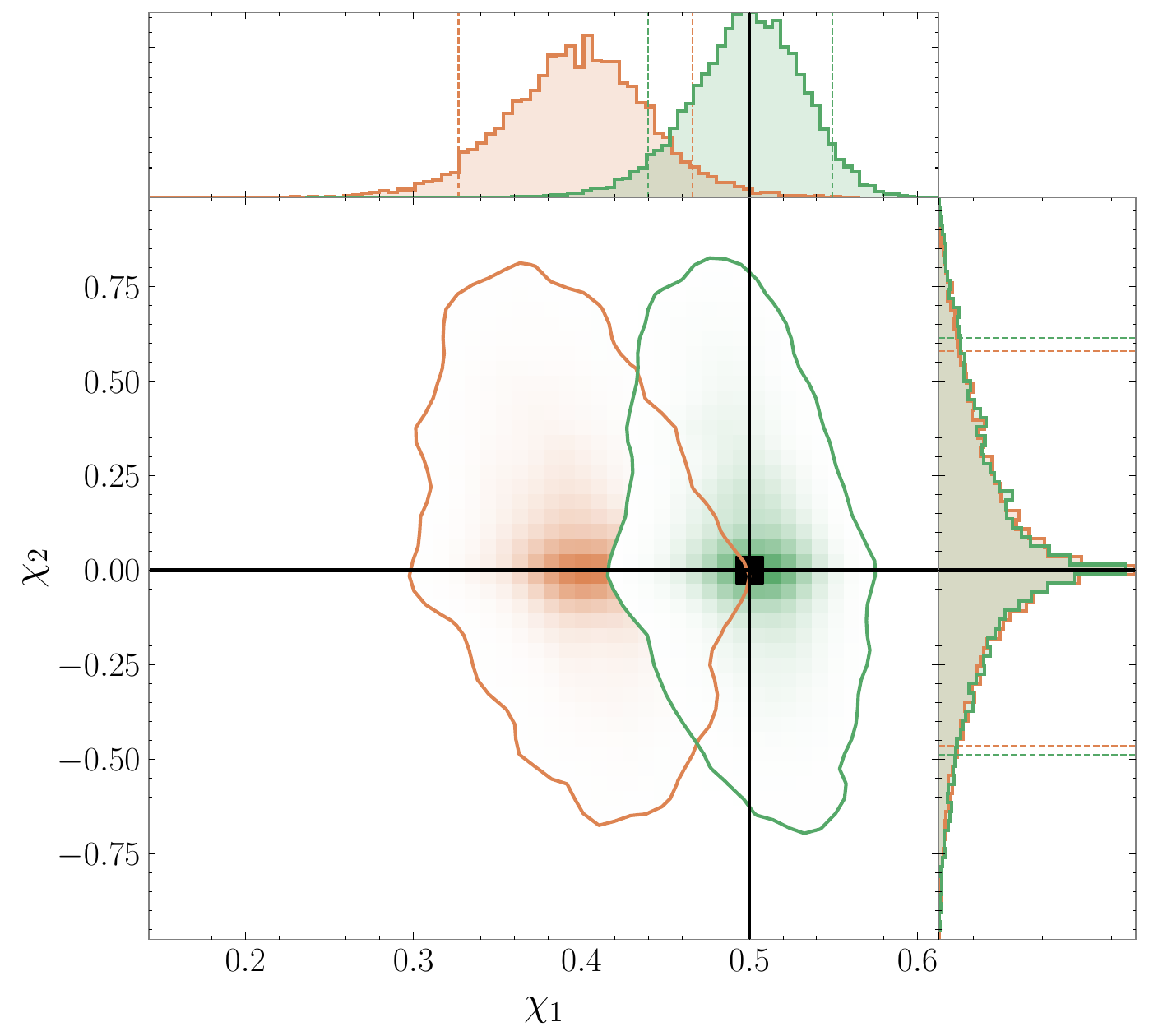}
\includegraphics[width=0.95\columnwidth]{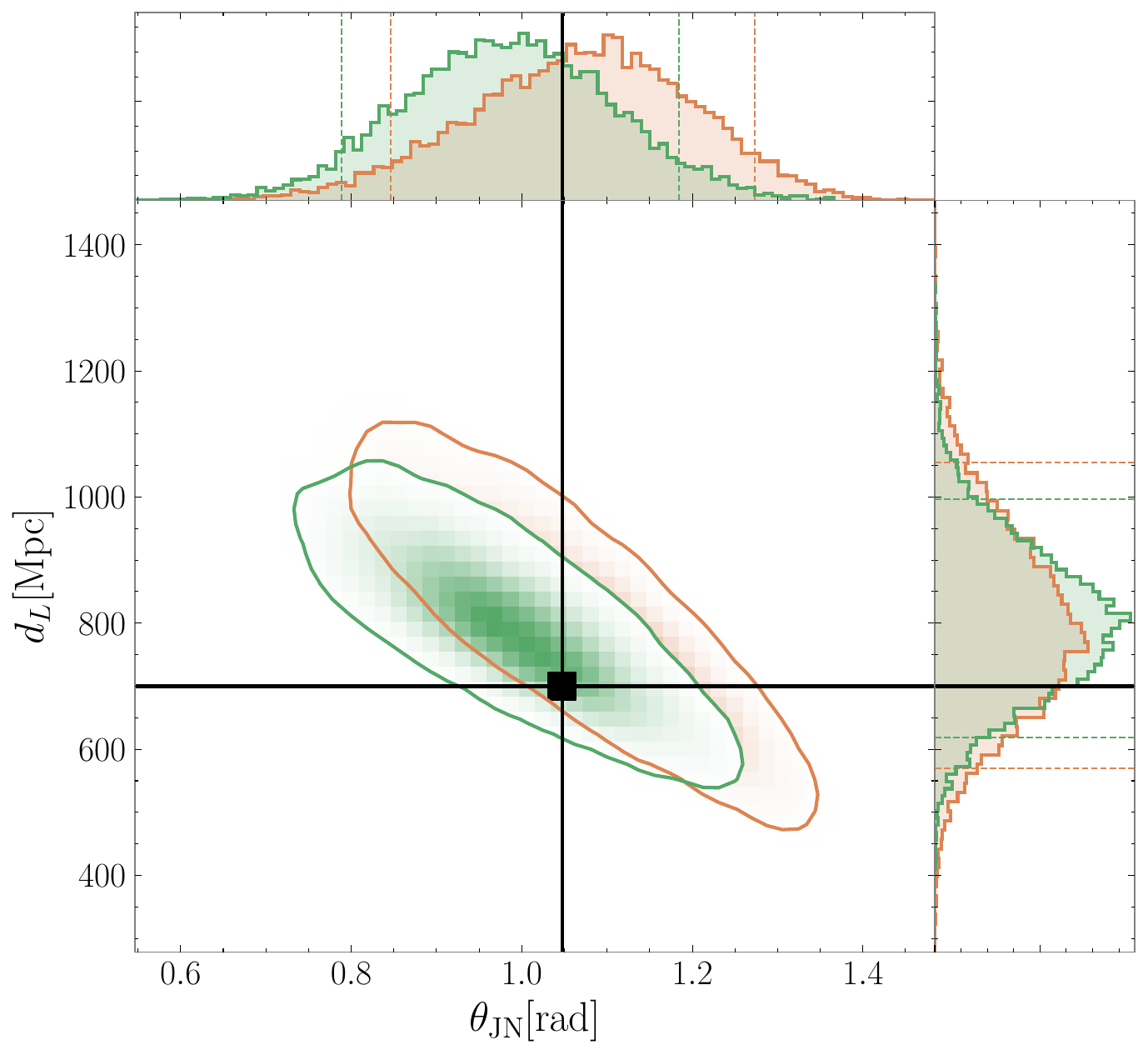}\\
 \caption{2D and 1D posterior distributions for some relevant parameters measured from the synthetic BBH signal with mass ratio  
 $q=15$, total source-frame mass of $162 M_\odot$, dimensionless spins $\chi_{1z} = 0.5$ and 
 $\chi_{2z} = 0.0$. The inclination with respect to the line of sight of the binary is $\iota=\pi/3$ rad. 
 The other parameters are specified in the text and in Table \ref{tab:injection_settings}. 
 The injected signal is the \texttt{SXS} NR waveform {\tt SXS:BBH:2464}. 
 In the 2D posteriors the solid contours represent the $90\%$ credible intervals and black dots show the values of the parameters of the injected signal. 
 In the 1D posteriors they are represented by dashed and solid vertical lines, respectively. 
 The parameter estimation is performed with the \texttt{SEOBNRv5HM} model (green) and the \texttt{IMRPhenomXHM} model (orange). }
\label{fig:sxsq15corner}
\end{figure*}

We perform two parameter estimation runs on this injected signal, one with the \texttt{SEOBNRv5HM} model presented in this paper, 
and a run with the state-of-the-art waveform model \texttt{IMRPhenomXHM} from the 4th generation of Fourier-domain phenomenological waveform models, to crosscheck the results. 
For \texttt{SEOBNRv5HM}, we employ the conditioning routine implemented in \pySEOBNR, which closely mimics the procedure of \texttt{LALSimulation} \cite{lalsuite}. 

The median recovered values for both models, and the $90\%$ confidence intervals, are listed in Table \ref{tab:injection_settings}, 
and some relevant 2D contours are highlighted in Fig.~\ref{fig:sxsq15corner}. The results show that the {\tt SEOBNRv5HM} model is able to 
recover better the synthetic signal, especially for the intrinsic parameters, with the injected value of all the parameters inside the $90\%$ 
confidence intervals and very small deviations between the median values of the posterior distributions and the actual injected values 
(the main deviation is in the reference phase parameter, whose recovered distribution is prior-dominated). 
On the contrary, the results inferred by the {\tt IMRPhenomXHM} model contain important biases in most of the intrinsic parameters, 
with the injected values outside the $90\%$ confidence intervals for the component masses, the total mass, the mass-ratio, 
and the effective-spin parameter $\chi_{\mathrm{eff}}$. For the extrinsic parameters, both models recover the injected 
values within the $90\%$ confidence intervals, with small but similar deviations in the median values for the distance and the inclination. 
The improved accuracy of {\tt SEOBNRv5HM} in this challenging region of parameter space (high asymmetric masses and spinning primary black hole) 
is also reflected in the recovered matched-filter SNR in the three detectors and the Bayes factor of the inference run, 
which are consistently higher than the corresponding values for {\tt IMRPhenomXHM}.
These results are consistent with the fact that {\tt SEOBNRv5HM} has lower unfaithfulness than {\tt IMRPhenomXHM} against this NR simulation, 
$0.5\%$ and $6.7 \%$ respectively, for the injected value of the total mass. 

\subsection{Inference of real gravitational-wave events}

We then perform parameter estimation on three real GW events: GW150914~\cite{LIGOScientific:2016aoc}, the first detection which has become a benchmark for testing new waveform models, 
GW170729~\cite{Chatziioannou:2019dsz}, an interesting event from O2, which has been analysed with multimode waveform models, and GW190412~\cite{LIGOScientific:2020zkf}, the first confident mass-asymmetric binary reported during O2. 
For each event, we employ the strain data, detector calibration uncertainties and PSD provided by the Gravitational Wave Open Science Center (GWOSC) \cite{LIGOScientific:2019lzm}.
We perform the runs using \texttt{Bilby} \cite{Ashton:2018jfp} version $2.0.0$ with the nested sampler \texttt{dynesty} \cite{Speagle:2019ivv}, and we employ the same settings as in the previous section, except for the number of accepted jumps during each MCMC chain that we set to $\mathrm{naccept}=60$. 
For each GW signal, we perform a run with \texttt{SEOBNRv5HM}, employing the PA approximation, and a crosschecking run with the \texttt{IMRPhenomXHM} waveform model.

In Fig.~\ref{fig:realevents} we show some relevant 2D posterior distribution for the parameters, and observe good agreement between waveform models.  
These results are also consistent with the published results for the events, taking into account that LVK catalog results employ precessing-spin waveform models and therefore minor differences are expected. 
The good agreement between the \texttt{SEOBNRv5HM} and \texttt{IMRPhenomXHM} posteriors is consistent with the fact that, 
for the events considered here, the recovered parameters are within the NR calibration region of both models. 
As in the case of the NR-injected signal, we observe a slight improvement in matched-filter SNR and Bayes factor for \texttt{SEOBNRv5HM} with respect to \texttt{IMRPhenomXHM}, more pronounced for the two more massive events, as seen in Fig~\ref{fig:snrevents}. 
Although the improvement is not drastic, these results suggest that \texttt{SEOBNRv5HM} describes the data more accurately, which is consistent with the unfaithfulness results discussed in Sec. \ref{subsec:faithfulness}.

\begin{table}[t!]
\centering
\begin{ruledtabular}
	\begin{tabular}{l c c c}
 Run/event & GW150914 & GW170729 & GW190412  \\
 \hline
 \smallskip
 \makecell[cl]{SEOBNRv5HM \\ Bilby (64 cores)} & 23h & 20h & 1d 18h  \\
\end{tabular}
\end{ruledtabular}
\caption{Evaluation time for the different parameter estimation runs on real GW events with the \texttt{SEOBNRv5HM} model. The time reported is actual real-time, while the total computational cost in CPU hours can be obtained by multiplying this time by the reported number of CPU cores employed.}
\label{tab:peruntime}
\end{table}

In Table \ref{tab:peruntime} we report the real-time spent on the inference for the parameters of these events for the 
waveform model \texttt{SEOBNRv5HM}. Employing \texttt{Bilby} on a single computing node (of 64 cores) requires 
less than a day for GW150914 and GW170729, and less than two days for GW190412, with a moderately low chirp mass. 
Therefore, the model is sufficiently efficient to be employed with the preferred parameter estimation pipeline by the LVK Collaboration. 


\begin{figure*}[htpb!]
     \centering
     \subfloat[GW150914.\label{fig:GW150914corner}]{
     \includegraphics[width=0.65\columnwidth]{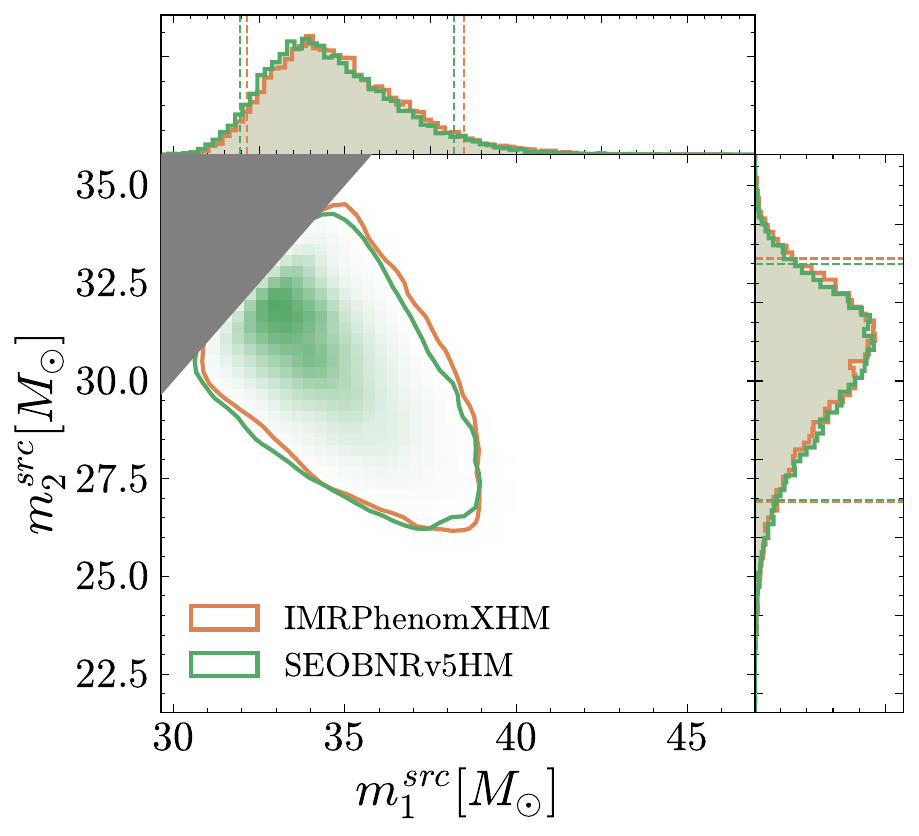}
     \includegraphics[width=0.65\columnwidth]{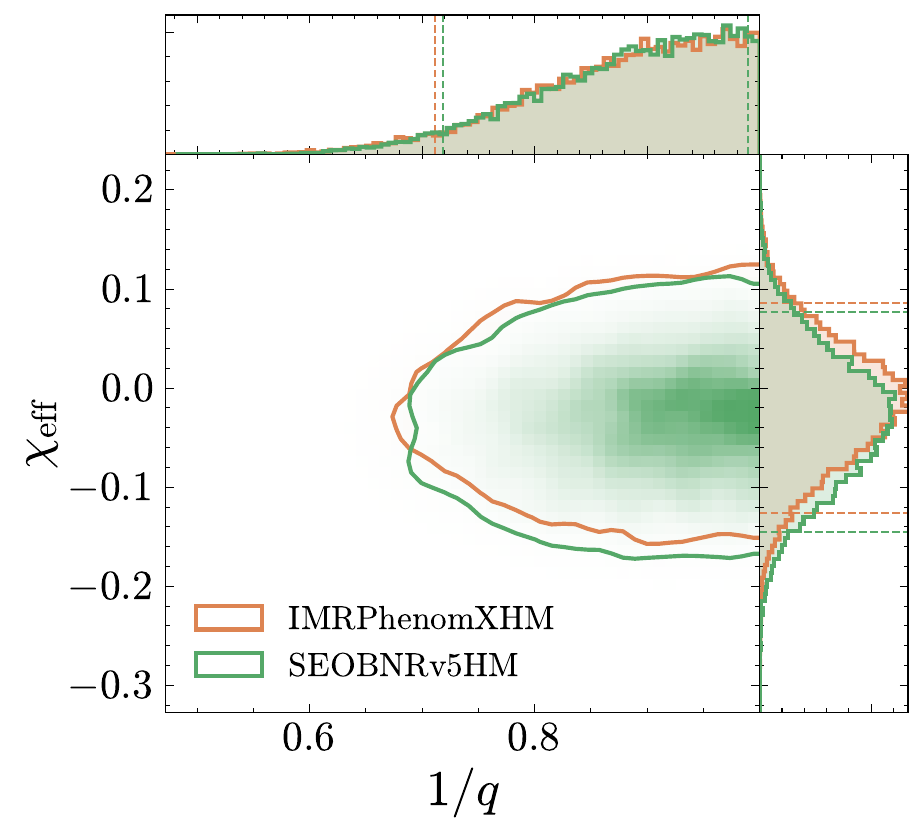}			         
	 \includegraphics[width=0.65\columnwidth]{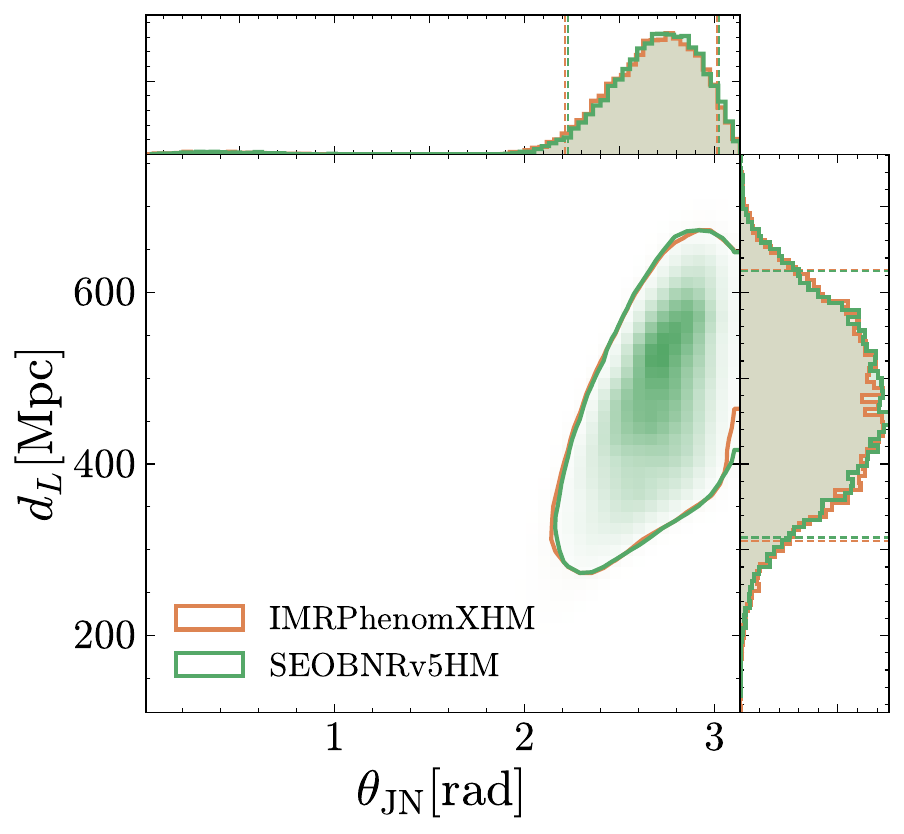}}
     \hfill
     \subfloat[GW170729.\label{fig:GW170729corner}]{
		\includegraphics[width=0.65\columnwidth]{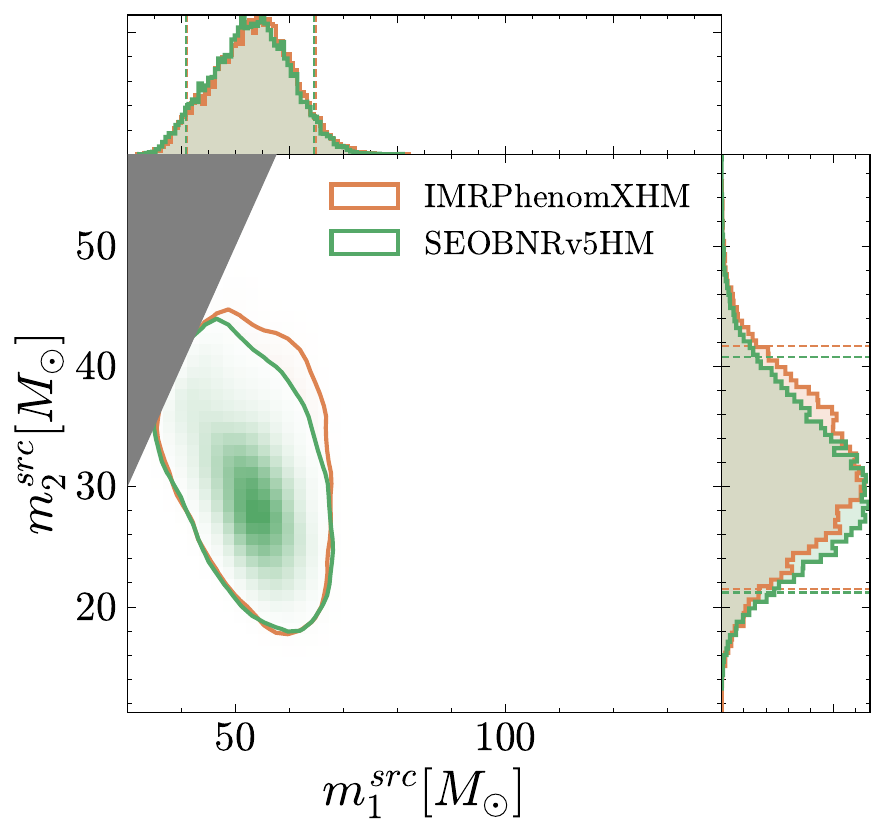}
		\includegraphics[width=0.65\columnwidth]{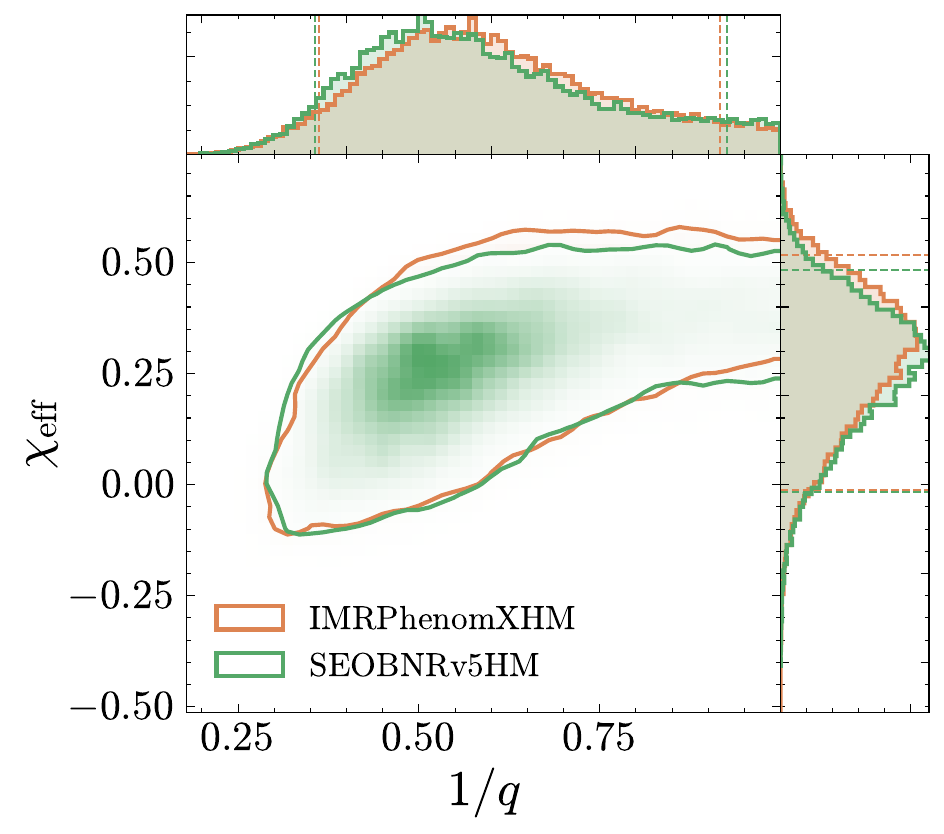}			         
		\includegraphics[width=0.65\columnwidth]{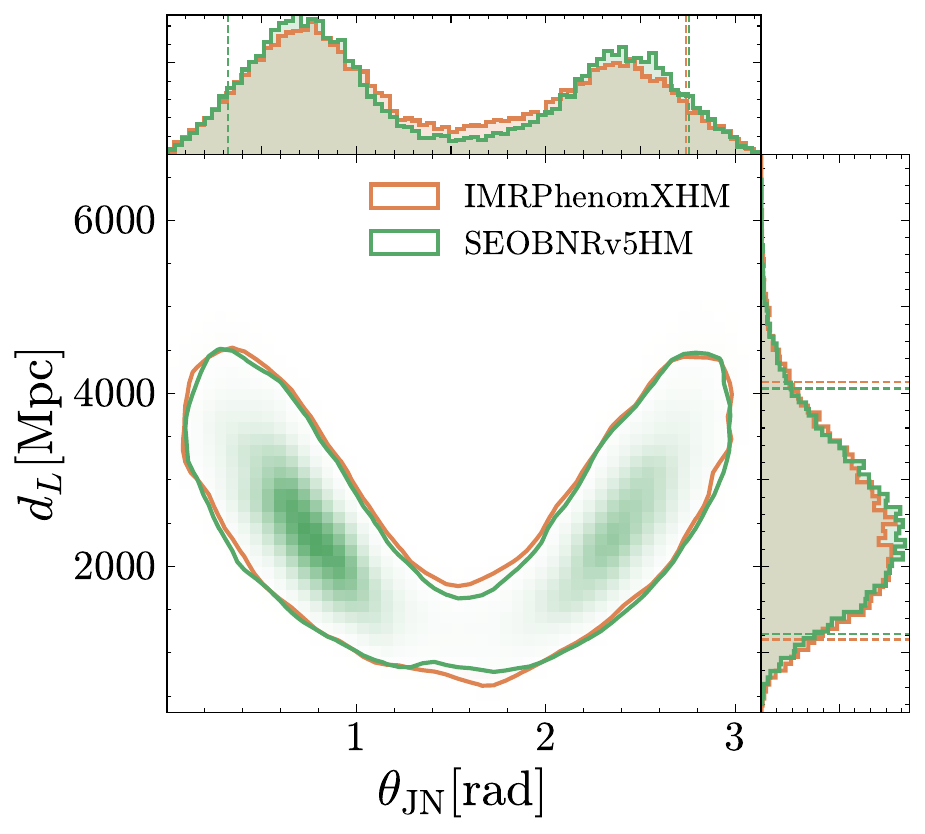}}
     \hfill
     \subfloat[GW190412.\label{fig:GW190412corner}]{
		\includegraphics[width=0.65\columnwidth]{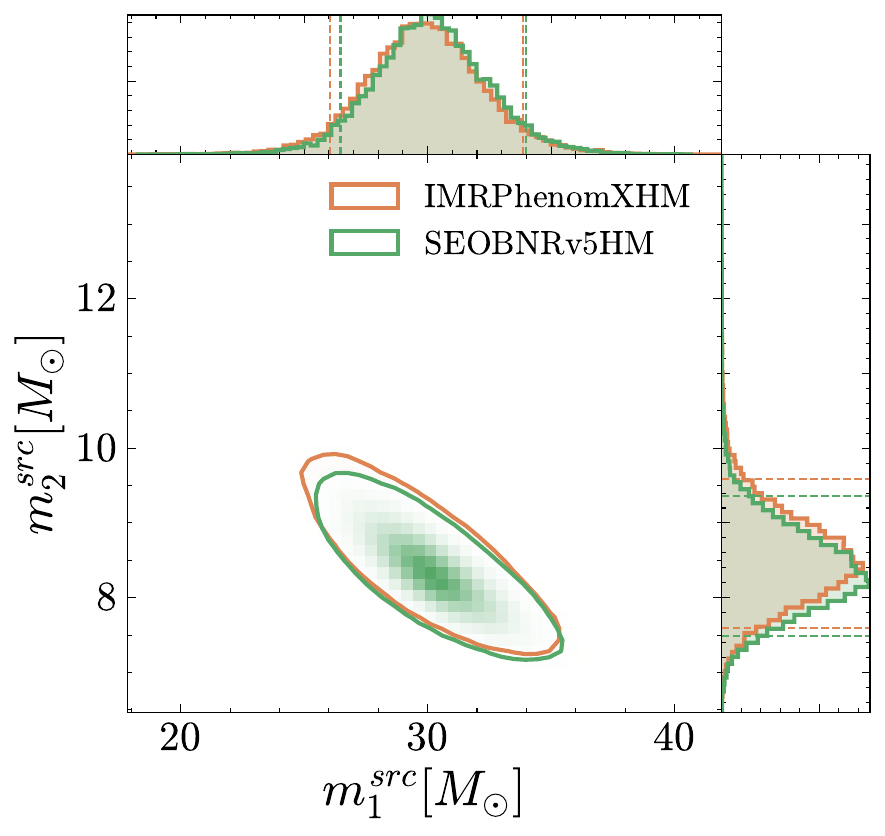}
		\includegraphics[width=0.65\columnwidth]{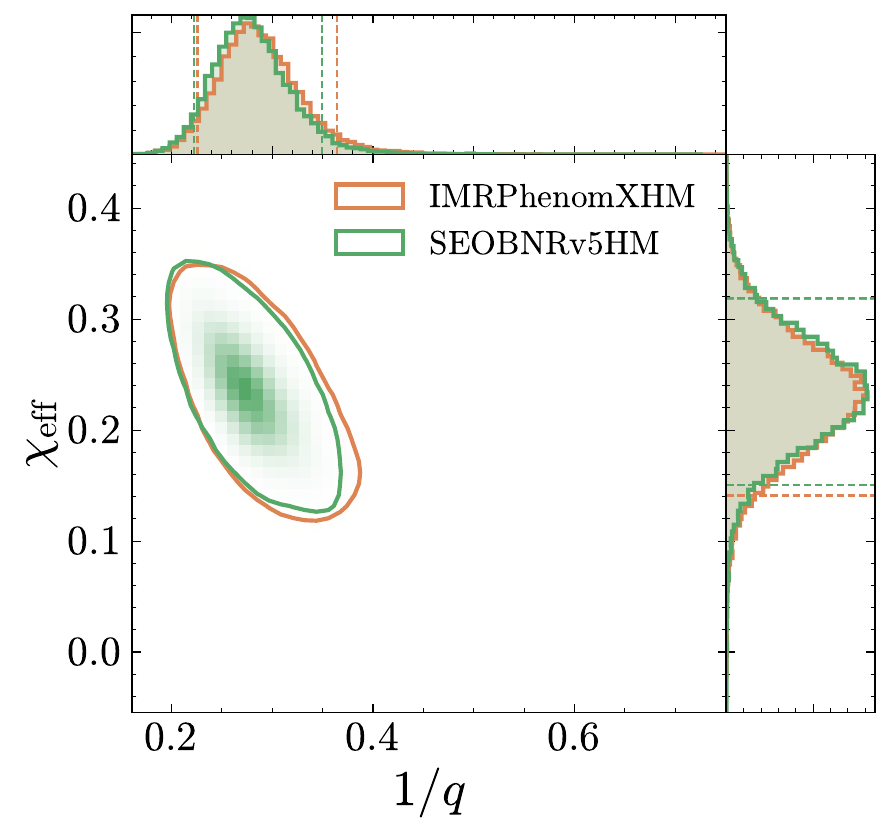}			         
		\includegraphics[width=0.65\columnwidth]{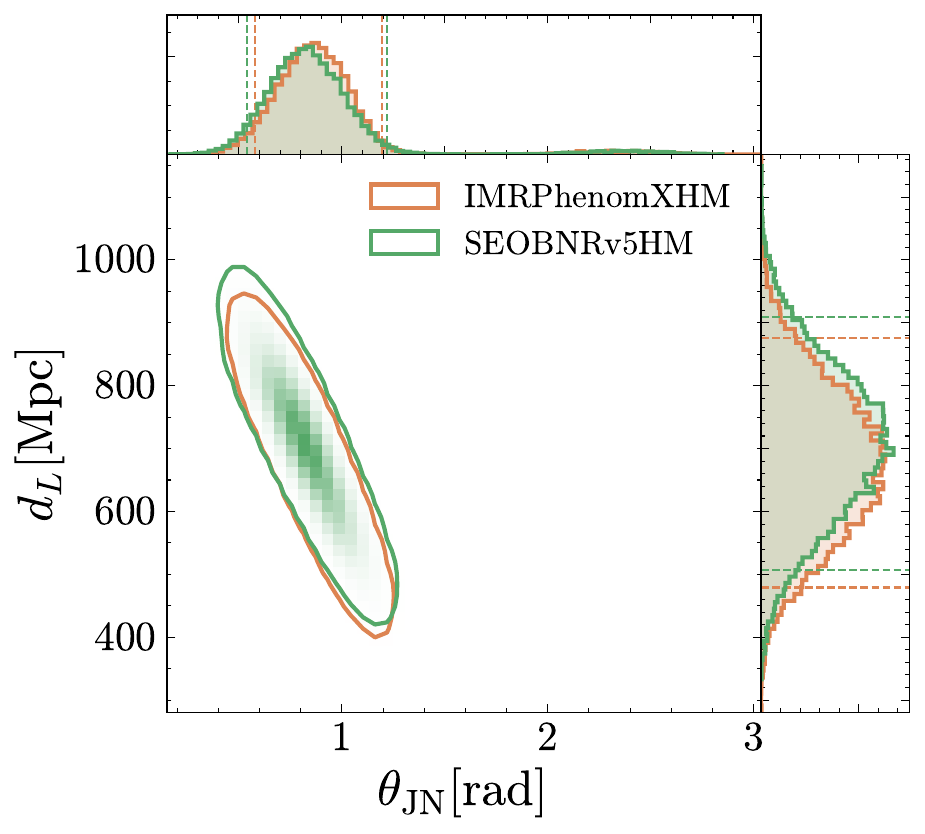}}
        \caption{1D and 2D posterior distributions for several parameters for the GW events GW150914, GW170729 and GW190412.}
        \label{fig:realevents}
\end{figure*}

\begin{figure*}[htpb!]
\includegraphics[width=0.65\columnwidth]{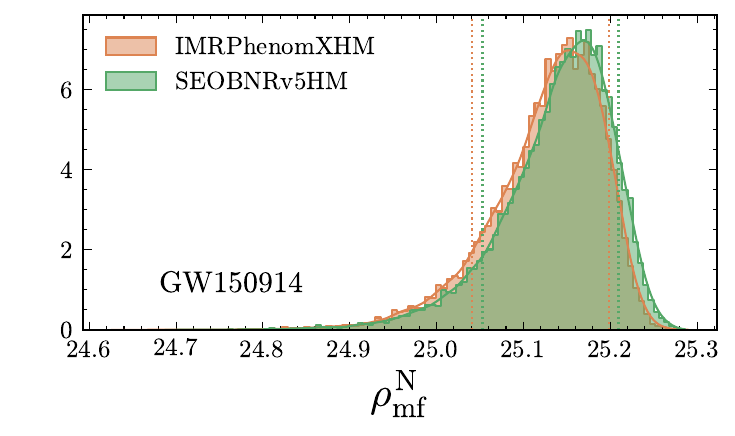}
\includegraphics[width=0.65\columnwidth]{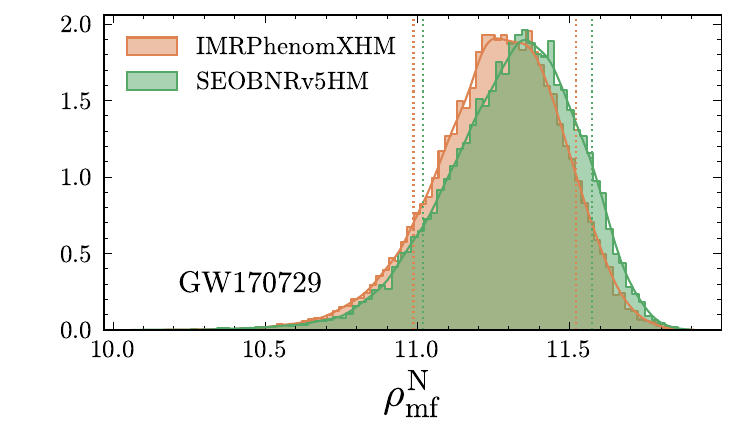}
\includegraphics[width=0.65\columnwidth]{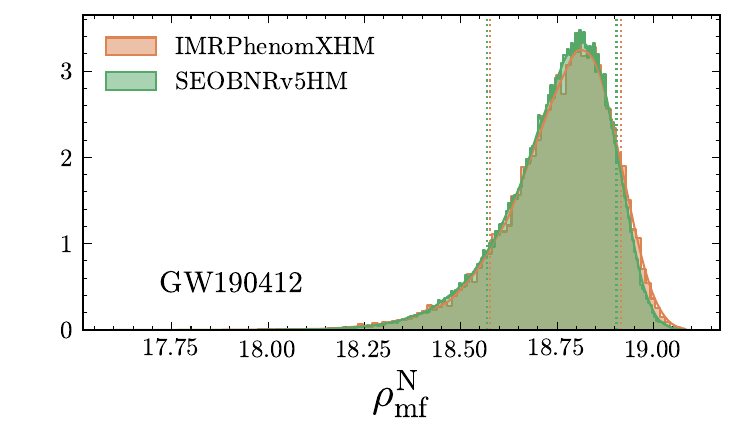}\\
 \caption{Network matched-filter SNR recovered by \texttt{IMRPhenomXHM} and \texttt{SEOBNRv5HM} for the three analyzed GW events.}
\label{fig:snrevents}
\end{figure*}

\section{Frequency domain reduced order model}
\label{sec:ROM}


The requirement of integrating a system of ODEs to solve for the dynamics of the binary in EOB models increases the time needed for generating a waveform.
Surrogate or reduced order modeling (ROM) techniques
~\cite{Field:2013cfa,Purrer:2014fza,Purrer:2015tud,Blackman:2015pia,Blackman:2017pcm,Blackman:2017dfb,Lackey:2018zvw,Doctor:2017csx,Setyawati:2019xzw,Varma:2018mmi,Varma:2019csw,Cotesta:2020qhw,Gadre:2022sed,Thomas:2022rmc}
have been applied in several contexts to accelerate slow waveform computation, in both EOB and NR models. 
These techniques involve decomposing the waveforms from a training set in orthonormal bases on sparse grids in time or frequency and then 
interpolating or fitting the resulting waveform data pieces over the binary parameter space. The result is a highly accurate, yet fast, method 
for generating waveforms for data analysis applications, which can reduce computational time by orders of magnitude compared to ODE-based waveform models.

A frequency domain (FD) ROM of \texttt{SEOBNRv4HM} was built in Ref.~\cite{Cotesta:2020qhw}, with modeling error introduced in building the
ROM below the unfaithfulness of \texttt{SEOBNRv4HM} against NR simulations used to calibrate the model, and waveform evaluation times reduced by two orders of magnitude. 
In this section we show the performance of \texttt{SEOBNRv5\_ROM}, a FD ROM of \texttt{SEOBNRv5}, 
built following the same techniques of \texttt{SEOBNRv4HM\_ROM} \cite{Cotesta:2020qhw,Purrer:2015tud,Purrer:2014fza}.
These mostly involve modeling in FD the phase of a carrier signal, based on
the time-domain orbital phase, and the “coorbital modes” obtained after extracting the carrier phasing from each FD mode.
The coorbital modes have an almost constant phase in the inspiral, and allow us to avoid zero-crossings in the subdominant harmonics which would 
complicate the interpolation of the training data.
As for \texttt{SEOBNRv4HM\_ROM}, the \texttt{SEOBNRv5\_ROM} model combines a higher resolution high-frequency ROM, starting from 
$20~\text{Hz}$ for binaries with total mass of $50 M_{\odot}$, and a lower resolution low-frequency ROM, starting from 
$20~\text{Hz}$ for binaries with total mass of $5 M_{\odot}$, 
and can be extended to arbitrarily low frequencies by hybridizing it with multipolar PN waveforms.
\texttt{SEOBNRv5\_ROM} can be generated for mass-ratios between $1$ and $100$, dimensionless spins between $[-0.998, 0.998]$, and includes only the dominant $(\ell, |m|)=(2,2)$ mode.
A multipolar reduced order model of \texttt{SEOBNRv5HM} (\texttt{SEOBNRv5HM\_ROM}), including the $(\ell, |m|)=(2,2),(3,3),(2,1),(4,4),(5,5),(3,2),(4,3)$ modes, 
is also under development, and will be presented in near future work. 
Despite the speed of \texttt{SEOBNRv5HM} being sufficient for many GW data analysis applications, 
using a ROM can still lead to a significant increase in efficiency. 
Additionally, there are several applications for which it is desirable to be able to generate clean FD waveforms of any length.

In Fig.~\ref{fig:mismatch_ROM} we show a histogram of the unfaithfulness between \texttt{SEOBNRv5\_ROM} and  \texttt{SEOBNRv5}, 
for different values of the total mass, for $10^5$ configurations with mass-ratios between $1$ and $100$ and 
dimensionless spins between $[-0.998, 0.998]$. We observe an excellent agreement, with median values $\lesssim 10^{-5}$. 
The unfaithfulness increases with the total mass of the system, as in previous ROM models \cite{Cotesta:2020qhw,Bohe:2016gbl}, 
because the ROM modes are generated up to a maximum frequency that scales with the inverse of the total mass.
In particular, the mismatch is larger for cases with high mass ratio and negative spins, as 
the maximum frequency of each mode is proportional to its least damped QNM frequency \cite{Cotesta:2020qhw}, 
which decreases in this region of the parameter space. Nonetheless, the modeling error introduced 
in the construction of the ROM is negligible compared to the inaccuracy of the \texttt{SEOBNRv5} 
waveforms with respect to the NR simulations.

Figure~\ref{fig:benchamrks_ROM} highlights the speedup of the ROM with respect to \texttt{SEOBNRv5}, 
by comparing walltimes of the two models for generating a FD waveform with the same parameters ($q=1,~\chi_1=0.8,~\chi_2=0.3$), 
as a function of the total mass $M$. For \texttt{SEOBNRv5} we also employ the PA approximation. 
As for Fig.~\ref{fig:benchamrks}, we use a starting frequency $f_{\rm{start}}=10~\rm{Hz}$ 
and we choose the sampling rate for the time domain model such that the Nyquist criterion is satisfied for the $\ell = 2$ multipoles. 
The FD \texttt{SEOBNRv5\_ROM} model is instead generated up to a maximum frequency equal to the corresponding Nyquist frequency.
Notably, we obtain an improvement from a factor $\sim 7$ for low total-mass binaries, to more than $\sim 20$ for $M \sim 100 M_{\odot}$. 
Overall, we can appreciate that \texttt{SEOBNRv5\_ROM} can be generated in less than $10~\text{ms}$ for $M \gtrsim 20 M_{\odot}$.

\begin{figure}
	\includegraphics[width=\linewidth]{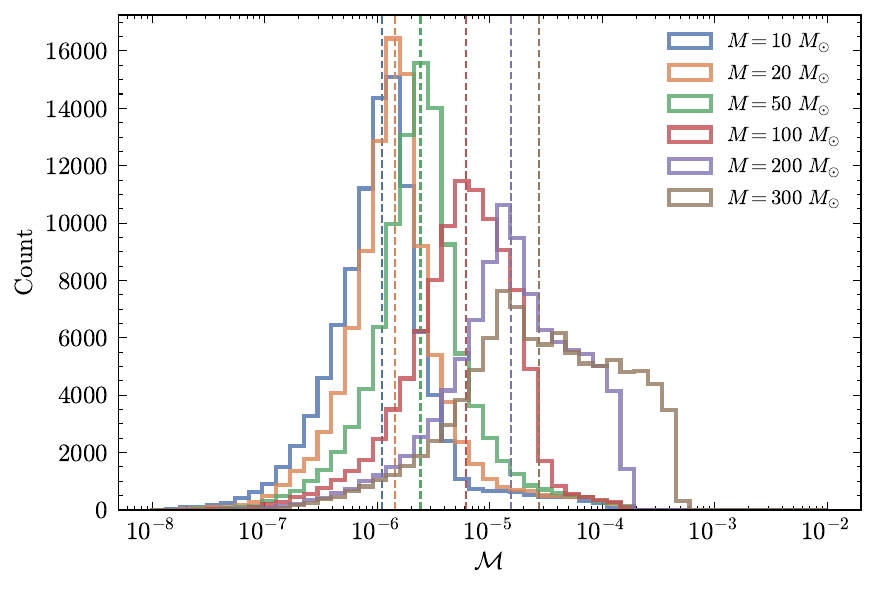}
	\caption{Mismatch of \texttt{SEOBNRv5\_ROM} against \texttt{SEOBNRv5} for different values of the total mass, for $10^{5}$ random configurations. 
	The dashed vertical lines show the medians.}
	\label{fig:mismatch_ROM}
\end{figure}

\begin{figure}
	\includegraphics[width=\linewidth]{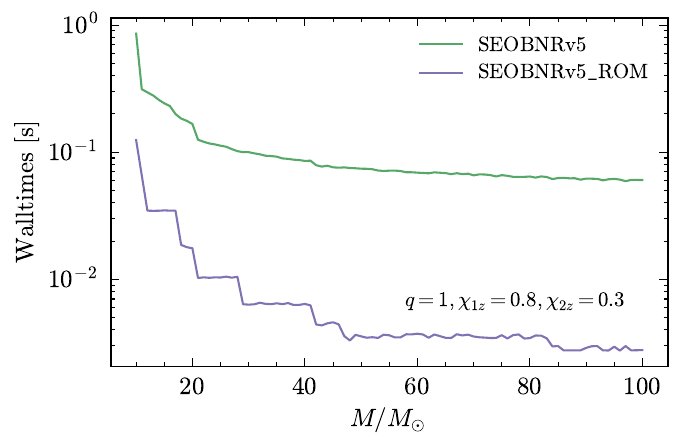}
	\caption{Walltime comparison between \texttt{SEOBNRv5} and \texttt{SEOBNRv5\_ROM}, using a strarting frequency $f_{\rm{start}}=10~\rm{Hz}$, 
	as a function of the total mass $M$. For the time domain \texttt{SEOBNRv5} model this also includes the conversion in Fourier domain.
	}
	\label{fig:benchamrks_ROM}
\end{figure}

\section{Conclusions}
\label{sec:conclusions}


In this paper we have presented \texttt{SEOBNRv5HM}, a new EOBNR waveform model for quasi-circular, spinning, nonprecessing BBHs, which improves the previous generation, 
\texttt{SEOBNRv4HM}~\cite{Cotesta:2018fcv}, on both speed and accuracy against NR simulations. 
The waveform model includes the modes $(\ell, |m|)=(2,2),(3,3),(2,1),(4,4),(5,5),(3,2),(4,3)$, and models the 
mode-mixing in the merger-ringdown for the modes $(3,2),(4,3)$, which were not included in \texttt{SEOBNRv4HM}.

Sections~\ref{sec:ham_eom} and~\ref{sec:waveforms} have outlined the building blocks of the waveform model. The aligned-spin \texttt{SEOBNRv5} Hamiltonian 
is based on a deformation of the equatorial Kerr Hamiltonian, and includes most of the known 5PN nonspinning and full 4PN information for spinning binaries, with improved resummations~\cite{Khalilv5}.
The factorized waveform modes and RR force of \texttt{SEOBNRv4HM} have been enhanced by additional PN information (as well as corrections to some of the terms) from Ref.~\cite{Henry:2022dzx}, 
and have been calibrated to 2GSF fluxes in the nonspinning limit in Ref.~\cite{VandeMeentv5}.
To improve the accuracy of the model in the inspiral, we have refined the calibration pipeline employed by Ref.~\cite{Bohe:2016gbl}, 
and largely upgraded its efficiency, in order to be able to tune the model to a large catalog of 442 NR simulations as shown in Sec.~\ref{sec:calibration}. 
We have also improved the modeling of the merger and ringdown using the full NR dataset at our disposal, as well as 13 waveforms from BH perturbation theory. 

In Sec.~\ref{sec:performance} we have compared \texttt{SEOBNRv5HM} and other state-of-the-art waveform approximants to NR simulations, using mismatch calculations. 
Results showed that the dominant (2,2) mode of \texttt{SEOBNRv5} is, on average, more accurate than \texttt{SEOBNRv4}~\cite{Bohe:2016gbl} by an order of magnitude, it is 
more accurate than the other state-of-the-art EOB model, \texttt{TEOBResumS-GIOTTO}~\cite{Riemenschneider:2021ppj,Nagar:2020pcj,Nagar:2019wds,Nagar:2018zoe}, by more than a factor 2, 
and is overall comparable to the Fourier-domain phenomenological model \texttt{IMRPhenomXAS}~\cite{Pratten:2020fqn}. 
By computing mismatches of the full polarizations at inclination $\iota = \pi/3$, we noted that all models become less accurate, 
nonetheless \texttt{SEOBNRv5HM} outperforms both \texttt{SEOBNRv4HM}~\cite{Cotesta:2018fcv} and \texttt{TEOBResumS-GIOTTO}, as well as the phenomenological model \texttt{IMRPhenomXHM}~\cite{Garcia-Quiros:2020qpx}, 
both considering average values and number of cases above $0.01$.
We have validated the model against the NR surrogate models \texttt{NRHybSur3dq8}~\cite{Varma:2018mmi} and \texttt{NRHybSur2dq15}~\cite{Yoo:2022erv} and found results 
consistent with the NR comparison, demonstrating the robustness of our calibration procedure. 
Further tests of the calibration pipeline are described in Appendix~\ref{appx:robustness}. 
In particular, we show that the accuracy of the model against the entire NR dataset of 442 waveforms does 
not change appreciably when using only 137 to calibrate the model. 
This confirms that the significant improvement over \texttt{SEOBNRv4} is largely due to the improved analytical prescriptions, particularly for the EOB Hamiltonian, 
but even if not all NR data are used directly, they remain extremely valuable for a detailed assessment of the effectiveness of different analytical choices. 
Direct use of the new NR data is particularly useful when considering high mass ratio, high spin configurations in the region of parameter space previously uncovered by simulations, 
and in modeling higher modes in the merger-ringdown phase, which benefits greatly from simulations with higher resultion.
We also show that our calibration pipeline is robust with respect to changes in the shape of the PSD used, 
as the unfaithfulness of the model against NR remains very similar when using a white noise curve, 
the Einsten Telescope \cite{Hild:2010id} and Cosmic Explorer \cite{Evans:2021gyd} PSDs. 
Nonetheless, such a comparison is limited by the length of the available NR simulations, 
which do not cover the entire bandwidth of next-generation GW detectors, 
and more detailed studies will be needed to assess the accuracy \texttt{SEOBNRv5HM} in such a context.
Comparing \texttt{SEOBNRv5HM} and \texttt{IMRPhenomXHM} against each other in a larger parameter space, we have seen 
instead that the mismatches can become very large in the region where both models are not calibrated to NR waveforms, 
in particular for both high mass ratio, say $\geq 5$, and high positive spin, say $\geq 0.8$, configurations. 
Thus, producing new NR simulations for these parameters would be critical to reduce modeling systematics. 
Comparing the angular-momentum flux and binding energy of \texttt{SEOBNRv5} and \texttt{SEOBNRv4} against NR, we 
have highlighted a similar improvement, also thanks to the 2GSF information~\cite{VandeMeentv5}, despite the fact that these quantities do not enter directly the calibration. 
This is a powerful check of the physical robustness of the model, 
and provides confidence in its reliability when extrapolating outside of the NR calibration region.
We have implemented \texttt{SEOBNRv5HM} in a flexible, high-performance, Python package \texttt{pySEOBNR}, 
and we have shown that the model is fast enough for typical GW data-analysis applications: 
it is more than 10 times faster than \texttt{SEOBNRv4HM} without using the PA approximation, up to two times faster than \texttt{SEOBNRv4HM\_PA} when using it, 
and overall close to \texttt{TEOBResumS-GIOTTO}. 

In Sec.~\ref{sec:PE} we have demonstrated that the model can be reliably used for GW parameter estimation, 
by performing a recovery on a NR injection, and by analyzing 3 events observed by LIGO and Virgo, GW150914, GW170729, GW190412.
For the 3 events, we have found consistent results when comparing the parameters recovered by \texttt{SEOBNRv5HM} and by \texttt{IMRPhenomXHM}, 
while still observing a slight improvement in matched-filter SNR and Bayes factor for \texttt{SEOBNRv5HM} with respect to \texttt{IMRPhenomXHM}.
On the other hand, the NR injection in a challenging region of parameter space (high asymmetric masses and spinning primary black hole) 
shows more significant differences. \texttt{SEOBNRv5HM} accurately recovers all the binary parameters, while 
the results inferred by the {\tt IMRPhenomXHM} model contain biases in most of the intrinsic parameters due to larger modeling errors. 
The use of waveform models that include both higher modes and spin-precession is now the standard in GW parameter estimation
~\cite{LIGOScientific:2020ibl,LIGOScientific:2021usb,LIGOScientific:2021djp}.
\texttt{SEOBNRv5HM} would still be useful, for example, to study evidence for spin-precession, by analyzing the data including or not this effect in the model,
and in tests of general relativity (GR), in which the additional computational cost of accounting for beyond-GR parameters often makes it convenient 
to employ a simpler and more efficient aligned-spin model~\cite{LIGOScientific:2020tif,LIGOScientific:2021sio,Ghosh:2021mrv,Mehta:2022pcn,Maggio:2022hre}.

In Sec.~\ref{sec:ROM} we have finally shown the performance of a FD ROM model 
(\texttt{SEOBNRv5\_ROM}) developed following the techniques used in Ref.~\cite{Cotesta:2020qhw},
which allows for a significant speedup in evaluation time, while retaining identical accuracy against NR.
Matched filter GW searches often use FD aligned-spin models such as \texttt{SEOBNRv4\_ROM}~\cite{LIGOScientific:2018mvr,LIGOScientific:2020ibl,LIGOScientific:2021usb,LIGOScientific:2021djp}
and \texttt{SEOBNRv5\_ROM} could be a valuable replacement for such applications.

The \texttt{pySEOBNR} code infrastructure~\cite{Mihaylovv5} is a framework developed with the goal of facilitating 
the development of future \texttt{SEOBNR} waveform models, and upcoming work would naturally revolve around 
adding more physical effects to the \texttt{SEOBNRv5} family, as well as improving its efficiency and accuracy by 
including ever more information from both NR simulations and different analytical frameworks. 
The first extension of \texttt{SEOBNRv5HM}, as far as additional physical effects are concerned, involves modeling
spin precession, and such a model (\texttt{SEOBNRv5PHM}) has been developed in parallel to this work in Ref.~\cite{RamosBuadesv5}.

An upcoming extension would also involve eccentric and hyperbolic orbits (\texttt{SEOBNRv5EHM}), following similar strategies 
adopted in the \texttt{SEOBNRv4EHM} model of Ref.~\cite{Ramos-Buades:2021adz}. 
The more efficient, flexible and parallelized calibration pipeline described in this work would allow having a more accurate eccentric model, with a consistent treatment of eccentric corrections in the waveform modes and RR force, after appropriately recalibrating the quasi-circular limit of the model.
Moreover, the efficiency of \texttt{SEOBNRv5HM} compared to \texttt{SEOBNRv4HM} without PA approximation, which is slow to use
when adding eccentricity, means that one could expect a significant speed-up of 
\texttt{SEOBNRv5EHM} compared to \texttt{SEOBNRv4EHM}. 
Further ongoing developments in the \texttt{SEOBNRv5} family involve the modeling of tidal effects, already incorporated in \texttt{SEOBNRv4} models \cite{Hinderer:2016eia,Steinhoff:2016rfi,Lackey:2018zvw,Matas:2020wab}, 
and the addition of parametrized GR deviations to perform theory agnostic tests of GR \cite{Ghosh:2021mrv,Mehta:2022pcn,Maggio:2022hre}. 

Another direction for improvements revolves around pushing the accuracy of the model against NR even further.
Exploring different ways to incorporate and resum information from the PN, PM and 2GSF approximations, 
while still retaining flexibility in the calibration, would be crucial, 
and an efficient calibration code (\pySEOBNR~\cite{Mihaylovv5}) is essential to understand the impact of different modeling and resummation choices. 
At the same time, more and better NR simulations, especially in currently unexplored regions of the binary parameter space, 
are also critical to reach the accuracy requirements of next-generation detectors \cite{Purrer:2019jcp, Ferguson:2020xnm}.
A limitation of all state-of-the-art approximants is the modeling of the higher modes, and 
a straightforward improvement to be done in future work would be to add all terms through 3.5PN to the waveform modes and RR force from Refs.~\cite{Henry:2022dzx,Henry:2022ccf}.
Further work should also go into improving the modeling of the HMs through merger and ringdown, as well as including additional modes. 
Finally, a calibration pipeline similar to the one developed here could be used to calibrate \texttt{SEOBNRv5PHM} \cite{RamosBuadesv5} to spin-precessing NR simulations.



\section*{Acknowledgments}

It is our pleasure to thank Geraint Pratten, Stanislav Babak, Alice Bonino, Eleanor Hamilton, N.V. Krishnendu, Piero Rettegno, Riccardo Sturani and Jooheon Yoo for performing the LIGO-Virgo-KAGRA review of the \texttt{SEOBNRv5} models. 

Part of M.K.'s work on this paper is supported by Perimeter Institute for Theoretical Physics. Research at Perimeter Institute is supported in part by the Government of Canada through the Department of Innovation, Science and Economic Development and by the Province of Ontario through the Ministry of Colleges and Universities.
M.v.d.M. is supported by VILLUM FONDEN (grant no. 37766), and the Danish Research Foundation.
R.C. is supported by NSF Grants No. AST-2006538, PHY-2207502, PHY-090003 and PHY-20043, and NASA Grants No. 19-ATP19-0051, 20-LPS20-0011 and 21-ATP21-0010.
H.R. is supported by the Funda\c c\~ao para a Ci\^encia e Tecnologia (FCT)  within the projects UID/04564/2021, UIDB/04564/2020, UIDP/04564/2020 and EXPL/FIS-AST/0735/2021. 
This work was supported in part by the Sherman Fairchild Foundation, 
by NSF Grants No.\ PHY-2011961, No.\ PHY-2011968, and No.\ OAC-1931266 at Caltech, 
by NSF Grants No.\ PHY-1912081, , No.\ PHY-2207342, No.\ OAC-1931280 and No.\ OAC-2209655 at Cornell, 
by NSF awards No.\ AST-1559694 and No.\ PHY-1654359, by Nicholas and Lee Begovich, and by the Dan Black Family Trust.

The computational work for this manuscript was carried out on the \texttt{Hypatia} computer cluster at the Max Planck Institute for Gravitational Physics in Potsdam.

\texttt{SEOBNRv5HM} is publicly available through the python package \texttt{pySEOBNR} \href{https://git.ligo.org/waveforms/software/pyseobnr}{\texttt{git.ligo.org/waveforms/software/pyseobnr}}. Stable versions of \texttt{pySEOBNR} are published through the Python Package Index (PyPI), and can be installed via ~\texttt{pip install pyseobnr}.
\texttt{SEOBNRv5\_ROM} is publicly available through LALSuite \cite{lalsuite}.

This research has made use of data or software obtained from the Gravitational Wave Open Science Center (gwosc.org), a service of LIGO Laboratory, the LIGO Scientific Collaboration, the Virgo Collaboration, and KAGRA. 
LIGO Laboratory and Advanced LIGO are funded by the United States National Science Foundation (NSF) as well as the Science and Technology Facilities Council (STFC) of the United Kingdom, the Max-Planck-Society (MPS), 
and the State of Niedersachsen/Germany for support of the construction of Advanced LIGO and construction and operation of the GEO600 detector. Additional support for Advanced LIGO was provided by the Australian Research Council. 
Virgo is funded, through the European Gravitational Observatory (EGO), by the French Centre National de Recherche Scientifique (CNRS), the Italian Istituto Nazionale di Fisica Nucleare (INFN) and the Dutch Nikhef, with contributions 
by institutions from Belgium, Germany, Greece, Hungary, Ireland, Japan, Monaco, Poland, Portugal, Spain. KAGRA is supported by Ministry of Education, Culture, Sports, Science and Technology (MEXT), Japan Society for the Promotion of 
Science (JSPS) in Japan; National Research Foundation (NRF) and Ministry of Science and ICT (MSIT) in Korea; Academia Sinica (AS) and National Science and Technology Council (NSTC) in Taiwan.

\clearpage

\appendix

\section{Hamiltonian coefficients}
\label{app:HamCoeffs}
We summarize here results that were derived in Ref.~\cite{Khalilv5}. In the nonspinning limit, the Hamiltonian is given by Eq.~\eqref{Heffzero}. The 5PN-expanded $\bar{D}_\text{noS}$, which enters the Hamiltonian through Eq.~\eqref{prstar}, is given by~\cite{Bini:2019nra,Bini:2020wpo}
\begin{widetext}
\begin{align}
\label{DpmTay}
\bar{D}_\text{noS}^\text{Tay}(u) &= 1 + 6 \nu u^2 + \left(52 \nu -6 \nu ^2\right)u^3 + \bigg[\nu  \left(-\frac{533}{45}-\frac{23761 \pi ^2}{1536}+\frac{1184 \gamma_E}{15}-\frac{6496 \ln 2}{15}+\frac{2916 \ln 3}{5}\right) \nonumber\\
&\qquad
+\left(\frac{123 \pi ^2}{16}-260\right) \nu ^2+\frac{592 \nu }{15} \ln u\bigg] u^4 
+ \left(-\frac{3392 \nu ^2}{15}-\frac{1420 \nu }{7}\right)  u^5 \ln u \nonumber\\
&\quad
+ \bigg[
\nu \left(\frac{294464}{175}-\frac{2840 \gamma_E}{7}-\frac{63707 \pi ^2}{512}+\frac{120648 \ln 2}{35}-\frac{19683 \ln 3}{7}\right)
+ \left(\frac{1069}{3}-\frac{205 \pi ^2}{16}\right) \nu^3 \nonumber\\
&\qquad
+ \left(d_5^{\nu^2}-\frac{6784 \gamma_E}{15}+\frac{67736}{105}+\frac{58320 \ln 3}{7}-\frac{326656 \ln 2}{21}\right) \nu^2
\bigg] u^5, 
\end{align}
where we set the remaining unknown coefficient $d_5^{\nu^2}$ to zero.
To improve agreement with NR, we perform a (2,3) Pad\'e resummation of $\bar{D}_\text{noS}^\text{Tay}(u)$.

For $Q_\text{noS}$, we use the full 5.5PN expansion derived in Refs.~\cite{Bini:2020wpo,Bini:2020hmy}, which is expanded in eccentricity to $\Order(p_r^8)$.
Instead of using $p_r$, we write $Q_\text{noS}$ in terms of $p_{r_*}$ using Eq.~\eqref{prstar}, then PN expand to 5.5PN order, leading to
\begin{align}
\label{QpmTay}
Q_\text{noS} &= \frac{p_{r_*}^4}{\mu^2} \Bigg\lbrace 2 (4 - 3 \nu) \nu u^2 
+u^3 \left[10 \nu ^3-131 \nu ^2+\nu  \left(-\frac{4348}{15}+\frac{496256 \ln 2}{45}-\frac{33048 \ln 3}{5}\right)\right]
+ u^4 \Bigg[\left(792-\frac{615 \pi ^2}{32}\right) \nu ^3  \nonumber\\
&\qquad
+\nu ^2 \left(-\frac{592 \ln u}{5}+\frac{31633 \pi ^2}{512}-\frac{1184 \gamma_E}{5}+\frac{45683}{105}+\frac{33693536 \ln 2}{105}-\frac{6396489 \ln 3}{70}-\frac{9765625 \ln 5}{126}\right) \nonumber\\
&\qquad
+\nu  \left(\frac{5428 \ln u}{105}+\frac{1249177}{1050}-\frac{93031 \pi ^2}{1536}+\frac{10856 \gamma_E }{105}-\frac{4396376}{105} \ln 2+\frac{9765625 \ln 5}{504}-\frac{601911 \ln 3}{280}\right)\Bigg] 
+ \frac{88703 \pi  \nu  u^{9/2}}{1890} \Bigg\rbrace 
\nonumber\\
&\quad 
+ \frac{p_{r_*}^6}{\mu^4} \Bigg\lbrace
u^2 \left[6 \nu ^3-\frac{27 \nu ^2}{5}+\nu  \left(-\frac{827}{3}-\frac{1}{25} 2358912 \ln 2+\frac{1399437 \ln 3}{50}+\frac{390625 \ln 5}{18}\right)\right] \nonumber\\
&\qquad
+ u^3 \Bigg[-14 \nu ^4+188 \nu ^3+\nu ^2 \left(\frac{154229}{75}-\frac{4998308864 \ln 2}{1575}+\frac{26171875 \ln 5}{18}-\frac{45409167 \ln 3}{350}\right) \nonumber\\
&\quad\qquad
+\nu  \left(-\frac{860317}{1050}+\frac{305146624 \ln 2}{945}+\frac{35643726 \ln 3}{175}-\frac{52468750 \ln 5}{189}\right)\Bigg]
-\frac{2723471 \pi  \nu  u^{7/2}}{756000} \Bigg\rbrace 
\nonumber\\
&\quad
+ \frac{p_{r_*}^8}{\mu^6} \Bigg\lbrace
u \nu  \left[-\frac{35772}{175}+\frac{21668992 \ln 2}{45}+\frac{6591861 \ln 3}{350}-\frac{27734375 \ln 5}{126}\right]
+\frac{5994461 \pi  \nu  u^{5/2}}{12700800} + u^2 \Bigg[-6 \nu ^4+\frac{24 \nu ^3}{7} \nonumber\\
&\quad\qquad
+\nu ^2 \left(\frac{870976}{525}+\frac{703189497728 \ln 2}{33075}+\frac{332067403089 \ln 3}{39200}-\frac{13841287201 \ln 7}{4320}-\frac{468490234375 \ln 5}{42336}\right) \nonumber\\
&\quad\qquad
+\nu  \left(-\frac{2222547}{2450}-\frac{1347019456}{525}  \ln 2+\frac{278690984375 \ln 5}{169344}+\frac{13841287201 \ln 7}{17280}-\frac{346536085761 \ln 3}{156800}\right)\Bigg]
\Bigg\rbrace.
\end{align}

In the aligned-spin Hamiltonian, the 3.5PN SO gyro-gravitomagnetic factors in Eq.~\eqref{HeffAnzAlign} are given by
\begin{subequations}
\label{gyros}
\begin{align}
g_{a_+}^\text{3.5PN} &= \frac{7}{4} 
+ \left[\tilde{L}^2 u^2 \left(-\frac{45 \nu }{32}-\frac{15}{32}\right)
+u\left(\frac{23 \nu }{32}-\frac{3}{32}\right)\right] 
\nonumber\\
&\quad
+ \Bigg[\tilde{L}^4u^4 \left(\frac{345 \nu ^2}{256}+\frac{75 \nu }{128}+\frac{105}{256}\right)
+\tilde{L}^2u^3 \left(-\frac{1591 \nu ^2}{768}-\frac{267 \nu }{128}+\frac{59}{256}\right)
+u^2\left(\frac{109 \nu ^2}{192}-\frac{177 \nu }{32}-\frac{5}{64}\right)\Bigg], \\
g_{a_-}^\text{3.5PN} &= \frac{1}{4} 
+ \left[\tilde{L}^2u^2 \left(\frac{15}{32}-\frac{9 \nu }{32}\right)
+u\left(\frac{11 \nu }{32}+\frac{3}{32}\right)\right] 
\nonumber\\
&\quad
+\Bigg[\tilde{L}^4u^4 \left(\frac{75 \nu ^2}{256}-\frac{45 \nu }{128}-\frac{105}{256}\right)
+\tilde{L}^2u^3 \left(-\frac{613 \nu ^2}{768}-\frac{35 \nu }{128}-\frac{59}{256}\right)
+u^2\left(\frac{103 \nu ^2}{192}-\frac{\nu }{32}+\frac{5}{64}\right)\Bigg],
\end{align}
\end{subequations}
\end{widetext}
where the square brackets collect different PN orders, and we defined $\tilde{L}\equiv L/(M\mu) \equiv p_\phi/(M\mu)$.
The cubic-in-spin term $G_{a^3}$ reads
\begin{equation}
G_{a^3}^\text{align} = \frac{Mp_\phi}{4r^2} \left(\delta a_- a_+^2-a_+^3\right).
\end{equation}

The potentials in the even-in-spin part of the effective Hamiltonian in Eq.~\eqref{HeffAnzAlign} include the 4PN SS information, and are given by
\begin{subequations}
\begin{align}
A^\text{align} &= \frac{a_+^2/r^2+A_\text{noS}+A_\text{SS}^\text{align}}{1+(1+2M/r)a_+^2/r^2}, \\
B_{np}^\text{align} &= -1 + \frac{a_+^2}{r^2} + A_\text{noS} \bar{D}_\text{noS} + B_{np,\text{SS}}^\text{align}, \\
B_{npa}^\text{Kerr\,eq} &= -\frac{1+2M/r}{r^2+a_+^2 (1+2M/r)},\\
Q^\text{align} &= Q_\text{noS} + Q_\text{SS}^\text{align},
\end{align}
\end{subequations}
where
\begin{subequations}
\label{SSalign}
\begin{align}
A_\text{SS}^\text{align} &= \frac{M^2}{r^4}\left[
\frac{9 a_+^2}{8}-\frac{5}{4} \delta a_- a_+ +a_-^2 \left(\frac{\nu }{2}+\frac{1}{8}\right)\right] \nonumber\\
&\quad
+\frac{M^3}{r^5}\bigg[
a_+^2 \left(-\frac{175 \nu }{64}-\frac{225}{64}\right)
+\delta a_- a_+ \left(\frac{117}{32}-\frac{39 \nu }{16}\right)\nonumber\\
&\quad\qquad
+a_-^2 \left(\frac{21 \nu ^2}{16}-\frac{81 \nu }{64}-\frac{9}{64}\right)\bigg], \\
B_{np,\text{SS}}^\text{align} &= \frac{M}{r^3} \left[a_+^2 \left(3 \nu +\frac{45}{16}\right)-\frac{21}{8}\delta a_- a_+ +a_-^2 \left(\frac{3 \nu }{4}-\frac{3}{16}\right)\right]\nonumber\\
&\quad
+ \frac{M^2}{r^4} \bigg[
a_+^2 \left(-\frac{1171 \nu }{64}-\frac{861}{64}\right)
+\delta a_- a_+ \left(\frac{13 \nu }{16}+\frac{449}{32}\right) \nonumber\\
&\quad\qquad
+a_-^2 \left(\frac{\nu ^2}{16}+\frac{115 \nu }{64}-\frac{37}{64}\right)
\bigg], \\
Q_\text{SS}^\text{align} &= \frac{Mp_r^4}{\mu^2r^3} \bigg[
a_+^2 \left(\frac{25}{32}-5 \nu ^2+\frac{165 \nu }{32}\right)
+\delta a_- a_+ \left(\frac{45 \nu }{8}-\frac{5}{16}\right) \nonumber\\
&\quad\qquad
+a_-^2 \left(-\frac{15 \nu ^2}{8}+\frac{75 \nu }{32}-\frac{15}{32}\right)
\bigg].
\end{align}
\end{subequations}

\section{Expressions for the factorized waveform modes}
\label{app:modes}
In this Appendix, we list the expressions for $\rho_{\ell m}$, $f_{\ell m}$ and $\delta_{\ell m}$, which are used in the factorized modes (see Eqs.~\eqref{hlmFactorized} and \eqref{frholm}).

In the $(2,2)$ mode, $\rho_{22}$ and $\delta_{22}$ are given by
\begin{widetext}
\begin{subequations}
\begin{align}
\rho_{22} &= 1
+ v_\Omega^2 \left(\tfrac{55}{84}\nu-\tfrac{43}{42}\right)
+ v_\Omega^3 \left[\left(\tfrac{2}{3}\nu-\tfrac{2}{3}\right) \chi _S-\tfrac{2}{3}\delta\chi _A\right]
+ v_\Omega^4 \Big[
\tfrac{19583}{42336} \nu^2-\tfrac{33025}{21168}\nu-\tfrac{20555}{10584}
+ \left(\tfrac{1}{2}-2 \nu \right) \chi _A^2+\delta  \chi _A \chi _S+\tfrac{1}{2}\chi _S^2\Big] \nonumber\\
&\quad
+ v_\Omega^5 \left[\delta  \left(-\highlight{\tfrac{19}{42}}\nu-\tfrac{34}{21}\right) \chi _A+\left(\tfrac{209}{126}\nu^2+\tfrac{49}{18}\nu-\tfrac{34}{21}\right) \chi _S\right] \nonumber \\
&\quad
+ v_\Omega^6 \Big[
\tfrac{10620745 \nu^3}{39118464}-\tfrac{6292061 \nu^2}{3259872}+\tfrac{41 \pi^2 \nu}{192}-\tfrac{48993925 \nu}{9779616}-\tfrac{428}{105}\text{eulerlog}(2,v_\Omega)+\tfrac{1556919113}{122245200} \nonumber \\
&\qquad 
\highlight{ + \delta  \left(\tfrac{89}{126}-\tfrac{781}{252}\nu\right) \chi _A \chi _S
+\left(-\tfrac{27}{14}\nu^2-\tfrac{457}{504}\nu+\tfrac{89}{252}\right) \chi _A^2
+\left(\tfrac{10}{9}\nu^2 -\tfrac{1817}{504}\nu +\tfrac{89}{252}\right) \chi _S^2} \Big] \nonumber\\
&\quad
+ v_\Omega^7 \Big[
\delta  \left(\tfrac{97865}{63504}\nu^2 +\tfrac{50140}{3969}\nu +\tfrac{18733}{15876}\right) \chi _A
+\left(\tfrac{50803}{63504}\nu^3 -\tfrac{245717}{63504}\nu^2 +\tfrac{74749 }{5292}\nu +\tfrac{18733}{15876}\right) \chi _S
\highlight{+ \delta \chi _A^3 \left(\tfrac{1}{3}-\tfrac{4}{3}\nu\right)+\delta  (2 \nu +1) \chi _A \chi _S^2} \nonumber\\
&\qquad
\highlight{+ \left(-4 \nu ^2-3 \nu +1\right) \chi _A^2 \chi _S+\left(\nu +\tfrac{1}{3}\right) \chi _S^3}\Big] 
+ v_\Omega^8 \left[\tfrac{9202}{2205}\text{eulerlog}(2,v_\Omega)-\tfrac{387216563023}{160190110080} + \frac{18353}{21168} a^2 - \frac{1}{8} a^4 \right]\nonumber \\
&\quad 
+ v_\Omega^{10} \left[\tfrac{439877}{55566}\text {eulerlog}(2,v_\Omega)-\tfrac{16094530514677}{533967033600}\right], \\
\delta_{22} &=\tfrac{7 }{3}\Omega H_\text{EOB}
+ \big(\Omega H_\text{EOB}\big)^2 \left[\left(\tfrac{8}{3}\nu-\tfrac{4}{3}\right) \chi _S-\tfrac{4}{3}\delta \chi _A + \tfrac{428}{105} \pi\right] + \big(\Omega H_\text{EOB}\big)^3 \left[\tfrac{1712}{315} \pi^2-\tfrac{2203}{81} \right] - 24 \nu v_\Omega^5,
\end{align}
\end{subequations}
where $\text {eulerlog}(m,v_\Omega)$ is defined by Eq.~\eqref{eulerlog}.
The coefficient $19/42$ of $\mathcal{O}(v_\Omega^5 \delta \chi_A \nu)$ in $\rho_{22}$ corrects a typo in the \texttt{SEOBNRv4} code, and we added in $\rho_{22}$ the NLO spin-squared and LO spin-cubed contributions, which are given by Eq. (4.11a) of Ref.~\cite{Henry:2022dzx}.
The spinning test-mass terms in the coefficient of $v_\Omega^8$ are parametrized in terms of $a = (1 - 2 \nu) \chi$, where $\chi =\chi_{S}+\chi_{A} \frac{\delta}{1-2 \nu}$

The $(2,1)$ mode reads
\begin{subequations}
\begin{align}
\rho_{21}^\text{NS} &=1
+v_\Omega^2 \left(\tfrac{23}{84}\nu -\tfrac{59}{56}\right) 
+v_\Omega^4 \left(\tfrac{617}{4704}\nu^2 -\tfrac{10993}{14112}\nu -\tfrac{47009}{56448}\right)  +v_\Omega^6 \left[\tfrac{7613184941}{2607897600}-\tfrac{107}{105} \text { eulerlog }\left(1, v_\Omega\right)\right]  \nonumber \\
&\quad+ v_\Omega^8\left[-\tfrac{1168617463883}{911303737344}+\tfrac{6313}{5880}\text { eulerlog }\left(1, v_\Omega\right)\right]
+v_\Omega^{10}\left[-\tfrac{63735873771463}{16569158860800} + \tfrac{5029963}{5927040} \text{ eulerlog }\left(1, v_\Omega\right)\right] ,
\\
f_{21}^\text{S} &= -\tfrac{3}{2} v_\Omega \left(\frac{\chi _A}{\delta} +\chi _S\right)
+ v_\Omega^3 \Big[\left(\tfrac{131}{84}\nu +\tfrac{61}{12}\right) \frac{\chi _A}{\delta} +\left(\tfrac{79}{84}\nu +\tfrac{61}{12}\right) \chi _S\Big]
+ v_\Omega^4 \left[(-2 \nu -3) \chi _A^2+ \left(\tfrac{21}{2}\nu -6\right) \frac{\chi _A \chi _S}{\delta}+\left(\tfrac{1}{2}\nu -3\right) \chi _S^2  \right] \nonumber\\
&\quad
+ v_\Omega^5 \Big[
\left(-\tfrac{703}{112}\nu^2 +\tfrac{8797}{1008}\nu -\tfrac{81}{16}\right) \frac{\chi _A}{\delta }+\left(\tfrac{613}{1008}\nu^2 +\tfrac{1709}{1008}\nu -\tfrac{81}{16}\right) \chi _S
+ \left(\tfrac{3}{4}-3 \nu \right) \frac{\chi _A^3}{\delta }+ \left(\tfrac{9}{4}-6 \nu \right) \frac{\chi _A \chi _S^2}{\delta }+\left(\tfrac{9}{4}-3 \nu \right) \chi _A^2 \chi _S+\tfrac{3}{4} \chi _S^3 \Big] \nonumber\\
&\quad
+ \highlight{v_\Omega^6 \Big[
\left(\tfrac{5}{7}\nu ^2 -\tfrac{9287}{1008}\nu +\tfrac{4163}{252}\right) \chi _A^2
+\left(\tfrac{139}{72} \nu ^2-\tfrac{2633 }{1008}\nu+\tfrac{4163}{252}\right) \chi _S^2
+ \left(\tfrac{9487}{504} \nu ^2 -\tfrac{1636}{21}\nu +\tfrac{4163}{126}\right) \frac{ \chi _A \chi _S}{\delta } \Big]},
\\
\label{eq:delta21}\delta_{21} &= \tfrac{2}{3}\Omega H_\text{EOB} +\tfrac{107}{105} \pi \big(\Omega H_\text{EOB}\big)^2 +\left(\tfrac{214}{315} \pi^2-\tfrac{272}{81}\right) \big(\Omega H_\text{EOB}\big)^3  - \highlight{\tfrac{25}{2}} \nu  v_\Omega^5,
\end{align}
\end{subequations}
where the $\mathcal{O}(v_\Omega^6 \chi^2 \nu^2)$ terms in $f_{21}^\text{S}$ correct those used in the \texttt{SEOBNRv4HM} model~\cite{Cotesta:2018fcv}.
We also fixed the coefficient $-25/2$ of $\mathcal{O}(\nu v_\Omega^5)$ in $\delta_{21}$, which was the result of an error in Ref.~\cite{Blanchet:2008je}, which was later corrected in an erratum, as noted in Ref.~\cite{Henry:2022dzx}.

The $(3,3)$ mode is given by
\begin{subequations}
\begin{align}
\rho_{33}^\text{NS}&= 1+v_\Omega^2 \left(\tfrac{2}{3}\nu-\tfrac{7}{6}\right)  + v_\Omega^4\left(-\tfrac{6719}{3960}-\tfrac{1861}{990}\nu+\tfrac{149 }{330}\nu^2\right) \nonumber \\
&\quad
+v_\Omega^6 \left[\tfrac{3203101567}{227026800}+\left(-\tfrac{129509}{25740}+\tfrac{41 \pi^2}{192}\right) \nu-\tfrac{274621}{154440} \nu^2+\tfrac{12011 }{46332}\nu^3-\tfrac{26}{7} \text{eulerlog}\left(3, v_\Omega\right)\right]  \nonumber \\
&\quad
+v_\Omega^8\left[-\tfrac{57566572157}{8562153600}+\tfrac{13}{3} \text { eulerlog }\left(3, v_\Omega\right)\right] 
+v_\Omega^{10}\left[-\tfrac{903823148417327}{30566888352000}+\tfrac{87347}{13860}\text{eulerlog}\left(3, v_\Omega\right)\right], \\
f_{33}^\text{S}&= v_\Omega^3 \left[\left(\tfrac{19}{2}\nu-2\right) \frac{\chi _A}{\delta }+\left(\tfrac{5}{2}\nu-2\right) \chi _S\right] 
+ v_\Omega^4 \bigg[
\left(\tfrac{3}{2}-6 \nu \right) \chi _A^2+(3-12 \nu ) \frac{\chi _A \chi _S}{\delta }+\tfrac{3 }{2}\chi _S^2 \bigg] \nonumber \\
&\quad
+ v_\Omega^5 \left[\left(\tfrac{407}{30}\nu ^2-\tfrac{593}{60}\nu+\tfrac{2}{3}\right) \frac{ \chi _A}{\delta }+\left(\tfrac{241}{30}\nu ^2 +\tfrac{11}{20}\nu +\tfrac{2}{3}\right) \chi _S\right] \nonumber\\
&\quad
+ v_\Omega^6 \bigg[
\left(-12 \nu ^2+\tfrac{11}{2}\nu -\tfrac{7}{4}\right) \chi _A^2
+\left(44 \nu ^2-\nu -\tfrac{7}{2}\right) \frac{\chi _A \chi _S}{\delta }
+\left(6 \nu ^2-\tfrac{27}{2}\nu -\tfrac{7}{4}\right) \chi _S^2 
\bigg] \nonumber\\
&\quad
+ i \big(\Omega H_{\rm{EOB}}\big)^2 \left[\left(\tfrac{7339}{540}\nu -\tfrac{81}{20}\right) \frac{\chi _A}{\delta }+\left(\tfrac{593}{108}\nu -\tfrac{81}{20}\right) \chi _S\right]
,\\
\delta_{33} &=\tfrac{13}{10}\left(H_{\mathrm{EOB}} \Omega\right)+\tfrac{39 \pi}{7}\left(H_{\mathrm{EOB}} \Omega\right)^2+\left(-\tfrac{227827}{3000}+\tfrac{78 \pi^2}{7}\right)\left(H_{\mathrm{EOB}} \Omega\right)^3-\tfrac{80897}{2430}\nu v_\Omega^5,
\end{align}
\end{subequations}
where the imaginary part of $f_{33}^\text{S}$ is included in $\delta_{33}$ in Ref.~\cite{Henry:2022dzx}, but we moved it to $f_{33}^\text{S}$ to facilitate the implementation in the equal-mass limit, for which we pull the factor $\delta$ from the leading order $h_{33}^\text{N}$ into $f_{\ell m}$ to cancel the divergent $1/\delta$.

For the $(4,4)$ mode, we use
\begin{subequations}
\begin{align}
\rho_{44} &= 1+v_\Omega^2\left[\tfrac{1614-5870 \nu +2625 \nu ^2}{1320 (-1+3 \nu )}\right] + v_\Omega^3 \left[\left(\tfrac{2}{3}-\tfrac{41 \nu }{15}+\tfrac{14 \nu ^2}{5}\right) \tfrac{1}{(-1+3 \nu )}\chi_S 
+\delta \left(\tfrac{2}{3}-\tfrac{13 \nu }{5}\right) \tfrac{1}{(-1+3 \nu )}\chi _A\right]  \nonumber \\
&\quad
+v_\Omega^4\left[-\tfrac{14210377}{8808800 (1-3 \nu )^2}+\tfrac{32485357 \nu
}{4404400 (1-3 \nu )^2}-\tfrac{1401149 \nu ^2}{1415700 (1-3 \nu )^2} 
-\tfrac{801565 \nu ^3}{37752 (1-3 \nu )^2}+\tfrac{3976393 \nu ^4}{1006720 (1-3 \nu )^2}+\tfrac{1}{2}\chi _A^2-2 \nu  \chi _A^2+\delta \chi _A \chi _S+\tfrac{1}{2}\chi _S^2\right]  \nonumber \\
&\quad+ v_\Omega^5\Bigg[\left(-\tfrac{69}{55}+\tfrac{16571 \nu }{1650}-\tfrac{2673 \nu ^2}{100}+\tfrac{8539 \nu ^3}{440}+\tfrac{591 \nu ^4}{44}\right) \tfrac{1}{(1-3 \nu)^2}\chi _S
+\delta
\left(-\tfrac{69}{55}+\tfrac{10679 \nu }{1650}-\tfrac{1933 \nu ^2}{220}+\tfrac{597 \nu ^3}{440}\right) \tfrac{1}{ (1-3 \nu )^2}\chi_A\Bigg]  \nonumber \\
&\quad+v_\Omega^6 \left[\tfrac{16600939332793}{1098809712000}-\tfrac{12568 }{3465}\text{eulerlog}\left(4,v_\Omega\right)\right]  +v_\Omega^8\left[-\tfrac{172066910136202271}{19426955708160000}+\tfrac{845198 }{190575}\text{eulerlog}\left(4,v_\Omega\right)\right] \nonumber \\
&\quad+ v_\Omega^{10}\left[-\tfrac{17154485653213713419357}{568432724020761600000}+\tfrac{22324502267}{3815311500} \text{eulerlog}\left(4,v_\Omega\right)\right]\,, \\
\delta_{44} &= \tfrac{(112+219 \nu )}{120 (1-3 \nu )}(\Omega  H_{\mathrm{EOB}})+\tfrac{25136 \pi}{3465}(\Omega  H_{\mathrm{EOB}})^2 +\left(\tfrac{201088}{10395}\pi^2 - \tfrac{55144}{375} \right) (\Omega  H_{\mathrm{EOB}})^3\,, 
\end{align}
\end{subequations}
and for the $(5,5)$ mode
\begin{subequations}
\begin{align}
	\rho_{55}^{\mathrm{NS}} =& 1+v_\Omega^2\left[\tfrac{487}{390 (-1+2 \nu )}-\tfrac{649 \nu }{195 (-1+2 \nu )}+\tfrac{256 \nu ^2}{195 (-1+2 \nu )}\right] -\tfrac{3353747}{2129400} v_\Omega^4 \nonumber \\
   &+v_\Omega^6\left[\tfrac{190606537999247}{11957879934000}-\tfrac{1546}{429}\text{eulerlog}\left(5,v_\Omega\right) \right] +v_\Omega^8 \left[-\tfrac{1213641959949291437}{118143853747920000}+\tfrac{376451 }{83655}\text{eulerlog}\left(5,v_\Omega\right)\right] \nonumber \\
   &+v_\Omega^{10}\left[-\tfrac{150082616449726042201261}{4837990810977324000000}+\tfrac{2592446431}{456756300}\text{eulerlog}\left(5,v_\Omega\right)\right]\,, \\
   f_{55}^{\mathrm{S}} & = v_\Omega^3 \left[\left(-\tfrac{70 \nu }{3 (-1+2 \nu )}+\tfrac{110 \nu ^2}{3 (-1+2 \nu )}+\tfrac{10}{3 (-1+2 \nu )}\right) \frac{\chi _A}{\delta}+\left(\tfrac{10}{3
   (-1+2 \nu )}-\tfrac{10 \nu }{-1+2 \nu }+\tfrac{10 \nu ^2}{-1+2 \nu }\right) \chi _S\right] \nonumber \\
   & \quad + v_\Omega^4 \left[\left(-\tfrac{5}{2} + 5 \nu\right)\tfrac{1}{(-1+2 \nu)}\chi_S^2 + \left(-5 + 30 \nu - 40 \nu^2\right) \tfrac{1}{(-1+2 \nu)} \frac{\chi_S \chi_A}{\delta} + \left(-\tfrac{5}{2}+15 \nu - 20 \nu^2\right) \tfrac{1}{(-1+2 \nu)} \chi_A^2 \right], \\
	\delta_{55} =& \tfrac{(96875+857528 \nu )}{131250 (1-2 \nu )}(\Omega  H_{\mathrm{EOB}}) + \tfrac{3865\pi}{429}(\Omega  H_{\mathrm{EOB}})^2 + \tfrac{-7686949127 + 954500400\pi^2}{31783752}(\Omega  H_{\mathrm{EOB}})^3\,,
\end{align}
\end{subequations}
which are both the same as in \texttt{SEOBNRv4HM}~\cite{Cotesta:2018fcv}.

The (3,2) mode is given by
\begin{subequations}
\begin{align}
\rho_{32} &= 1 + v_\Omega \frac{4 \nu \chi _S}{3 (1-3 \nu )}
+ v_\Omega^2 \left[
\frac{-\tfrac{32}{27}\nu ^2+\tfrac{223}{54}\nu-\tfrac{164}{135}}{1-3 \nu }
\highlight{-\frac{16 \nu ^2 \chi _S^2}{9 (1-3 \nu )^2}}
\right] \nonumber\\
&\quad
+ v_\Omega^3 \left[\highlight{\left(\tfrac{13 }{9}\nu +\tfrac{2}{9}\right) \frac{\delta \chi _A}{1-3 \nu }
+\left(\tfrac{607}{81}\nu ^3 +\tfrac{503}{81}\nu ^2 -\tfrac{1478}{405}\nu +\tfrac{2}{9}\right)  \frac{\chi _S}{(1-3 \nu )^2} } 
\highlight{+\frac{320 \nu ^3 \chi _S^3}{81 (1-3 \nu )^3}}\right]\nonumber\\
&\quad
+ v_\Omega^4 \Bigg[
\frac{\tfrac{77141}{40095} \nu ^4 -\tfrac{508474 }{40095}\nu ^3 -\tfrac{945121 }{320760}\nu ^2 +\tfrac{1610009 \nu }{320760}-\tfrac{180566}{200475}}{(1-3 \nu )^2}
+ \highlight{\left(4 \nu ^2-3 \nu +\tfrac{1}{3}\right) \frac{\chi _A^2}{1-3 \nu }
+ \left(-\tfrac{50}{27}\nu ^2-\tfrac{88}{27}\nu+\tfrac{2}{3}\right) \frac{\delta \chi _A \chi _S}{(1-3 \nu )^2} } \nonumber\\
&\qquad
\highlight{
+ \left(-\tfrac{2452}{243} \nu ^4 -\tfrac{1997}{243} \nu ^3 +\tfrac{1435}{243}\nu ^2 -\tfrac{43}{27}\nu +\tfrac{1}{3}\right)  \frac{\chi _S^2}{(1-3 \nu )^3}}
\Bigg] \nonumber\\
&\quad
+ \highlight{v_\Omega^5 \Bigg[
\left(-\tfrac{1184225 }{96228}\nu ^5 -\tfrac{40204523}{962280} \nu ^4 +\tfrac{101706029 }{962280}\nu ^3 -\tfrac{14103833 }{192456}\nu ^2 +\tfrac{20471053}{962280}\nu -\tfrac{2788}{1215}\right)  \frac{\chi _S}{(1-3 \nu )^3}
+ \left(\tfrac{608}{81} \nu ^3+\tfrac{736}{81} \nu ^2 -\tfrac{16  }{9}\nu\right)  \frac{\delta \chi _A \chi _S^2}{(1-3 \nu )^3}} \nonumber\\
&\qquad
\highlight{+\left(\tfrac{889673}{106920}\nu ^3-\tfrac{75737}{5346} \nu ^2+\tfrac{376177 }{35640}\nu -\tfrac{2788}{1215}\right)  \frac{\delta  \chi _A}{(1-3 \nu )^2}
+\left(\tfrac{96176 }{2187}\nu ^5 +\tfrac{43528}{2187} \nu ^4-\tfrac{40232 }{2187}\nu ^3 +\tfrac{376 }{81}\nu ^2 -\tfrac{8 \nu }{9}\right)  \frac{\chi _S^3}{(1-3 \nu )^4}} \nonumber\\
&\qquad
\highlight{+\left(-\tfrac{32 }{3}\nu ^3+8 \nu ^2-\tfrac{8}{9}\nu\right)  \frac{\chi _A^2 \chi _S}{(1-3 \nu )^2} \Bigg]}
+v_\Omega^6\left[\tfrac{5849948554}{940355325}-\tfrac{104 \text { eulerlog}(2,v_\Omega)}{63}\right] 
+v_\Omega^8 \left[\tfrac{17056 \text { eulerlog}(2,v_\Omega)}{8505}-\tfrac{10607269449358}{3072140846775}\right] \nonumber \\
&\qquad
+ \highlight{v_\Omega^{10}\left[- \tfrac{1312549797426453052}{176264081083715625} +\tfrac{18778864 \text { eulerlog}(2,v_{\Omega} )}{12629925}   \right]}, \\
\delta_{32} &= \left(\tfrac{11}{5}\nu+\tfrac{2}{3}\right)  \frac{\Omega H_\text{EOB}}{1-3 \nu } +\tfrac{52}{21} \pi(\Omega  H_{\mathrm{EOB}})^2+\left(\tfrac{208}{63} \pi^2-\tfrac{9112}{405}\right) (\Omega  H_{\mathrm{EOB}})^3,
\end{align}
\end{subequations}
where we added all spin contributions beyond the LO spin-orbit in $\rho_{32}$, as well as the test-mass limit terms given in Eq.~(\ref{eq:rho32tml}).

The (4,3) mode is given by
\begin{subequations}
\begin{align}
\rho_{43}^\text{NS} &= 1 + \frac{v_\Omega^2}{1-2 \nu } \left(-\tfrac{10}{11} \nu ^2+\tfrac{547}{176}\nu-\tfrac{111}{88}\right) -\tfrac{6894273}{7047040} v_\Omega^4
+v_\Omega^6\left[\tfrac{1664224207351}{195343948800}-\tfrac{1571}{770} \text { eulerlog}(3,v_\Omega)\right] \nonumber \\
&\quad
+ \highlight v_\Omega^8\left[-\tfrac{2465107182496333}{460490801971200}  + \tfrac{174381}{67760} \text { eulerlog}(3,v_{\Omega} )   \right] \\
f_{43}^\text{S} &= \frac{v_\Omega}{1-2 \nu}\left(\tfrac{5}{2} \nu  \chi _S- \tfrac{5}{2}\nu\frac{\chi _A}{\delta}\right)
\highlight{+ \frac{v_\Omega^3}{1-2\nu} \left[
\left(\tfrac{887}{44}\nu -\tfrac{3143}{132} \nu ^2\right) \frac{\chi _A}{\delta}
+ \left(-\tfrac{529}{132} \nu ^2-\tfrac{667}{44}\nu\right) \chi _S
\right]}\nonumber\\
&\quad
+ \highlight{\frac{v_\Omega^4}{1-2\nu} \Big[
\left(12 \nu ^2-\tfrac{37}{3}\nu +\tfrac{3}{2}\right) \chi _A^2
+ \left(\tfrac{137}{6} \nu ^2-18 \nu +3\right) \frac{\chi _A \chi _S}{\delta}
+ \left(\tfrac{35}{6} \nu ^2+\tfrac{1 }{3}\nu+\tfrac{3}{2}\right) \chi _S^2
\Big]},\\
\delta_{43} &= \left(\tfrac{4961}{810}\nu+\tfrac{3}{5}\right) \frac{\Omega  H_\text{EOB}}{1-2 \nu } +\tfrac{1571}{385} \pi (\Omega  H_{\mathrm{EOB}})^2,
\end{align}
\end{subequations}
where we added all spin contributions beyond the LO spin-orbit in $f_{43}^\text{S}$, and the test-mass limit terms of Eq.~(\ref{eq:rho43tml}) in $\rho_{43}^\text{NS}$.

All other modes, which are used in the RR force in Eq.~\eqref{RRforce}, are the same as in \texttt{SEOBNRv4HM}.
They are written in Refs.~\cite{Taracchini:2012ig,Pan:2010hz}, but we also list them here for completeness:
\begin{align}
\rho_{31}^\text{NS} &= 1-v_\Omega^2\left[\frac{2}{9}\nu+\frac{13}{18}\right] 
+ v_\Omega^4  \left[-\frac{829}{1782}\nu^2-\frac{1685}{1782}\nu+\frac{101}{7128}\right] 
+ v_\Omega^6 \left[\frac{11706720301}{6129723600}-\frac{26}{63}\text{eulerlog}(1,v_\Omega)\right] \nonumber\\
&\quad
+ v_\Omega^8 \left[\frac{169}{567}\text{eulerlog}(1,v_\Omega) +\frac{2606097992581}{4854741091200}\right],  \\
f_{31}^\text{S} &= v_\Omega^3 \left[\left(\frac{11}{2}\nu -2\right) \frac{\chi _A}{\delta }+\left(\frac{13}{2}\nu-2\right) \chi _S\right], \\
\delta_{31} &= \frac{13}{30} \Omega  H_{\mathrm{EOB}} + \frac{13}{21}\pi (\Omega  H_{\mathrm{EOB}})^2 + \left(\frac{26}{63}\pi^2 - \frac{227827}{81000}\right)(\Omega  H_{\mathrm{EOB}})^3 - \frac{17}{10} \nu v_\Omega^5, \\
\rho_{42}&=1+\frac{285\,\nu^2-3530\,\nu+1146}{1320\,(3\,\nu-1)} v_\Omega^2 - \frac{v_\Omega^3}{15(1-3\nu)} \left[(78\nu^2 - 59\nu + 10)\chi_S +(10-21\nu)\delta\,\chi_A \right] \nonumber \\
&\quad+\frac{-379526805\,\nu^4-3047981160\,\nu^3+1204388696\,\nu^2+295834536\,\nu-114859044}{317116800\,(1-3\,\nu)^2} v_\Omega^4 \nonumber\\
&\quad+ \left[\frac{848238724511}{219761942400}-\frac{3142}{3465}\text{eulerlog}(2,v_\Omega)\right] v_\Omega^6\,,\\
\delta_{42} &= \left(\frac{7}{15} + \frac{14}{5}\nu\right) \frac{\Omega  H_{\mathrm{EOB}}}{1-3\nu} + \frac{6284}{3465}\pi (\Omega  H_{\mathrm{EOB}})^2,\\
\rho_{41}^\text{NS} &= 1 +\frac{288\,\nu^2-1385\,\nu+602}{528\,(2\,\nu-1)}v_\Omega^2
-\frac{7775491}{21141120} v_\Omega^4 + \left[\frac{1227423222031}{1758095539200}-\frac{1571}{6930}\text{eulerlog}(1,v_\Omega)\right]\, v_\Omega^6 \,,\\
f_{41}^\text{S} &= \frac{5}{2} \nu  \frac{v_\Omega}{1-2 \nu } \left(\chi _S-\frac{\chi _A}{\delta }\right), \\
\delta_{41} &= \left(\frac{1}{5}+\frac{507}{10}\nu\right)\frac{\Omega  H_{\mathrm{EOB}}}{1-2\nu} + \frac{1571}{3465} \pi(\Omega  H_{\mathrm{EOB}})^2,\\
\rho_{54}&=1+\frac{33320\,\nu^3-127610\,\nu^2+96019\,\nu-17448}{13650\,(5\,\nu^2-5\,\nu+1)} v_\Omega^2 -\frac{16213384}{15526875} v_\Omega^4 \,, \\
\delta_{54}&= \frac{8}{15} \Omega H_\text{EOB},\\
\rho_{53}&=1+\frac{176\,\nu^2-850\,\nu+375}{390\,(2\,\nu-1)} v_\Omega^2 -\frac{410833}{709800} v_\Omega^4\,, \\
\delta_{53}&= \frac{31}{70} \Omega H_\text{EOB}, \\
\rho_{52}&=1+\frac{21980\,\nu^3-104930\,\nu^2+84679\,\nu-15828}{13650\,(5\,\nu^2-5\,\nu+1)} v_\Omega^2 -\frac{7187914}{15526875} v_\Omega^4 \,, \\
\delta_{52}&= \frac{4}{15} \Omega H_\text{EOB}, \\
\rho_{51}&=1+\frac{8\,\nu^2-626\,\nu+319}{390\,(2\,\nu-1)}v_\Omega^2 -\frac{31877}{304200} v_\Omega^4\,, \\
\delta_{51}&= \frac{31}{210} \Omega H_\text{EOB}, \\
\rho_{66} &= 1+\frac{273\,\nu^3-861\,\nu^2+602\,\nu-106}{84\,(5\,\nu^2-5\,\nu+1)} v_\Omega^2 - \frac{1025435}{659736} v_\Omega^4 \,, \\
\delta_{66}&= \frac{43}{70} \Omega H_\text{EOB}, \\
\rho_{65}& = 1 + \frac{220\,\nu^3-910\,\nu^2+838\,\nu-185}{144\,(3\,\nu^2-4\,\nu+1)} v_\Omega^2\,, \\
\delta_{65}&= \frac{10}{21} \Omega H_\text{EOB}, \\
\rho_{64} &= 1 + \frac{133\,\nu^3-581\,\nu^2+462\,\nu-86}{84\,(5\,\nu^2-5\,\nu+1)} v_\Omega^2 - \frac{476887}{659736} v_\Omega^4 \,, \\
\delta_{64}&= \frac{43}{105} \Omega H_\text{EOB}, \\
\rho_{63} &= 1 + \frac{156\,\nu^3-750\,\nu^2+742\,\nu-169}{144\,(3\,\nu^2-4\,\nu+1)} v_\Omega^2\,, \\
\delta_{63}&= \frac{2}{7} \Omega H_\text{EOB},\\
\rho_{62} &= 1 + \frac{49\,\nu^3-413\,\nu^2+378\,\nu-74}{84\,(5\,\nu^2-5\,\nu+1)} v_\Omega^2 - \frac{817991}{3298680} v_\Omega^4 \,, \\
\delta_{62}&= \frac{43}{210} \Omega H_\text{EOB},\\
\rho_{61} &= 1 + \frac{124\,\nu^3-670\,\nu^2+694\,\nu-161}{144\,(3\,\nu^2-4\,\nu+1)} v_\Omega^2\,, \\
\delta_{61}&= \frac{2}{21} \Omega H_\text{EOB},\\
\rho_{77} &= 1 + \frac{1380 \nu ^3-4963 \nu ^2+4246 \nu -906}{714 \left(3 \nu ^2-4 \nu +1\right)} v_{\Omega }^2 \,, \\
\delta_{77}&= \frac{19}{36} \Omega H_\text{EOB},\\
\rho_{76} &= 1 + \frac{6104 \nu ^4-29351 \nu ^3+37828 \nu^2-16185 \nu +2144} {1666 \left(7 \nu ^3-14 \nu ^2+7 \nu-1\right)} v_{\Omega }^2 \,, \\
\rho_{75} &= 1 + \frac{804 \nu ^3-3523 \nu ^2+3382 \nu -762}{714 \left(3 \nu ^2-4 \nu +1\right)} v_{\Omega }^2 \,, \\
\delta_{75}&= \frac{95}{252} \Omega H_\text{EOB},\\
\rho_{74} &= 1 + \frac{41076 \nu ^4-217959 \nu ^3+298872 \nu^2-131805 \nu +17756} {14994 \left(7 \nu ^3-14 \nu ^2+7 \nu-1\right)} v_{\Omega }^2 \,, \\
\rho_{73} &= 1 + \frac{420 \nu ^3-2563 \nu ^2+2806 \nu -666}{714 \left(3 \nu ^2-4 \nu +1\right)} v_{\Omega }^2 \,, \\
\delta_{73}&= \frac{19}{84} \Omega H_\text{EOB},\\
\rho_{72} &= 1 + \frac{32760 \nu ^4-190239 \nu ^3+273924 \nu^2-123489 \nu +16832} {14994 \left(7 \nu ^3-14 \nu ^2+7 \nu-1\right)} v_{\Omega }^2 \,, \\
\rho_{71} &= 1 + \frac{228 \nu ^3-2083 \nu ^2+2518 \nu -618}{714 \left(3 \nu ^2-4 \nu +1\right)} v_{\Omega }^2 \,, \\
\delta_{71}&= \frac{19}{252} \Omega H_\text{EOB},\\
\rho_{88} &= 1+ \frac{3482 - 26778\nu + 64659\nu^2 - 53445\nu^3 + 12243\nu^4}{2736 (-1 + 7 \nu- 14 \nu^2 + 7 \nu^3)} v_\Omega^2\,, \\
\rho_{87} &= 1+ \frac{23478 - 154099\nu + 309498\nu^2 - 207550\nu^3 + 38920\nu^4}{18240 (-1 + 6\nu - 10\nu^2 + 4\nu^3)} v_\Omega^2\,, \\
\rho_{86} &= 1+ \frac{1002 - 7498\nu + 17269\nu^2 - 13055\nu^3 + 2653\nu^4}{912(-1 + 7\nu - 14\nu^2 + 7\nu^3)} v_\Omega^2\,, \\
\rho_{85} &= 1+ \frac{4350 - 28055\nu + 54642\nu^2 - 34598\nu^3 + 6056\nu^4}{3648 (-1 + 6\nu - 10\nu^2 + 4\nu^3)} v_\Omega^2\,, \\
\rho_{84} &= 1+ \frac{2666 - 19434\nu + 42627\nu^2 - 28965\nu^3 + 4899\nu^4}{2736(-1 + 7\nu - 14\nu^2 + 7\nu^3)} v_\Omega^2\,, \\
\rho_{83} &= 1+ \frac{20598 - 131059\nu + 249018\nu^2 - 149950\nu^3 + 24520\nu^4}{18240(-1 + 6\nu - 10\nu^2 + 4\nu^3)} v_\Omega^2\,, \\
\rho_{82} &= 1+ \frac{2462 - 17598\nu + 37119\nu^2 - 22845\nu^3 + 3063\nu^4}{2736(-1 + 7\nu - 14\nu^2 + 7\nu^3)} v_\Omega^2\,, \\
\rho_{81} &= 1+ \frac{20022 - 126451\nu + 236922\nu^2 - 138430\nu^3 + 21640\nu^4}{18240(-1 + 6\nu - 10\nu^2 + 4\nu^3)} v_\Omega^2\,.
\end{align}
\end{widetext}


\section{Fits of nonquasicircular input values}

\label{appx:NQC_fits}
In this appendix we provide fits for the nonquasicircular (NQC) input values, $\left|h_{\ell m}\left(t_{\text {match }}^{\ell m}\right)\right|$,
$\partial_{t}\left|h_{\ell m}\left(t_{\text {match }}^{\ell m}\right)\right|$,
$\partial_{t}^2\left|h_{\ell m}\left(t_{\text {match }}^{\ell m}\right)\right|$,
$\omega_{\ell m}\left(t_{\text {match }}^{\ell m}\right)$,
$\partial_{t}\omega_{\ell m}\left(t_{\text {match }}^{\ell m}\right)$.
To produce the fits we used NR simulations with the highest level of resolution available and extrapolation order $N=2$.
Depending on the mode, we excluded a different number of NR waveforms from the fits, where numerical errors prevented us from fitting them accurately.
As in Ref.~\cite{Cotesta:2018fcv} we define the following combinations of $m_1$, $m_2$, $\chi_1$, $\chi_2$ to be used in the fits.
\begin{align}
\delta &= \dfrac{(m_1-m_2)}{(m_1+m_2)}, \\
\chi_{33} &=\chi_{S} \delta +\chi_{A} \\
\chi_{21 A} &=\frac{\chi_{S}}{1-1.3 \nu} \delta+\chi_{A} \\
\chi_{44 A} &=(1-5 \nu) \chi_{S}+\chi_{A} \delta \\
\chi_{21 D} &=\frac{\chi_{S}}{1-2 \nu} \delta +\chi_{A} \\
\chi_{44 D} &=(1-7 \nu) \chi_{S}+\chi_{A} \delta \\
\chi &=\chi_{S}+\chi_{A} \frac{\delta}{1-2 \nu}
\end{align}
The variables $\chi_{33}$, $\chi_{21A}$, $\chi_{21D}$ vanish by construction for equal-mass equal-spin configurations,
and are used to enforce that the odd-$m$ modes also vanish in the same limit as required by symmetry.

\subsection{Amplitude's fits}
\begin{widetext}
\begin{align}
\dfrac{|h_{22}^{\rm{NR}} (t_{22}^{\rm{match}}) |}{\nu} = &|0.430147 \chi^{3}\nu - 0.084939 \chi^{3} + 0.619889 \chi^{2}\nu^{2} - 0.020826 \chi^{2} - 13.357614 \chi\nu^{3} \nonumber \\
&+ 7.194264 \chi\nu^{2} - 1.743135 \chi\nu + 0.18694 \chi + 71.979698 \nu^{4} - 46.87586 \nu^{3} \nonumber \\
&+ 12.440405 \nu^{2} - 0.868289 \nu + 1.467097| \\
\dfrac{|h_{33}^{\rm{NR}} (t_{33}^{\rm{match}}) |}{\nu} = &|-0.088371 \chi_{33}^{2}\delta\nu + 0.036258 \chi_{33}^{2}\delta + 1.057731 \chi_{33}\nu^{2} - 0.466709 \chi_{33}\nu \nonumber \\
&+ 0.099543 \chi_{33} + 1.96267 \delta\nu^{2} + 0.027833 \delta\nu + 0.558808 \delta| \\
\dfrac{|h_{21}^{\rm{NR}} (t_{21}^{\rm{match}}) |}{\nu} = &|-0.033175 \chi_{21A}^{3}\delta + 0.086356 \chi_{21A}^{2}\delta\nu - 0.049897 \chi_{21A}^{2}\delta + 0.012706 \chi_{21A}\delta \nonumber \\
&+ 0.168668 \chi_{21A}\nu - 0.285597 \chi_{21A} + 1.067921 \delta\nu^{2} - 0.189346 \delta\nu + 0.431426 \delta| \\
\dfrac{|h_{44}^{\rm{NR}} (t_{44}^{\rm{match}}) |}{\nu} = &|0.031483 \chi_{44A}^{2} - 0.180165 \chi_{44A}\nu + 0.063931 \chi_{44A} + 6.239418 \nu^{3} - 1.947473 \nu^{2} \nonumber \\
&- 0.615307 \nu + 0.262533| \\
\dfrac{|h_{55}^{\rm{NR}} (t_{55}^{\rm{match}}) |}{\nu} = &|-7.402839 \chi_{33}\nu^{3} + 3.965852 \chi_{33}\nu^{2} - 0.762776 \chi_{33}\nu + 0.062757 \chi_{33} \nonumber \\
&+ 1.093812 \delta\nu^{2} - 0.462142 \delta\nu + 0.125468 \delta| \\
\dfrac{|h_{32}^{\rm{NR}} (t_{32}^{\rm{match}}) |}{\nu} = &|0.022598 \chi^{2} + 0.307803 \chi\nu - 0.020771 \chi + 8.917771 \nu^{3} - 2.194506 \nu^{2} \nonumber \\
&- 0.387911 \nu + 0.155446| \\
\dfrac{|h_{43}^{\rm{NR}} (t_{43}^{\rm{match}}) |}{\nu} = &|-0.071554 \chi_{33}^{2}\delta\nu + 0.021932 \chi_{33}^{2}\delta - 1.738079 \chi_{33}\nu^{2} + 0.436576 \chi_{33}\nu \nonumber \\
&- 0.020081 \chi_{33} + 0.809615 \delta\nu^{2} - 0.273364 \delta\nu + 0.07442 \delta|
\end{align}
\end{widetext}

\subsection{Amplitude-first-derivative's fits}
\begin{widetext}
\begin{align}
\dfrac{1}{\nu} \dfrac{d |h_{22}^{\rm{NR}} (t)|} {dt}\bigg|_{t=t^{\rm{match}}_{22}} \equiv &0 \\
\dfrac{1}{\nu} \dfrac{d |h_{33}^{\rm{NR}} (t)|} {dt}\bigg|_{t=t^{\rm{match}}_{33}} = &\chi_{33}^{2} \delta \left(0.004941 \nu - 0.002094\right) \nonumber \\
&+ 0.001781 \left|{\chi_{33}^{2} + \chi_{33} \delta \left(39.247538 \nu - 2.986889\right) + \delta^{2} \left(85.173306 \nu + 4.637906\right)}\right|^{1/2} \\
\dfrac{1}{\nu} \dfrac{d |h_{21}^{\rm{NR}} (t)|} {dt}\bigg|_{t=t^{\rm{match}}_{21}} = &\chi_{21D} \delta \left(0.023534 \nu - 0.008064\right) + \delta \left(0.006743 - 0.0297 \nu\right) \nonumber \\
&+ 0.008256 \left|{\chi_{21D} - \delta \left(5.471011 \nu^{2} + 1.235589 \nu + 0.815482\right)}\right| \\
\dfrac{1}{\nu} \dfrac{d |h_{44}^{\rm{NR}} (t)|} {dt}\bigg|_{t=t^{\rm{match}}_{44}} = &-0.001251 \chi_{44D}^{3} + 0.006387 \chi_{44D}^{2}\nu - 0.001223 \chi_{44D}^{2} - 0.034308 \chi_{44D}\nu^{2} \nonumber \\
&+ 0.014373 \chi_{44D}\nu - 0.000681 \chi_{44D} + 1.134679 \nu^{3} - 0.417056 \nu^{2} \nonumber \\
&+ 0.024004 \nu + 0.003498 \\
\dfrac{1}{\nu} \dfrac{d |h_{55}^{\rm{NR}} (t)|} {dt}\bigg|_{t=t^{\rm{match}}_{55}} = &\chi_{33}^{2} \delta \left(0.008568 \nu - 0.00155\right) + \chi_{33} \delta \left(0.002705 \nu - 0.001015\right) \nonumber \\
&+ \delta \left(0.002563 - 0.010891 \nu\right) + 0.000284 \left|{\chi_{33} + \delta \left(32.459725 \nu + 0.165336\right)}\right| \\
\dfrac{1}{\nu} \dfrac{d |h_{32}^{\rm{NR}} (t)|} {dt}\bigg|_{t=t^{\rm{match}}_{32}} = &-0.000806 \chi^{3} - 0.011027 \chi^{2}\nu + 0.002999 \chi^{2} - 0.14087 \chi\nu^{2} + 0.063211 \chi\nu \nonumber \\
&- 0.006783 \chi + 1.693423 \nu^{3} - 0.510999 \nu^{2} + 0.020607 \nu + 0.003674 \\
\dfrac{1}{\nu} \dfrac{d |h_{43}^{\rm{NR}} (t)|} {dt}\bigg|_{t=t^{\rm{match}}_{43}} = &\chi_{33}^{2} \delta \left(0.001773 - 0.012159 \nu\right) + \chi_{33} \delta \left(0.022249 \nu - 0.004295\right) \nonumber \\
&+ \delta \left(0.012043 \nu - 0.001067\right) + 0.00082 \left|{\chi_{33} + \delta \left(3.880171 - 20.015436 \nu\right)}\right|
\end{align}
\end{widetext}

\subsection{Amplitude-second-derivative's fits}
\begin{widetext}
\begin{align}
\dfrac{1}{\nu} \dfrac{d^2 |h_{22}^{\rm{NR}} (t)|} {dt^2}\bigg|_{t=t^{\rm{match}}_{22}} = &0.000386 \chi^{2} + 0.003589 \chi\nu + 0.001326 \chi - 0.003353 \nu^{2} - 0.005615 \nu - 0.002457 \\
\dfrac{1}{\nu} \dfrac{d^2 |h_{33}^{\rm{NR}} (t)|} {dt^2}\bigg|_{t=t^{\rm{match}}_{33}} = &\chi_{33} \delta \left(0.000552 \nu + 0.001029\right) - 0.000218 \nonumber \\
& \cdot \left|{\chi_{33} + \delta \left(- 2188.340923 \nu^{4} + 1331.981345 \nu^{3} - 289.772357 \nu^{2} + 32.212775 \nu + 3.396168\right)}\right|\\
\dfrac{1}{\nu} \dfrac{d^2 |h_{21}^{\rm{NR}} (t)|} {dt^2}\bigg|_{t=t^{\rm{match}}_{21}} = &0.00015 \delta - \left|0.000316 \chi_{21D}^{3} - \chi_{21D}^{2} \delta \left(- 0.043291 \nu^{2} + 0.005682 \nu + 0.000502\right) \right. \nonumber \\
&\left.+ 0.000372 \chi_{21D} \delta - \delta \left(0.003643 \nu + 2.8 \cdot 10^{-5}\right)\right| \\
\dfrac{1}{\nu} \dfrac{d^2 |h_{44}^{\rm{NR}} (t)|} {dt^2}\bigg|_{t=t^{\rm{match}}_{44}} = &-0.000591 \chi^{2}\nu + 0.000174 \chi^{2} - 0.000501 \chi\nu + 0.000318 \chi + 0.138496 \nu^{3} \nonumber \\
&- 0.047008 \nu^{2} + 0.003899 \nu - 0.000451 \\
\dfrac{1}{\nu} \dfrac{d^2 |h_{55}^{\rm{NR}} (t)|} {dt^2}\bigg|_{t=t^{\rm{match}}_{55}} = &\chi_{33}^{2} \cdot \left(0.000278 \nu - 5.6 \cdot 10^{-5}\right) + \chi_{33} \delta \left(0.000246 \nu - 6.8 \cdot 10^{-5}\right) + \delta \left(0.000118 - 5.9 \cdot 10^{-5} \nu\right) \\
\dfrac{1}{\nu} \dfrac{d^2 |h_{32}^{\rm{NR}} (t)|} {dt^2}\bigg|_{t=t^{\rm{match}}_{32}} = &-0.002882 \chi^{2}\nu + 0.000707 \chi^{2} - 0.027461 \chi\nu^{2} + 0.008481 \chi\nu - 0.000691 \chi \nonumber \\
&+ 0.20836 \nu^{3} - 0.053191 \nu^{2} + 0.001604 \nu - 5.6 \cdot 10^{-5}\\
\dfrac{1}{\nu} \dfrac{d^2 |h_{43}^{\rm{NR}} (t)|} {dt^2}\bigg|_{t=t^{\rm{match}}_{43}} = &\chi_{33} \delta \left(0.00291 \nu - 0.000348\right) - 5.0 \cdot 10^{-6} \nonumber \\
&\cdot \left|{\chi_{33} + \delta \left(- 25646.358742 \nu^{4} + 12647.805787 \nu^{3} + 291.751053 \nu^{2} - 531.965263 \nu + 23.849357\right)}\right|&
\end{align}
\end{widetext}

\subsection{Frequency and frequency-derivative fits}
\begin{widetext}
\begin{align}
\omega_{22}^{\rm{NR}}(t_{22}^{\rm{match}}) = &-0.015259 \chi^{4} + 0.241948 \chi^{3}\nu - 0.066927 \chi^{3} - 0.971409 \chi^{2}\nu^{2} + 0.518014 \chi^{2}\nu \nonumber\\
&- 0.087152 \chi^{2} + 3.751456 \chi\nu^{3} - 1.697343 \chi\nu^{2} + 0.250965 \chi\nu - 0.091339 \chi \nonumber \\
&+ 5.893523 \nu^{4} - 3.349305 \nu^{3} + 0.285392 \nu^{2} - 0.317096 \nu - 0.268541\\
\omega_{33}^{\rm{NR}}(t_{33}^{\rm{match}}) = &-0.045141 \chi^{3} + 0.346675 \chi^{2}\nu - 0.119419 \chi^{2} - 0.745924 \chi\nu^{2} + 0.478915 \chi\nu \nonumber \\
&- 0.17467 \chi + 8.887163 \nu^{3} - 4.226831 \nu^{2} - 0.427167\\
\omega_{21}^{\rm{NR}}(t_{21}^{\rm{match}}) = &-0.01009 \chi^{3} + 0.077343 \chi^{2}\nu - 0.02411 \chi^{2} - 0.168854 \chi\nu^{2} + 0.159382 \chi\nu \nonumber \\
&- 0.047635 \chi - 1.965157 \nu^{3} + 0.53085 \nu^{2} - 0.237904 \nu - 0.176526\\
\omega_{44}^{\rm{NR}}(t_{44}^{\rm{match}}) = &-0.042529 \chi^{3} + 0.415864 \chi^{2}\nu - 0.155222 \chi^{2} - 0.768712 \chi\nu^{2} + 0.592568 \chi\nu \nonumber \\
&- 0.244508 \chi + 13.651335 \nu^{3} - 5.490329 \nu^{2} - 0.574041\\
\omega_{55}^{\rm{NR}}(t_{55}^{\rm{match}}) = &-0.091629 \chi^{3} + 0.802759 \chi^{2}\nu - 0.246646 \chi^{2} - 3.04576 \chi\nu^{2} + 1.43471 \chi\nu \nonumber \\
&- 0.329591 \chi + 13.81386 \nu^{3} - 6.61611 \nu^{2} + 0.472474 \nu - 0.589341\\
\omega_{32}^{\rm{NR}}(t_{32}^{\rm{match}}) = &-0.045647 \chi^{2} - 2.758635 \chi\nu^{2} + 0.811353 \chi\nu - 0.112477 \chi - 2.346024 \nu^{3} \nonumber \\
&+ 1.57986 \nu^{2} - 0.317756 \nu - 0.331141\\
\omega_{43}^{\rm{NR}}(t_{43}^{\rm{match}}) = &-0.037919 \chi^{3} + 0.226903 \chi^{2}\nu - 0.087288 \chi^{2} - 0.905919 \chi\nu^{2} + 0.291092 \chi\nu \nonumber \\
&- 0.1198 \chi - 55.534105 \nu^{3} + 23.913277 \nu^{2} - 3.487986 \nu - 0.34306\\
\dot{\omega}_{22}^{\rm{NR}}(t_{22}^{\rm{match}}) = &0.000614 \chi^{3} - 0.008393 \chi^{2}\nu + 0.001948 \chi^{2} + 0.07799 \chi\nu^{2} - 0.028772 \chi\nu \nonumber \\
&+ 0.001705 \chi - 0.237126 \nu^{3} + 0.092215 \nu^{2} - 0.03104 \nu - 0.005484\\
\dot{\omega}_{33}^{\rm{NR}}(t_{33}^{\rm{match}}) = &0.001697 \chi^{3} - 0.016231 \chi^{2}\nu + 0.003985 \chi^{2} + 0.154378 \chi\nu^{2} - 0.050618 \chi\nu \nonumber \\
&+ 0.002721 \chi + 0.255402 \nu^{3} - 0.08663 \nu^{2} - 0.027405 \nu - 0.009736 \\
\dot{\omega}_{21}^{\rm{NR}}(t_{21}^{\rm{match}}) = &0.00149 \chi^{3} - 0.008965 \chi^{2}\nu + 0.002739 \chi^{2} + 0.033831 \chi\nu^{2} - 0.005752 \chi\nu \nonumber \\
&+ 0.002003 \chi - 0.204368 \nu^{3} + 0.120705 \nu^{2} - 0.035144 \nu - 0.006579\\
\dot{\omega}_{44}^{\rm{NR}}(t_{44}^{\rm{match}}) = &0.001812 \chi^{3} - 0.024687 \chi^{2}\nu + 0.00568 \chi^{2} + 0.162693 \chi\nu^{2} - 0.061205 \chi\nu \nonumber \\
&+ 0.003623 \chi + 0.536664 \nu^{3} - 0.094797 \nu^{2} - 0.045406 \nu - 0.013038\\
\dot{\omega}_{55}^{\rm{NR}}(t_{55}^{\rm{match}}) = &0.001509 \chi^{3} - 0.01547 \chi^{2}\nu + 0.002802 \chi^{2} + 0.164011 \chi\nu^{2} - 0.056516 \chi\nu \nonumber \\
&+ 0.002072 \chi + 0.043963 \nu^{3} + 0.048045 \nu^{2} - 0.045197 \nu - 0.008688\\
\dot{\omega}_{32}^{\rm{NR}}(t_{32}^{\rm{match}}) = &-0.036711 \chi^{2}\nu + 0.005532 \chi^{2} + 0.09192 \chi\nu^{2} - 0.030713 \chi\nu + 0.005927 \chi \nonumber \\
&- 2.494788 \nu^{3} + 0.995116 \nu^{2} - 0.10163 \nu - 0.010763\\
\dot{\omega}_{43}^{\rm{NR}}(t_{43}^{\rm{match}}) = &0.000537 \chi^{3} - 0.009876 \chi^{2}\nu + 0.003279 \chi^{2} + 0.13296 \chi\nu^{2} - 0.060884 \chi\nu \nonumber \\
&+ 0.008513 \chi - 5.160613 \nu^{3} + 2.180781 \nu^{2} - 0.292607 \nu - 0.005308
\end{align}
\end{widetext}

\section{Fits for amplitude and phase of merger-ringdown model}
\label{appx:coeff_fits}
In this appendix we provide fits across parameter space for the free coefficients in the merger-ringdown ansatz given by Eqs.~(\ref{eq:ansatz_a}) and (\ref{eq:ansatz_phi}).
To produce the fits we use NR simulations with the highest level of resolution available and extrapolation order $N=2$. They read:
\begin{widetext}
\begin{align}
c_{1,f}^{22} = &- 0.001777 \chi^{4} + 0.062842 \chi^{3} \nu - 0.018908 \chi^{3} + 0.013161 \chi^{2} \nu^{2} + 0.049388 \chi^{2} \nu \nonumber \\
&- 0.019314 \chi^{2} + 1.867978 \chi \nu^{3} - 0.702488 \chi \nu^{2} + 0.033885 \chi \nu - 0.011612 \chi \nonumber \\
&- 4.238246 \nu^{4} + 2.043712 \nu^{3} - 0.406992 \nu^{2} + 0.053589 \nu + 0.086254 \\
c_{2,f}^{22} = &1.021875 \chi^{3}\nu - 0.20348 \chi^{3} - 3.556173 \chi^{2}\nu^{2} + 1.970082 \chi^{2}\nu - 0.264297 \chi^{2} \nonumber \\
&+ 2.002947 \chi\nu^{3} - 5.585851 \chi\nu^{2} + 1.837724 \chi\nu - 0.27076 \chi - 63.286459 \nu^{4} \nonumber \\
&+ 44.331389 \nu^{3} - 9.529573 \nu^{2} + 1.155695 \nu - 0.528763 \\
d_{1,f}^{22} = &-0.013321 \chi^{4} + 0.047305 \chi^{3}\nu - 0.024203 \chi^{3} + 1.033352 \chi^{2}\nu^{2} - 0.254351 \chi^{2}\nu \nonumber \\
&- 0.007847 \chi^{2} + 4.113463 \chi\nu^{3} - 1.652924 \chi\nu^{2} + 0.090834 \chi\nu - 28.423701 \nu^{4} \nonumber \\
&+ 20.719874 \nu^{3} - 6.075679 \nu^{2} + 0.780093 \nu + 0.135758 \\
d_{2,f}^{22} = &\exp(-0.163113 \chi^{4} - 3.398858 \chi^{3}\nu + 0.728816 \chi^{3} + 23.975132 \chi^{2}\nu^{2} - 10.064954 \chi^{2}\nu \nonumber \\
&+ 1.2115 \chi^{2} + 9.057306 \chi\nu^{3} - 5.268296 \chi\nu^{2} + 0.464553 \chi\nu + 0.56269 \chi \nonumber\\
&- 352.249383 \nu^{4} + 275.843499 \nu^{3} - 81.483314 \nu^{2} + 11.184576 \nu + 0.03571)\\
c_{1,f}^{33} = &-0.00956 \chi^{3} + 0.029459 \chi^{2}\nu - 0.020264 \chi^{2} - 0.494524 \chi\nu^{2} + 0.169463 \chi\nu \nonumber \\
&- 0.026285 \chi - 5.847417 \nu^{3} + 1.957462 \nu^{2} - 0.171682 \nu + 0.093539 \\
c_{2,f}^{33} = &- 0.057346 \chi^{3} + 0.237107 \chi^{2} \nu - 0.094285 \chi^{2} - 4.250609 \chi \nu^{2} + 1.763105 \chi \nu \nonumber \\
&- 0.315826 \chi + 14.801916 \nu^{3} - 7.060581 \nu^{2} + 1.158627 \nu - 0.646888 \\
d_{1,f}^{33} = &-0.016524 \chi^{3} + 0.221466 \chi^{2}\nu - 0.066323 \chi^{2} + 0.678442 \chi\nu^{2} - 0.261264 \chi\nu \nonumber \\
&+ 0.006664 \chi + 2.316434 \nu^{3} - 2.192227 \nu^{2} + 0.424582 \nu + 0.161577 \\
d_{2,f}^{33} = &\exp(0.275999 \chi^{3} - 1.830695 \chi^{2}\nu + 0.512734 \chi^{2} + 29.072515 \chi\nu^{2} - 10.581319 \chi\nu \nonumber \\
&+ 1.310643 \chi + 324.310223 \nu^{3} - 124.681881 \nu^{2} + 13.200426 \nu + 0.410855) \\
c_{1,f}^{21} = &0.173462 \chi^{2}\nu - 0.028873 \chi^{2} + 0.197467 \chi\nu^{2} - 0.026139 \chi - 2.934735 \nu^{3} \nonumber \\
&+ 1.009106 \nu^{2} - 0.112721 \nu + 0.099889\\
c_{2,f}^{21} = &0.183489 \chi^{3} + 0.10573 \chi^{2} - 20.792825 \chi\nu^{2} + 6.867746 \chi\nu - 0.484948 \chi \nonumber \\
&- 54.917585 \nu^{3} + 16.466312 \nu^{2} + 0.426316 \nu - 0.92208 \\
d_{1,f}^{21} = &0.018467 \chi^{4} + 0.398621 \chi^{3}\nu - 0.050499 \chi^{3} - 0.877201 \chi^{2}\nu^{2} + 0.414553 \chi^{2}\nu \nonumber \\
&- 0.068277 \chi^{2} - 10.648526 \chi\nu^{3} + 4.104918 \chi\nu^{2} - 0.723576 \chi\nu + 0.039227 \chi \nonumber \\
&+ 42.715534 \nu^{4} - 18.280603 \nu^{3} + 2.236592 \nu^{2} - 0.048094 \nu + 0.16335 \\
d_{2,f}^{21} = &\exp(0.814085 \chi^{3} - 1.197363 \chi^{2}\nu + 0.560622 \chi^{2} + 6.44667 \chi\nu^{2} - 5.630563 \chi\nu \nonumber \\
&+ 0.949586 \chi + 91.269183 \nu^{3} - 27.329751 \nu^{2} + 1.101262 \nu + 1.040761) \\
c_{1,f}^{44} = &4.519504 \chi\nu^{2} - 1.489036 \chi\nu + 0.068403 \chi - 1656.065439 \nu^{4} + 817.835726 \nu^{3} \nonumber \\
&- 127.055379 \nu^{2} + 6.921968 \nu + 0.009386\\
c_{2,f}^{44} = &0.964861 \chi^{3}\nu - 0.185226 \chi^{3} - 12.647814 \chi^{2}\nu^{2} + 5.264969 \chi^{2}\nu - 0.539721 \chi^{2} \nonumber \\
&- 254.719552 \chi\nu^{3} + 105.698791 \chi\nu^{2} - 12.107281 \chi\nu + 0.2244 \chi - 393.727702 \nu^{4} \nonumber \\
&+ 145.32788 \nu^{3} - 15.556222 \nu^{2} + 1.592449 \nu - 0.677664 \\
d_{1,f}^{44} = &-0.020644 \chi^{3} + 0.494221 \chi^{2}\nu - 0.127074 \chi^{2} + 4.297985 \chi\nu^{2} - 1.284386 \chi\nu \nonumber \\
&+ 0.062684 \chi - 44.280815 \nu^{3} + 11.021482 \nu^{2} - 0.162943 \nu + 0.166018 \\
d_{2,f}^{44} = &\exp(37.735116 \chi\nu^{2} - 12.516669 \chi\nu + 1.309868 \chi - 528.368915 \nu^{3} + 155.115196 \nu^{2} \nonumber \\
&- 6.612448 \nu + 0.787726) \\
c_{1,f}^{55} = &-0.009957 \chi^{3} + 0.059748 \chi^{2}\nu - 0.02146 \chi^{2} - 0.206811 \chi\nu^{2} + 0.055078 \chi\nu \nonumber \\
&- 0.014528 \chi - 5.966891 \nu^{3} + 1.76928 \nu^{2} - 0.055272 \nu + 0.080368 \\
c_{2,f}^{55} =  &0.119703 \chi^{4} + 1.638345 \chi^{2}\nu^{2} - 0.064725 \chi^{2} - 28.499278 \chi\nu^{3} + 3.73034 \chi\nu^{2} \nonumber \\
&+ 1.853723 \chi\nu - 0.225283 \chi - 1887.591102 \nu^{4} + 794.134711 \nu^{3} - 107.010824 \nu^{2} \nonumber \\
&+ 6.32117 \nu - 1.507483 \\
d_{1,f}^{55} = &-0.021537 \chi^{3} + 0.168071 \chi^{2}\nu - 0.050263 \chi^{2} + 0.871799 \chi\nu^{2} - 0.230057 \chi\nu \nonumber \\
&+ 9.018546 \nu^{3} - 5.009488 \nu^{2} + 0.606313 \nu + 0.150622 \\
d_{2,f}^{55} = &\exp( 28.839035 \chi\nu^{2} - 9.726025 \chi\nu + 0.901423 \chi + 143.745208 \nu^{3} - 64.478227 \nu^{2} \nonumber \\
&+ 6.223833 \nu + 2.058139) \\
c_{1,f}^{32} = &-0.133035 \chi^{3} + 0.641681 \chi^{2}\nu - 0.111865 \chi^{2} + 8.987763 \chi\nu^{2} - 1.582259 \chi\nu \nonumber \\
&+ 0.095604 \chi - 26.991806 \nu^{3} + 13.716801 \nu^{2} - 1.63083 \nu + 0.157543 \\
c_{2,f}^{32} = &0.121608 \chi^{3} - 1.590623 \chi^{2}\nu + 0.167231 \chi^{2} - 25.544931 \chi\nu^{2} + 10.127968 \chi\nu \nonumber \\
&- 0.999062 \chi - 51.469773 \nu^{3} + 46.209833 \nu^{2} - 6.484571 \nu - 0.716883 \\
d_{1,f}^{32} = &\exp( -0.764015 \chi^{3} - 8.684722 \chi^{2}\nu + 0.691946 \chi^{2} - 0.518291 \chi\nu^{2} - 1.407934 \chi\nu \nonumber \\
&+ 0.236427 \chi + 81.222175 \nu^{3} - 18.040529 \nu^{2} + 2.216406 \nu - 1.879455)\\
d_{2,f}^{32} = &\exp(-1.819822 \chi^{3} - 24.501503 \chi^{2}\nu + 3.287882 \chi^{2} - 39.324579 \chi\nu^{2} + 14.379901 \chi\nu \nonumber \\
&- 215.372399 \nu^{3} + 136.20936 \nu^{2} - 16.842816 \nu + 1.463485) \\
c_{1,f}^{43} = &0.041585 \chi^{3} + 4.188908 \chi\nu^{2} - 1.365732 \chi\nu + 0.058908 \chi + 44.311948 \nu^{3} \nonumber \\
&- 22.114177 \nu^{2} + 3.386082 \nu - 0.035315 \\
c_{2,f}^{43} = &0.125764 \chi^{3} + 0.337235 \chi^{2}\nu + 0.146202 \chi^{2} - 9.803187 \chi\nu^{2} + 3.995199 \chi\nu \nonumber \\
&- 0.240976 \chi - 57.968821 \nu^{3} + 7.820929 \nu^{2} + 3.364741 \nu - 1.121716 \\
d_{1,f}^{43} = &\exp( -0.888286 \chi^{3} + 3.97869 \chi^{2}\nu - 1.047181 \chi^{2} - 14.823391 \chi\nu^{2} + 6.940856 \chi\nu \nonumber \\
&- 0.367801 \chi + 366.645645 \nu^{3} - 161.732513 \nu^{2} + 19.564699 \nu - 2.29578)\\
d_{2,f}^{43} = &\exp(-0.950676 \chi^{3} - 0.31428 \chi^{2} + 39.21796 \chi\nu^{2} - 10.651167 \chi\nu + 1.339732 \chi \nonumber \\
&+ 730.42296 \nu^{3} - 312.960598 \nu^{2} + 37.402567 \nu - 0.061894)
\end{align}
\end{widetext}

\section{Robustness of the calibration pipeline}
\label{appx:robustness}

In this appendix we demonstrate that the calibration pipeline described in Sec. \ref{sec:calibration} is robust with regard 
to the number of NR waveforms and the PSD used in the calibration likelihood. 

For the first point, we repeat the procedure for the aligned-spin calibration parameters $\boldsymbol{\theta}_{\rm{S}} \equiv \{d_{\rm{SO}}, \Delta t^{22}_{\rm{ISCO, S}}\}$, by 
using a representative subset of 119 aligned-spin NR simulations, selected with a greedy algorithm following Ref.~\cite{Field:2013cfa}. We do not change the nonspinning fits for $\boldsymbol{\theta}_{\rm{noS}} \equiv \{ a_6, \Delta t^{22}_{\rm{ISCO, noS}}\}$ 
(which already used a subset of 18 simulations out of 39), and the fits for the merger-ringdown and NQC corrections. 
This brings the total number of waveforms used to 137, which is comparable to the 141 used in the calibration of \texttt{SEOBNRv4}.
As in Sec.~\ref{sec:performance}, we compute the (2,2)-mode mismatch of this model against the entire set of 442 cases summarized in Sec.~\ref{sec:calibration}. 
\begin{figure}
	\includegraphics[width=\linewidth]{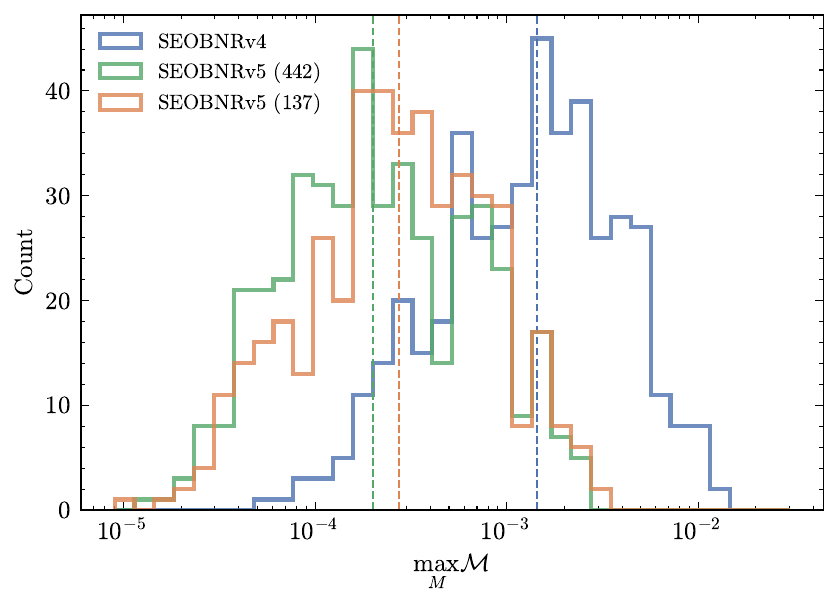}
	\caption{Histogram of the maximum (2,2)-mode mismatch over a range of total masses between 10 and 300 $M_{\odot}$, 
	between the 442 NR simulations used in this work and \texttt{SEOBNRv4} (blue), 
	\texttt{SEOBNRv5} calibrated to 442 NR simulations (green) and \texttt{SEOBNRv5} 
	calibrated to 137 NR simulations (orange). The vertical dashed lines show the medians.}
	\label{fig:mismatch_hist_137}
\end{figure}
Figure~\ref{fig:mismatch_hist_137} shows the maximum mismatch across a range of total masses between $[10, 300] M_{\odot}$ 
for \texttt{SEOBNRv5} calibrated to 442 simulations, \texttt{SEOBNRv5} calibrated to 137 simulations, and \texttt{SEOBNRv4}.
The median (dashed vertical line) goes only from $1.99 \times 10^{-4}$ for \texttt{SEOBNRv5} calibrated to 442 simulations,
to $2.74 \times 10^{-4}$ for \texttt{SEOBNRv5} calibrated to 137 simulations, 
which is more than $5$ times smaller than \texttt{SEOBNRv4} ($1.44 \times 10^{-3}$). 
Moreover for \texttt{SEOBNRv5} calibrated to 137 simulations there is only a single case with mismatch just above $0.003$ 
(\texttt{BFI:q8-7d:0080}, with parameters $(q, \chi_1, \chi_2) = (8.0, 0.0, -0.8)$). 
As already pointed out in Sec.~\ref{sec:performance}, 
even the default \texttt{SEOBNRv5} model can have mismatches slightly above $10^{-3}$ against similar
cases, with high mass ratio, small $a_{+}$, but large individual spins,
as the calibration term $\propto a_{+} d_{\rm{SO}}$ does not tend to be very effective.
This shows that the calibration pipeline does not rely on using the entire NR dataset described in Sec.~\ref{sec:calibration}. Although we 
added some critical new NR simulations, especially for high mass ratios, the improvement with respect to \texttt{SEOBNRv4} is also largely due to the improved analytical 
prescriptions, for the waveform modes, RR force and Hamiltonian, and the more effective calibration procedure.

The calibration likelihood of Eq.~(\ref{eq:calib_likelihood}) also depends on the Advanced LIGO \cite{Barsotti:2018} PSD. 
To show that our calibration pipeline is robust with respect to changes in the shape of the PSD used, we  
compute again (2,2)-mode mismatches against NR simulations (as this is the metric used in the likelihood) using a white noise curve, the Einsten Telescope \cite{Hild:2010id} and Cosmic Explorer \cite{Evans:2021gyd} PSDs. 
For this purpose we use the original fits of $\{d_{\rm{SO}}, \Delta t^{22}_{\rm{ISCO, S}}\}$ given in Eqs.~(\ref{eq:dt_fit_s}) and (\ref{eq:dSO_fit}).
\begin{figure}
	\includegraphics[width=\linewidth]{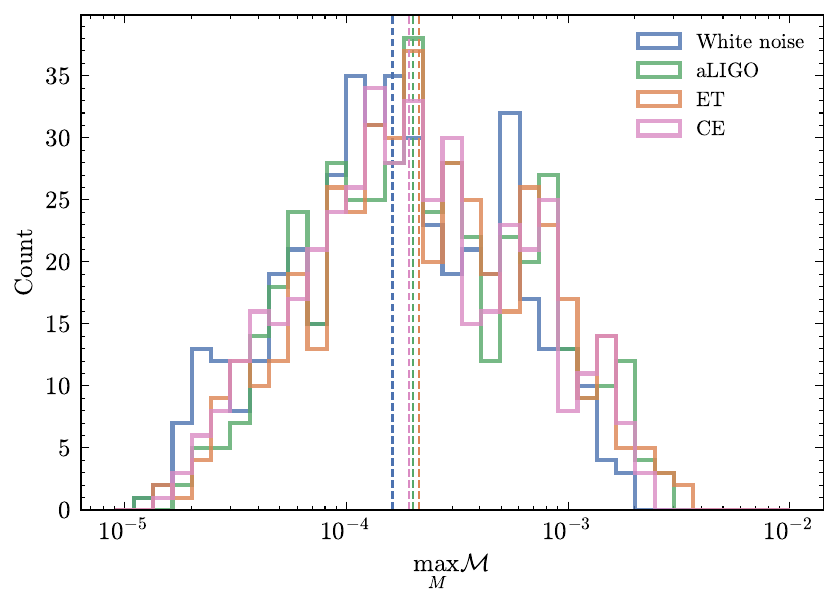}
	\caption{Histogram of the maximum (2,2)-mode mismatch over a range of total masses between 10 and 300 $M_{\odot}$, 
	between the 442 NR simulations used in this work and \texttt{SEOBNRv5}, using  
	a white noise curve and the PSDs of Advanced LIGO, Einstein Telescope and Cosmic Explorer. The vertical dashed lines show the medians. The (2,2)-mode 
	of \texttt{SEOBNRv5} is calibrated using the Advanced LIGO PSD, but performs as well using different noise curves.}
	\label{fig:mismatch_hist_ETCE}
\end{figure}
Figure~\ref{fig:mismatch_hist_ETCE} shows again the maximum mismatch, across a range of total masses between $[10, 300] M_{\odot}$,  
of \texttt{SEOBNRv5} against the 442 NR simulations used in this work, using 
a white noise curve and the PSDs of Advanced LIGO, Einstein Telescope and Cosmic Explorer. 
We see that the result is very similar for all the cases and, despite \texttt{SEOBNRv5} being calibrated 
using the Advanced LIGO PSD, it performs equally well using different noise curves.

\section{Impact of NQC corrections in the radiation-reaction force}
\label{appx:NQC_flux}

In Sec.~\ref{subsec:hlm_inspiral_plunge} we pointed out that we do not include the NQC corrections in the \texttt{SEOBNRv5} RR force. 
Recently, Refs.~\cite{Riemenschneider:2021ppj,Albertini:2021tbt} implemented a fast prescription in \texttt{TEOBResumS} to include 
fits of NQC corrections in both the waveform and RR force, without requiring an iterative procedure. 
A similar prescription could also be used in \texttt{SEOBNRv5HM}. Reference~\cite{VandeMeentv5} tested it 
in the nonspinning limit, finding that the inclusion of NQCs corrections has a smaller effect than the calibration to 2GSF data
in bringing the angular-momentum flux closer to the NR's one, except in the last fraction of a GW cycle before merger. 
Moreover, Ref.~\cite{VandeMeentv5} found that the inclusion of the nonspinning NQC corrections has a negligible effect on the 
waveform after recalibrating the conservative dynamics to NR, as the degeneracy between changes in the flux and in the 
Hamiltonian reabsorbs any imperfection in the flux with the calibration.
The nonspinning limit of the model is however already very close to the NR error, 
and the effect on more challenging aligned-spin cases could be larger.

While for the current work we did not do a systematic study in the entire aligned-spin parameter space 
to include the NQC corrections in the RR force, in this Appendix we try to understand the potential improvement 
for one binary configuration and compare the results with \texttt{SXS:BBH:1432}, having 
parameters $(q, \chi_1, \chi_2) = (5.839, 0.658, 0.793)$. Thus, we iteratively include the NQCs in the RR force 
and repeat the nested sampling procedure detailed in Sec.\ref{subsec:nested_sampling} to find new values for the 
aligned-spin calibration parameters $\{\Delta t^{22}_{\rm{ISCO, S}}, d_{\rm{SO}}\}$, 
which are going to slightly change compared to the default ones, given the different dissipative dynamics. 
To do a comparison without performing fits of the NQCs and of the calibration parameters across parameter space, 
we directly compare the maximum likelihood points of the calibration posteriors, 
corresponding to the values of the parameters giving the best unfaithfulness and time to merger. 

\begin{figure}
	\includegraphics[width=\linewidth]{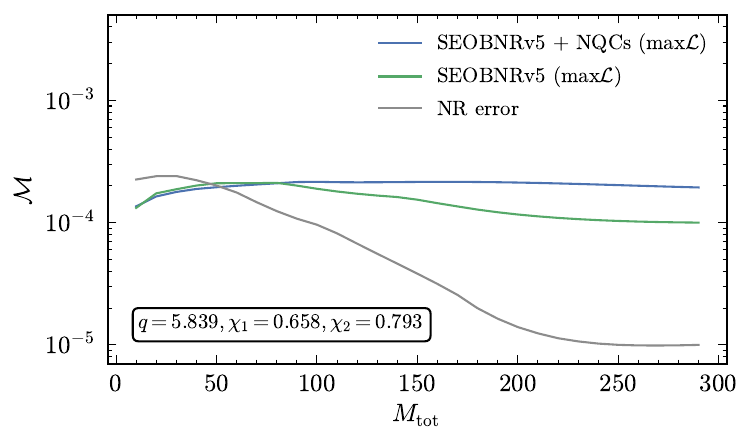}
	\caption{$(2,2)$-mode mode mismatch against the NR simulation \texttt{SXS:BBH:1432}
	for the maximum likelihood points of the \texttt{SEOBNRv5} calibration posteriors, with and without NQCs in the RR force.  
	The unfaithfulness is very similar, as the calibration of the Hamiltonian reabsorbs any difference in the dissipative dynamics.
	We also show an estimate of the NR error obtained by comparing simulations of different resolutions.}
	\label{fig:mm_comparison_NQC}
\end{figure}
\begin{figure}
	\includegraphics[width=\linewidth]{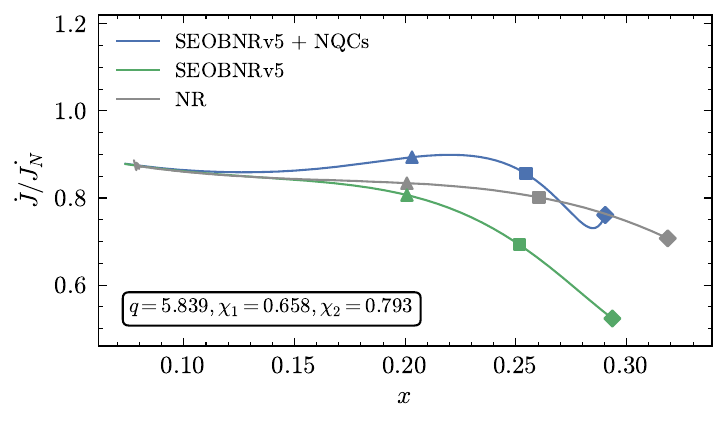}
	\caption{Comparison of the Newtonian normalized angular-momentum flux between \texttt{SEOBNRv5} with and without NQCs in the radiation-reaction, and
	the NR simulation \texttt{SXS:BBH:1432}. The triangle, square and diamond correspond respectively to 
	3, 1 and 0 cycles before merger, which for each model is taken as the peak of the (2,2)-mode amplitude. 
	\texttt{SEOBNRv5} with NQCs matches NR at merger as expected, but does not agree with NR as well as \texttt{SEOBNRv5} without NQCs at low frequencies.}
	\label{fig:flux_comparison_NQC}
\end{figure}

Figure~\ref{fig:mm_comparison_NQC} shows the $(2,2)$-mode mismatch
against NR for the maximum likelihood points of the \texttt{SEOBNRv5}
calibration posteriors with and without NQCs in the RR force, together
with an estimate of the NR error obtained by comparing different
resolutions.  The unfaithfulness is very similar, as the calibration
of the Hamiltonian reabsorbs any difference in the dissipative
dynamics.  Moreover, in Fig.~\ref{fig:flux_comparison_NQC} we compare to NR the
Newtonian normalized angular-momentum flux between \texttt{SEOBNRv5}
with and without NQCs in the RR force.  The triangle, square
and diamond correspond respectively to 3, 1 and 0 cycles before
merger, which for each model is taken as the peak of the $(2,2)$-mode
amplitude.  \texttt{SEOBNRv5} with NQCs matches NR at merger at
expected, but does not agree with NR as well as \texttt{SEOBNRv5}
without NQCs at low frequencies, showing that the addition of the NQCs
does not necessarily improve the flux.

This behavior could be potentially improved by doing more iterations,
finding new values for the NQCs given the corrected calibration
parameters, and repeating the calibration, but would be time-consuming
and not necessarily bring a significant improvement in the waveforms.
Nonetheless, a consistent treatment of the NQCs both in the waveform
and the RR force would most likely provide more faithful
representations of the angular-momentum and energy fluxes, and a more
systematic study across parameter space will be done in a future
update of the model, together with a recalibration of the conservative
dynamics.  Alternative ways to improve the waveform close to merger, 
and reduce the impact of the NQC corrections, should be
also investigated, especially in light of the upcoming eccentric
generalization of this model \texttt{SEOBNRv5EHM}, as past 
experience in developing \texttt{SEOBNRv4EHM} have demonstarted~\cite{Ramos-Buades:2021adz}. 

\section{Comparison against time-domain nonprecessing phenomenological models}
\label{appx:pheonthm}

In this appendix we compare the performance of \texttt{SEOBNRv5HM} and the Fourier domain \texttt{IMRPhenomXHM}~\cite{Pratten:2020fqn} 
against the time-domain nonprecessing phenomenological model \texttt{IMRPhenomTHM} \cite{Estelles:2020osj,Estelles:2020twz}, which
includes the $(\ell, |m|)=(2,2),(3,3),(2,1),(4,4),(5,5)$ modes. 
In particular, we repeat the mismatch calculation against the NR catalog detailed in Sec.~\ref{sec:performance}, 
both for the dominant mode only (Fig.~\ref{fig:mismatch_spaghetti_22_pheonom} and Fig.~\ref{fig:mismatch_hist_22_phenom}) 
and for the full polarizations, at inclination $\iota = \pi/3$ (Fig.~\ref{fig:mismatch_spaghetti_HM_phenom} and Fig.~\ref{fig:mismatch_hist_HM_phenom}). 

Considering the dominant mode mismatches, we see that \texttt{IMRPhenomT} performs slightly
worse than \texttt{SEOBNRv5}, considering both the median mismatch and the fraction of cases below 
$10^{-4}$, while \texttt{IMRPhenomXAS} achieves on average slightly lower values of the maximum unfaithfulness, 
as already noted in Sec.~\ref{sec:performance}. More quantitatively, \texttt{IMRPhenomT} features 
$91\%$ of cases with maximum unfaithfulness below $10^{-3}$, $5\%$ of cases below $10^{-4}$, and 
a median of $2.31 \times 10^{-4}$ (see also Table~\ref{tab:mm_22}).

Considering the mismatches of the full polarizations at inclination $\iota = \pi/3$ 
we note instead that \texttt{IMRPhenomTHM} is slightly more accurate than \texttt{SEOBNRv5HM}, 
and both models are appreciably more accurate than \texttt{IMRPhenomXHM}. 
More specifically, \texttt{IMRPhenomTHM} shows maximum unfaithfulness below $10^{-2}$ for $99\%$ of cases, 
below $10^{-3}$ for $57\%$ of cases, and a median of $7.49 \times 10^{-4}$ (see also Table~\ref{tab:mm_hm}).
The NR simulation \texttt{BFI:ExtremeAligned:0003}, with $q = 10.0, ~\chi_1 = 0.8, ~\chi_2 = 0.5$, 
is the only outlier reaching $3 \%$ mismatch at high total mass, for both 
\texttt{IMRPhenomTHM} and \texttt{SEOBNRv5HM}, while being at the level of $0.1 \%$ when considering only the $(2,2)$ mode.
This suggests common limitations in the modeling of the higher modes for such extreme configurations. 
In Fig.~\ref{fig:mm_l4_appendix} we show the unfaithfulness between \texttt{SEOBNRv5HM}, \texttt{IMRPhenomTHM} and \texttt{BFI:ExtremeAligned:0003},
using multipoles up to $\ell_{\rm{max}} = 5$ and $\ell_{\rm{max}} = 4$, in both the models and the NR waveform. 
We note that the mismatch for high total masses reduces significantly for both models when removing $\ell = 5$ multipoles, 
indicating that part of the disagreement is due to mismodelling of the $(5,5)$ mode, as well as missing contributions in the 
models from the $(\ell=5,m\neq5)$ multipoles, which can have a non-negligible impact in the ringdown of such high mass-ratio binaries.
From the behavior of the unfaithfulness as a function of the total mass in Fig.~\ref{fig:mismatch_spaghetti_HM_phenom} 
we appreciate that \texttt{SEOBNRv5HM} is overall closer to \texttt{IMRPhenomTHM} than to \texttt{IMRPhenomXHM}, 
as expected given the similar merger-ringdown model, and the comparable NR calibration coverage.

\begin{figure*}
	\includegraphics[width=\textwidth]{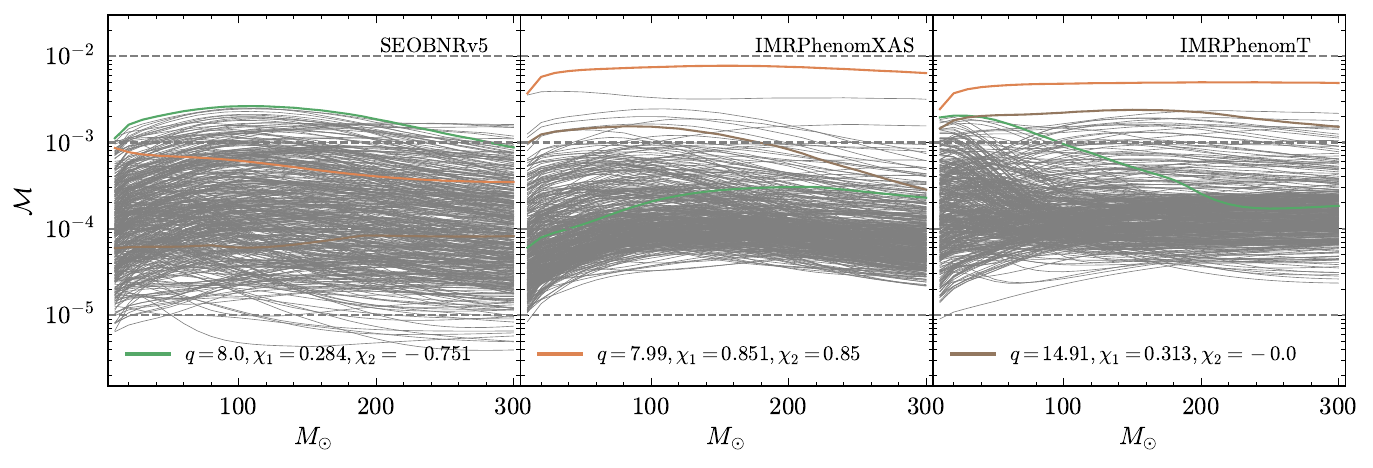}
	\caption{(2,2)-mode mismatch over a range of total masses between 10 and 300 $M_{\odot}$, 
	between \texttt{SEOBNRv5}, \texttt{IMRPhenomXAS}, \texttt{IMRPhenomT} and the 442 NR simulations used in this work.
	The colored lines highlight cases with the worst maximum mismatch for each model.}
	\label{fig:mismatch_spaghetti_22_pheonom}
\end{figure*}

\begin{figure}
	\includegraphics[width=\linewidth]{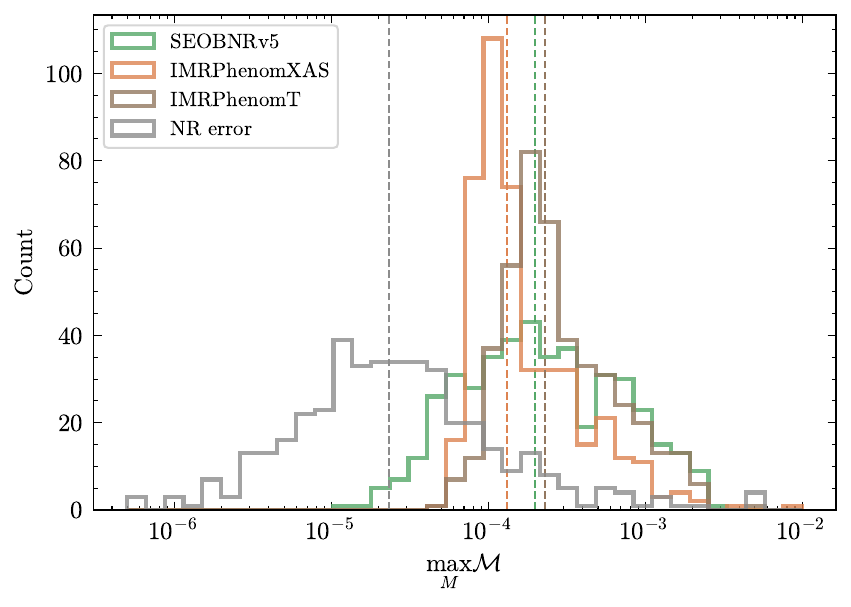}
	\caption{Histogram of the maximum (2,2)-mode mismatch over a range of total masses between 10 and 300 $M_{\odot}$, 
	between \texttt{SEOBNRv5}, \texttt{IMRPhenomXAS}, \texttt{IMRPhenomT} and the 442 NR simulations used in this work.
	The NR error is estimated by computing the mismatch between NR simulations with the highest and second-highest resolutions. 
	The vertical dashed lines show the medians. }
	\label{fig:mismatch_hist_22_phenom}
\end{figure}

\begin{figure*}
	\includegraphics[width=\textwidth]{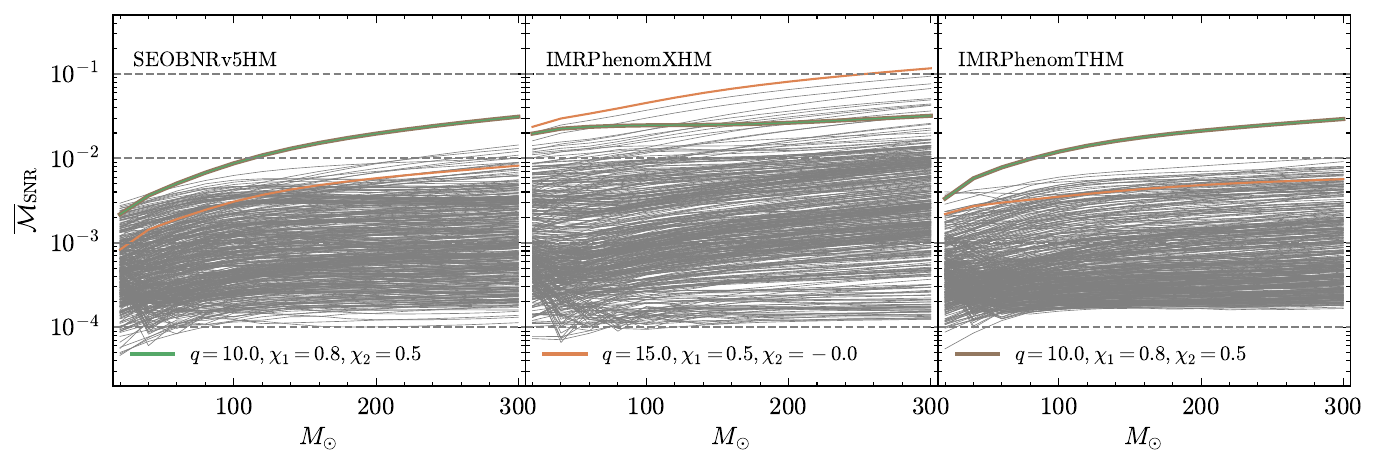}
	\caption{The sky-and-polarization averaged, SNR-weighted mismatch, for inclination $\iota = \pi/3$, over a range of total masses between 20 and 300 $M_\odot$ 
	between \texttt{SEOBNRv5HM}, \texttt{IMRPhenomXHM}, \texttt{IMRPhenomTHM} and the 441 \texttt{SXS} NR simulations used in this work.
	The colored lines highlight cases with the worst maximum mismatch for each model. 
	This comparison highlights the similarity in performance of the time-domain models \texttt{SEOBNRv5HM} and \texttt{IMRPhenomTHM}.}
	\label{fig:mismatch_spaghetti_HM_phenom}
\end{figure*}

\begin{figure}
	\includegraphics[width=\linewidth]{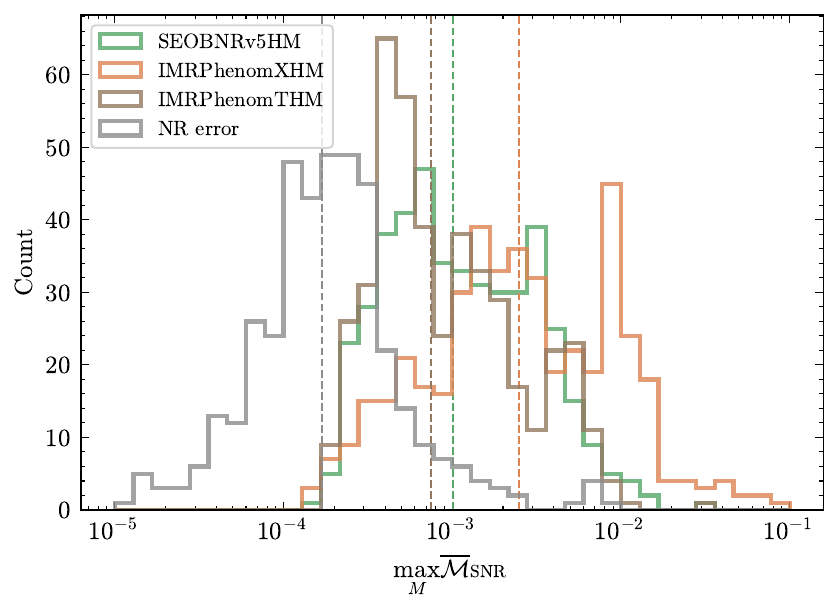}
	\caption{Histogram of the maximum sky-and-polarization averaged, SNR-weighted mismatch, for inclination $\iota = \pi/3$, over a range of total masses between 20 and 300 $M_{\odot}$, 
	between \texttt{SEOBNRv5HM}, \texttt{IMRPhenomXMH}, \texttt{IMRPhenomTHM} and the 441 \texttt{SXS} NR simulations used in this work.
	The NR error is estimated by computing the mismatch between NR simulations with the highest and second-highest resolutions. The vertical dashed lines show the medians.}
	\label{fig:mismatch_hist_HM_phenom}
\end{figure}

\begin{figure}
	\includegraphics[width=\linewidth]{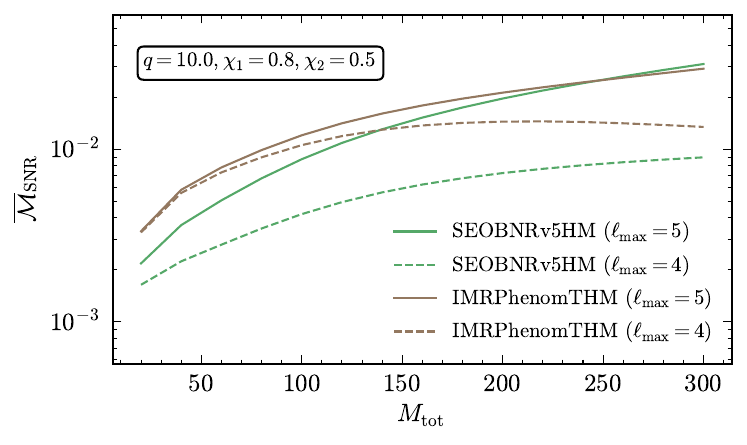}
	\caption{Sky-and-polarization averaged, SNR-weighted mismatch, for inclination $\iota = \pi/3$ as a function of the total mass
	between \texttt{SEOBNRv5HM}, \texttt{IMRPhenomTHM} and the NR simulation \texttt{BFI:ExtremeAligned:0003}, with $q = 10.0, ~\chi_1 = 0.8, ~\chi_2 = 0.5$, 
	using $\ell_{\rm{max}} = 5$ and $\ell_{\rm{max}} = 4$, in both the models and the NR waveform. }
	\label{fig:mm_l4_appendix}
\end{figure}

\bibliography{../references}

\end{document}